\documentclass{siamart190516}

\usepackage{amssymb}

\usepackage{amsmath}

\usepackage[left=2cm,right=2cm,top=2cm,bottom=2cm]{geometry}

\usepackage{color}

\usepackage{graphicx}

\usepackage{subcaption}
\usepackage{enumitem}
\usepackage{hyperref,enumerate}

\usepackage{float}

\newcommand{\bq}{{\bf q}}
\newcommand{\br}{{\bf r}}
\newcommand{\bp}{{\bf p}}
\newcommand{\bb}{{\bf b}}
\newcommand{\ba}{{\bf a}}

\newcommand{\eps}{\varepsilon}

\title{A Discrete-to-Continuum Model of Weakly Interacting Incommensurate Two-Dimensional Lattices: The hexagonal case}

\author{
    Malena I.~Espa\~ nol\footnotemark[1]
\and
    Dmitry Golovaty\footnotemark[2]
\and
    J. Patrick Wilber\footnotemark[2]}

\begin{document}
\maketitle

\renewcommand{\thefootnote}{\fnsymbol{footnote}}
\footnotetext[1]{
    School of Mathematical and Statistical Sciences,
    Arizona State University,
    Tempe,
    AZ 85281, USA.}
\footnotetext[2]{
    Department of Mathematics,
    University of Akron,
    Akron,
    OH 44325, USA.}

\begin{abstract}
In this paper, we extend the discrete-to-continuum procedure we developed in \cite{espanol2018discrete} to derive a continuum variational model for a hexagonal twisted bilayer material in which one layer is fixed. We use a discrete energy containing elastic terms
and a weak interaction term that could utilize either a Lennard-Jones potential or a Kolmogorov-Crespi potential.
To validate our modeling, we perform numerical simulations to compare
the predictions of the original discrete model and the proposed continuum
model, which also show an agreement with experimental findings for, e.g., twisted bilayer graphene.

\end{abstract}


\section{Introduction} \label{intro}

%%%%%%%%%%%%%%%%%%%%%%%%%%%%%%%%%%%%%%%

% brief description, then motivation via buzzwords

In this paper, we apply a discrete-to-continuum procedure to develop a
model that predicts relaxation in a twisted bilayer of hexagonal
atomic lattices.  Relaxation of bilayer graphene, other layered
two-dimensional materials, and van der Waals heterostructures has
attracted significant interest over the last several years
\cite{andrei2020graphene,sunku2018photonic,carr2020electronic}.
Predicting lattice reconstruction and equilibrium configurations is
crucial for understanding some of the fundamental physical phenomena
displayed by bilayers and heterostructures
\cite{carr2018relaxation,carr2019exact,cazeaux2020energy,zhang2018structural,weston2020atomic}.
The quasiperiodic relaxed moir\'{e} patterns that occur in slightly
misaligned or slightly incommensurate lattices induce superlattice
effects, which include superconductivity, strong interactions, and
other novel electronic and optical properties
\cite{cao2018unconventional,yoo2019atomic,li2021continuous,dong2021flat}.
More generally, the study of the mechanics of these nanoscale
structures is driven in part by the possibility of engineering
advanced materials with novel properties by stacking the same or
different types of individual layers in appropriate sequences
\cite{novoselov20162d,sulleirofabrication,xiang2020one,tran2019evidence}.

% Relaxed Moire patterns.

\begin{figure}[htb]
\centering
    \includegraphics[scale = 0.7]{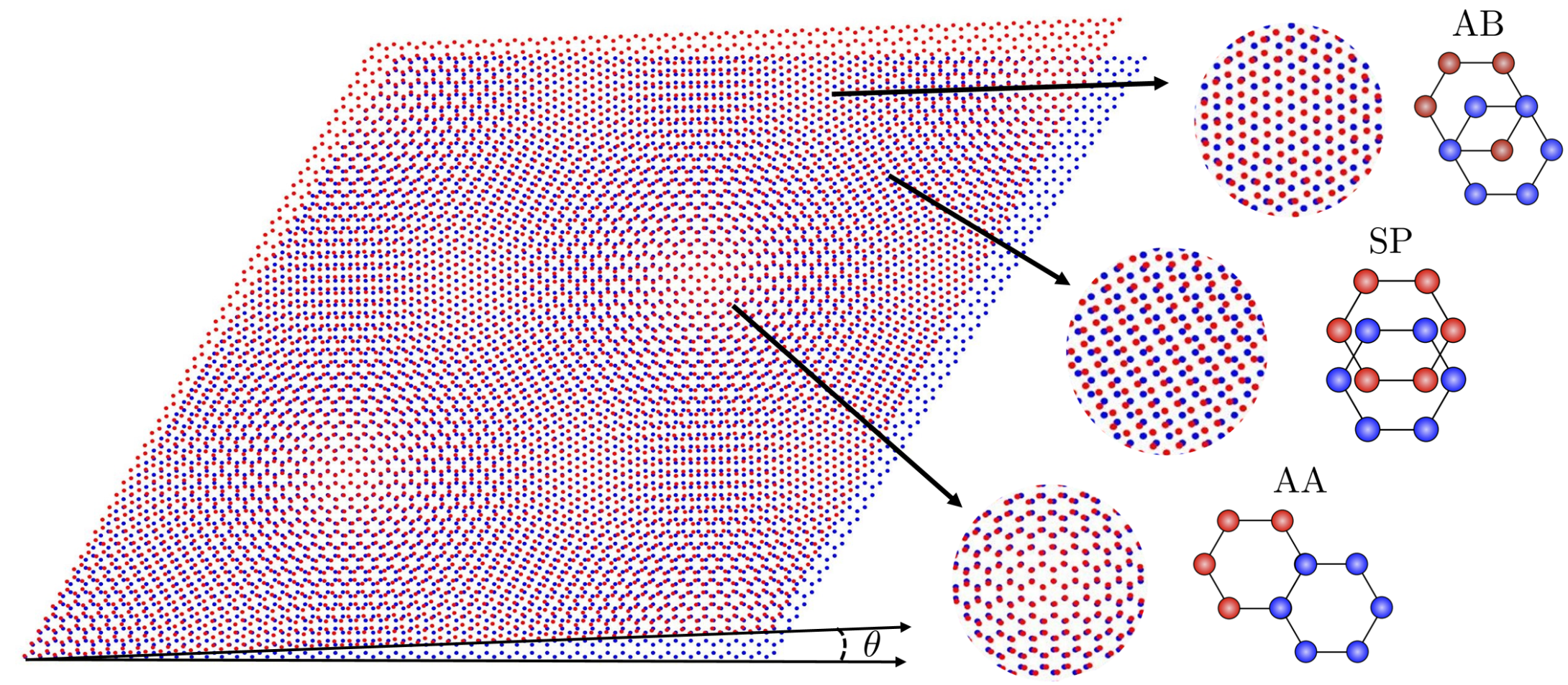}
  \caption{A moir\'{e} pattern formed by bilayer graphene with a twist of angle $\theta$, containing three main stacking configurations: AA, AB, and SP.}
  \label{fig:moire}
\end{figure}

Relaxation can be understood by considering a bilayer of graphene in
which the two layers are given an initial small relative rotation.
Variations in the local stacking between the layers generate a
quasiperiodic moir\'{e} pattern with a period that scales inversely
with the size of the angle of the relative rotation (see
Figure~\ref{fig:moire}).  Mechanically, the bilayer has strong
intralayer bonding and weak van der Waals interactions between the
layers.  As a consequence, the atoms in each lattice adjust through
in-plane and out-of-plane deformations.  As the lattice structure
relaxes, local regions with the energetically favorable AB of BA
alignment grow, while regions with AA alignment decrease in size.
Typically, the relaxed structure exhibits a network of narrow ridges
or wrinkles forming domain walls that separate the relatively large
commensurate regions with AB and BA stacking between the
lattices~\cite{van2015relaxation,van2014moire,PhysRevB.96.075311,jain2016structure,enaldiev2020stacking}.

% Nain results.  

We derive our continuum model starting from an expression for the
discrete energy for two interacting hexagonal lattices.  The lattices
may have different lattice parameters and there may be a slight
relative rotation between the lattices.  For simplicity, we keep the
atoms on one lattice fixed.  In our discrete energy, we model
intralayer interactions between neighboring atoms on the deformable
lattice with extensional, torsional, and dihedral springs.  These
describe how the bonds between neighboring atoms resist stretching and
bending and how the system maintains an hexagonal lattice structure.
The discrete energy also includes a term for the weak interaction
between atoms on different lattices.  This term is based on a pairwise
potential between non-bonded atoms, and we consider both a
Lennard-Jones-type potential and a version of the Kolmogorov-Crespi
potential~\cite{PhysRevB.71.235415}.

To move from the discrete to the continuum, we introduce a small
parameter defined as the ratio of the typical interlayer spacing to
the lateral extent of the parallel lattices.  Exploiting this small
parameter, we develop Taylor expansions of the terms in our discrete
energy.  After appropriately truncating these expansions, a Riemann
sum argument is used to replace sums over the lattice with integrals.
This procedure yields a continuum energy for the interacting bilayer.
The minimizers of this continuum energy represent equilibrium
configurations of the deformable lattice.  We note that our continuum
weak interlayer interaction energy is a version of the generalized
stacking fault energy \cite{vitek,zhou2015van}.  Although a continuum
description, the weak energy retains information about the local
mismatch between the original discrete lattices.

% Description of numerical results. 

We validate the discrete-to-continuum modeling procedure just
described through numerical simulations.  We use the open-source
discrete modeling software LAMMPS~\cite{PLIMPTON19951} to simulate the lattice relaxation
predicted by the discrete model.  Also, we use the multiphysics
software COMSOL~\cite{comsol} to solve numerically the Euler-Lagrange equations
derived from the continuum energy.  For both the Lennard-Jones and the
Kolmogorov-Crespi potentials, we obtain good agreement between the
results of the discrete and continuum numerical simulations for
several parameter regimes.  In particular, we show that our continuum
model can reproduce relaxed moir\'{e} patterns observed in the
simulations based on the discrete model and observed in other
studies~\cite{van2015relaxation,enaldiev2020stacking}.  For the
Lennard-Jones potential, there is poor agreement with the out-of plain
displacement of the domain walls in some parameter ranges.  However,
we show that decreasing the well depth of the potential yields a
better match between the discrete and continuum.  For the
Kolmogorov-Crespi potential, we get good agreement between the
discrete and the continuum.  Our numerical results indicate that this
agreement improves as $\varepsilon$ decreases, as expected.  How well
our continuum model works depends on the size of the elastic
constants.  When the elastic constants are relatively small, the
solutions to the discrete simulations exhibit small scale spatial
oscillations.  This suggests that one of the basic assumptions of our
discrete-to-continuum modeling procedure---that the atomic lattice can
be embedded in a smooth surface---is violated.  In this case, we
observe poor agreement between the predictions of the discrete and
continuum models.  On the other hand, when the elastic constants are
relatively large, small-scale spatial oscillations do not occur in the
discrete solutions.  The continuum model in this case predicts
solutions that agree well with the solutions predicted by the discrete
model.

% literature review

%% our earlier papers

This paper extends the discrete-to-continuum procedure developed in
our previous papers \cite{espanol2017discrete,espanol2018discrete}.
In \cite{espanol2017discrete}, we use a discrete-to-continuum
procedure similar to the procedure in this paper to derive a continuum
variational model for two chains of atoms with slightly incommensurate
lattices.  The continuum model recovers both qualitatively and
quantitatively the behavior observed in the corresponding discrete
model.  The numerical solutions for both models demonstrate the
presence of large commensurate regions separated by localized
incommensurate domain walls.  In \cite{espanol2018discrete}, we
develop a continuum variational model for a two-dimensional deformable square lattice of atoms interacting with a two-dimensional rigid
square lattice.  We use the same discrete-to-continuum procedure as in
this paper.  The two lattices have slightly different lattice
parameters and there is a small relative rotation between them. We show that the continuum model recovers both qualitatively and
quantitatively the behavior observed in the corresponding discrete
model.

% Other relevant modeling papers.

% Srolovitz

In \cite{dai2016structure,dai2016twisted}, the authors present a
multiscale model that predicts the deformation of bilayers of graphene
and bilayers of other two-dimensional materials. In their model, the
total energy of the bilayer has an elastic contribution, associated
with the stretching and bending of the individual layers, and a misfit
energy, which describes the van der Waals interactions between the two
layers. The misfit energy is defined using the generalized
stacking-fault energy for bilayers, which the authors develop in an
earlier publication \cite{zhou2015van} from density-functional theory
calculations.  The misfit energy is a function of the separation and
disregistry between layers. In \cite{dai2016structure}, the authors
use their model to explain the structure of deformed bilayer graphene
in terms of dislocation theory.  The model is applied to determine the
structure and energetics of four interlayer dislocations in bilayer
graphene, where the different cases are determined by the angle
between the Burgers vector and the line of dislocation.  In
\cite{dai2016twisted}, the authors use the model to study deformations that result from a small rotation between the layers. The model
predicts two distinct equilibrium structures, which the authors call a
breathing mode and a bending mode.  The latter, more stable at small
rotation angles, is characterized by a twist in the dislocation
structure near the dislocation nodes, at which there are also large
out-of-plane displacements.  The authors note that this newly
discovered structure has both different symmetry and period from the
classical moir\'{e} structure that is often assumed for rotated bilayer
graphene.

The continuum model we develop in this paper has essential elements in
common with the model presented in
\cite{dai2016structure,dai2016twisted}.  Specifically, our model
contains terms for the elastic energy of the deformable layer and a
term for the van der Waals interactions between the two layers.
However, we derive all terms in our continuum energy by upscaling from
an atomistic description of the problem.  Our upscaling procedure
introduces a small parameter that determines the relative size of the
various contributions to the continuum energy. Hence, we gain insight
into how the balance of these terms produces phenomena like relaxed
moir\'e patterns in interacting bilayers.  Furthermore, our modeling
sets the stage for additional analysis to rigorously determine the
relation between atomistic and continuum descriptions of the problem
\cite{braides2007derivation, braides2014discrete}.

% Tadmor papers

In \cite{zhang2018structural}, the authors use multiscale simulations
to study the structural relaxation in twisted graphene bilayers.  They
also study the electron diffraction patterns associated with the
relaxation.  Their simulations show that the relaxation exhibits a
localized rotation and shrinking of the AA domains.  For small
twisting angles, the localized rotation are approximately a constant,
while for large twisting angles, the rotation scales linearly with the
angle.  The authors use a continuum model to explain their results
theoretically.  A nonlinear elasticity model describes the mechanical
response within a graphene layer.  The elastic energy density consists
of a Saint Venant–Kirchhoff membrane term and a Helfrich bending term.
For the interlayer energy, the authors develop a discrete-continuum
approximation based on the Kolmogorov-Crespi potential.  The
interlayer energy is calculated in two parts.  Locally, the potential
is evaluated exactly over a short-range discrete region.  Outside this region, a continuum integral approximation is used~\cite{zhang2017energy}.

% Luskin papers

In \cite{carr2018relaxation}, the authors study bilayer relaxation by
minimizing a total continuum energy over a collection of all possible
local atomic environments, which they call configuration space.  For
the configuration-space approach, every atomic site in the bilayer is
associated with a vector that describes the local relative stacking
disregistry at that site.  The authors use this approach to study the
relaxation patterns in configuration space.  They present
computational results for small-angle twisted bilayer graphene and
molybdenum disulfide, and demonstrate the computational efficiency of
their method for computing relaxations.  In \cite{cazeaux2020energy},
these authors extend the configuration-space approach from bilayers to
general weakly coupled incommensurate deformable multi-layers.  Their
main result is the derivation of an elastic model for the relaxation
of vertical stacks of any number of incommensurate, weakly coupled
deformable layers.  When specialized to a bilayer heterostructure, the
model in \cite{cazeaux2020energy} reduces to a well-posed variational
problem for the continuum displacement field on a periodic moir\'e
domain even for aperiodic atomistic configurations.  Their model in
this case is similar to the continuum model we derive in this paper,
although our approach is entirely different from theirs.

% The couple of other papers I found

In \cite{halbertal2021moire}, a model similar to that developed in
\cite{carr2018relaxation} is used to describe how the shapes of
moir\'e domains and domain boundaries yield information about the
generalized stacking fault energy function at the low twist-angle
limit.

In \cite{enaldiev2020stacking}, the authors use multiscale modeling to
study lattice reconstruction in twisted bilayers of transition metal
dichalcogenides at low twist angles.  Using density functional theory,
they develop interpolation formulae for the interlayer adhesion
energies of the bilayers.  The authors combine the interlayer adhesion
energies with elasticity theory.  The resulting model is used to
analyze the mesoscale domain structures formed during lattice
relaxation.

This paper is organized as follows.  In Section~\ref{sec: atomistic model}, we formulate a
discrete energy of the system of a graphene sheet over a substrate.
In Section \ref{sec: continuum model} we derive a continuum energy that keeps track of the
mismatch of the spacing between the atoms on each curve.  Section \ref{s:results}
includes numerical results that compare the atomistic model with the
continuum model.  We summarize in Section \ref{s:conclusion}.

\section{Atomistic Model} \label{sec: atomistic model}

We consider a discrete system that consists of parallel
2-dimensional atomic lattices, $\hat{\mathcal A}_1$ and
$\hat{\mathcal A}_2$, both infinite in extent.  
The atoms in $\hat{\mathcal A}_2$ can move and each of these atoms
interacts with its neighbors within $\hat{\mathcal A}_2$ via a 
strong bond potential.  $\hat{\mathcal A}_2$ describes a layer of a
2-dimensional material that is nearly inextensible and has a finite
resistance to bending.  In the absence of interactions with atoms on
$\hat{\mathcal  A}_1$, the atoms in
$\hat{\mathcal  A}_2$ in equilibrium form a flat hexagonal lattice that has lattice parameter
$h_2$.
The atoms in $\hat{\mathcal A}_1$ are fixed and form a flat hexagonal lattice
with lattice parameter $h_1$.  In this work, $\hat{\mathcal A}_1$
describes a rigid substrate.  All atoms in $\hat{\mathcal A}_2$ are
assumed to interact with all atoms in $\hat{\mathcal A}_1$ via a weak
interatomic potential.  Below we refer to
$\hat{\mathcal A}_1$ as the rigid lattice and to $\hat{\mathcal A}_2$
as the deformable lattice.

\begin{figure}[htb]
\centering
    \includegraphics[scale = 0.6]{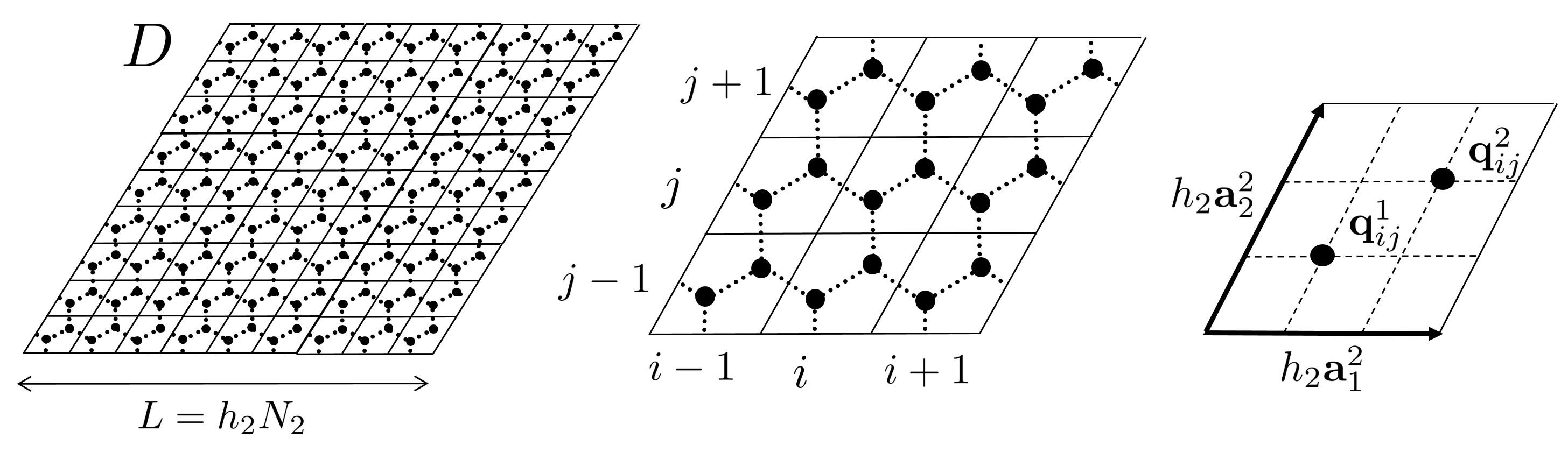}
  \caption{Lattice $\hat{\mathcal A}_2$. Left: $\hat{\mathcal A}_2$
    consists of $N_2^2$ cells. Middle: each cell is indexed by $i,j$. Right: the cell $i,j$ contains two atoms $\bq^1_{ij}$ and $\bq^2_{ij}$. The lattice vectors  $\ba^2_1$ and $\ba^2_2$ are also shown.}
  \label{fig:hex}
\end{figure}

We assume that $\hat{\mathcal A}_2$ deforms periodically and that, in
its reference configuration, one periodic cell of
$\hat{\mathcal A}_2$
occupies a parallelogram-shaped, planar domain $\mathcal A^{0}_2$ with sides
of length $L$.  To define $\mathcal A^{0}_2$,
we set
$\ba^2_1 = (1, 0)$ and $\ba^2_2 = (1/2,\sqrt{3}/2)$, we let
$A$ be the $2 \times 2$ matrix with 
$\ba^2_1$ and $\ba^2_2$ as its columns, and we define
\begin{equation}
  D
  =
  \left\{
  \mathbf{x} = A \mathbf{y}
  \mbox{ with }  \mathbf{y} \in [0,L]^2
  \right\}.
  \label{ee104}
\end{equation}
Then
$\mathcal A^{0}_2 =
\left\{
  \left({\bf x},\sigma\right)
  \colon
  {\bf x}=(x_1,x_2)\in D
\right\}$.
We set $N_2=L/h_2$.  The set $D$ can be divided into $N_2^2$
unit cells each containing two atoms.  See Figure~\ref{fig:hex}.
The deformed and reference positions of the $(2N_2)^2$ atoms in 
$\mathcal A^{0}_2$ are given by the set of vectors $${\bf
  Q}:=\left\{\bq^1_{ij}\right\}_{i,j=1}^{N_2}\cup
\left\{\bq^2_{ij}\right\}_{i,j=1}^{N_2} \subset \mathbb{R}^{3} \mbox{
  and } {\bf R}:=\left\{\br^1_{ij}\right\}_{i,j=1}^{N_2}\cup
\left\{\br^2_{ij}\right\}_{i,j=1}^{N_2} \subset \mathbb{R}^{3},$$
respectively.
Then the positions of the atoms in the reference configuration
are  
\begin{equation}
\label{eq:R}
\br^k_{ij}=\left(h_2\left((i+ k/3)\,{\bf a}^2_1+(j + k/3)\,{\bf
  a}^2_2\right),\sigma\right) \mbox{ for } k=1,2,
\end{equation}
and the positions of the atoms in the deformed configuration are
\begin{equation}
\label{eq:qs}
\bq^k_{ij}= \br^k_{ij} + ({\bf u}^k_{ij},v^k_{ij}) \mbox{ for } k=1,2,
\end{equation}
where ${\bf u}^k_{ij}$ is the displacement in the $x,y$ plane, and
$v^k_{ij}$ in the $z$-direction.  Because $\hat{\mathcal A}_2$ deforms
periodically, we identify $i=N_2+1$ with $i=1$ and $j=N_2+1$ with $j=1$.

For the rigid lattice, the current and the reference configurations
are the same.
To describe these, we define the rotation matrix
\[R(\theta)=
\left(
\begin{array}{ccc}
 \cos{\theta} & -\sin{\theta}   \\
 \sin{\theta} & \cos{\theta}
\end{array}
\right)
\]
and we define $\ba^1_i$ as the transpose of $R(\theta)(\ba^2_i)^{T}$ for $i=1,2$.
We assume that there is a unit cell 
of $\hat{\mathcal A}_2$ whose lower left corner
sits directly above the lower left corner
of a unit cell in $\hat{\mathcal A}_1$.
Then, the positions of atoms on the rigid lattice
are
${\bf P}:=\left\{{\bf p}^1_{\ell
  m}\right\}_{l,m=-\infty}^{\infty}\cup \left\{{\bf p}^2_{\ell
  m}\right\}_{\ell,m=-\infty}^{\infty}\subset \mathbb{R}^{3}$,
where  
\begin{equation}
\label{eq:P}
\bp^{\tilde k}_{\ell m}=\left(h_1\left((\ell + \tilde k/3)\,{\bf a}^1_1+(m + \tilde k/3)\,{\bf a}^1_2\right),0\right) \mbox{ for } \tilde k=1,2.
\end{equation}
Note that we assume that the rigid lattice is infinite in extent in
order to appropriately compute the nonlocal weak
energy.

We assume that, in the reference configuration, the lattices are
planar and parallel. 
Also, the lattices are separated by a distance
$\sigma$, where $\sigma$ is a length scale associated with the weak
potential.
When $h_1=h_2$ (e.g., bilayer graphene)
and $\theta=0$, the equilibrium configuration of the
system is $\hat{\mathcal A}_2$ shifted relatively to
$\hat{\mathcal A}_1$ such that half of the atoms in $\hat{\mathcal
  A}_2$ sit above the centers of the hexagons in
$\hat{\mathcal A}_1$ (AB stacking).
The system would be in {\it global registry}.  In this paper, we 
consider the situation where $\hat{\mathcal A}_1$
and $\hat{\mathcal A}_2$ in the reference configuration have slightly
different orientations (small values of $\theta$) and/or when they
have slightly different lattice parameter ($h_1\neq h_2$, but
$|h_1-h_2|/h_1\ll1$).  See Figure~\ref{fig:moire}.
%In this case, global registry is
%not attained in a flat undeformed configuration.  It follows that in
%order to achieve equilibrium, the deformable lattice would have to
%adjust and have different local regions formed by AA, AB, BA, and SP
%stackings (). 

%For instance, for the case of graphene on a MoS2 substrate, we have that $h_1=3.18$ and $h_2= 2.47$. 

We assume that the total energy of the discrete system depends on the
position of the atoms in the deformable lattice and is given
by  
\begin{equation}
E({\bf Q}): = E_s({\bf Q})  + E_t({\bf Q})  + E_d({\bf Q})  + E_w({\bf Q}).
\label{eq:entot}
\end{equation}
That is, the total energy is the sum of the intralayer energy---composed
of the \emph{stretching} $E_s$, \emph{torsional} $E_t$, and
\emph{dihedral} $E_d$ energies---and the interlayer (weak) energy $E_w$.

\begin{figure}
\centering
\includegraphics[scale=0.7] {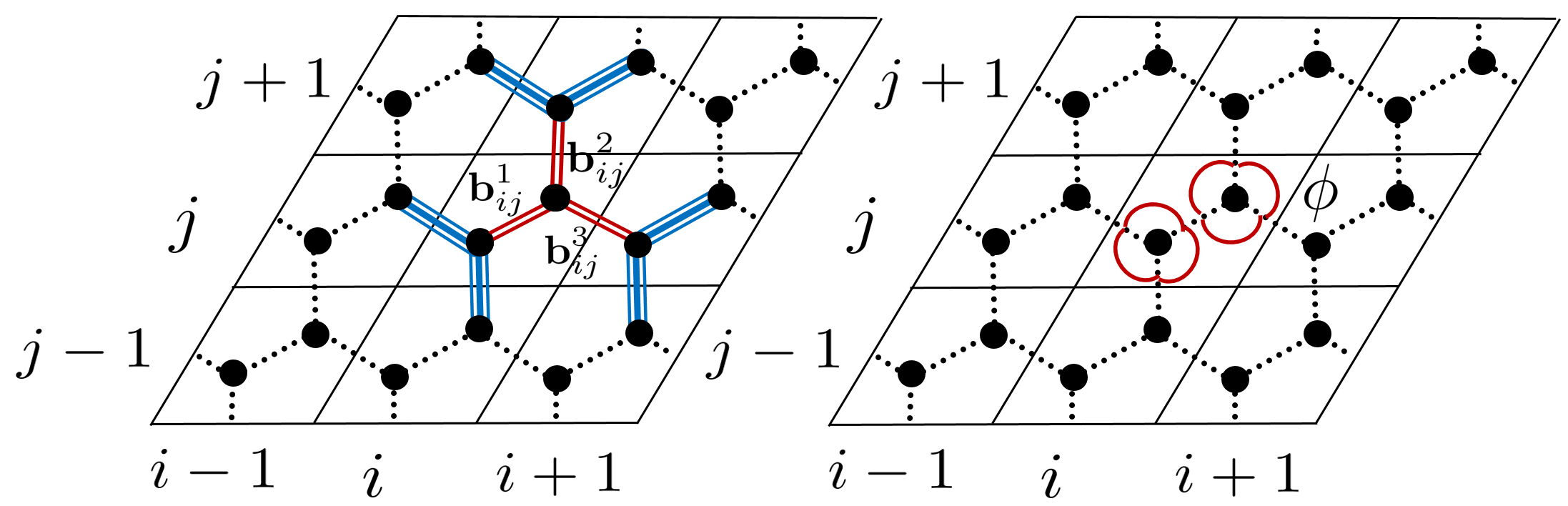}
  \caption{Left: The lattice with the main three bonds (red double lines) corresponding to the $i,j$ cell, used to define the stretching energy and the adjacent bonds (blue triple lines) used to defined the dihedral angles. Right: The six angles $\phi$ (red solid lines) corresponding to the $i,j$ cell used to define the torsional energy.}
  \label{fig:bonds}
\end{figure}

The stretching energy $E_s$ is the energy associated with stretching or
compressing bonds between neighboring atoms in $\hat{\mathcal A}_2$.  
Using a harmonic potential, we define 
\begin{equation}\label{eq:DiscreteE_s}
E_s({\bf Q}) : = \sum_{i,j=1}^{N_2} \frac{k_s}{2}\left[\left( \frac{\|\bb^1_{ij}\|-\frac{h_2}{\sqrt{3}}}{\frac{h_2}{\sqrt{3}}}\right)^2+\left( \frac{\|\bb^2_{ij}\|-\frac{h_2}{\sqrt{3}}}{\frac{h_2}{\sqrt{3}}}\right)^2 + \left( \frac{\|\bb^3_{ij}\|-\frac{h_2}{\sqrt{3}}}{\frac{h_2}{\sqrt{3}}}\right)^2\right], 
\end{equation}
where the vectors 
$$\bb^1_{ij} = \bq^1_{ij}-\bq^2_{ij},\quad \bb^2_{ij} = \bq^1_{ij+1}-\bq^2_{ij},\quad\mbox{and}\quad \bb^3_{ij} = \bq^1_{i+1j}-\bq^2_{ij}$$
represent the bonds between the atom $\bq^2_{ij}$ and its neighbors for every cell
$i,j=1,\ldots,N_2$ (see Figure~\ref{fig:bonds} (left)).  In \eqref{eq:DiscreteE_s}, $k_s$ is
the spring constant.  

The torsional energy is the energy associated with
changing the angle between adjacent bonds.
We model this energy by assuming we have torsional springs
between adjacent bonds.  Each atom is related to three torsional
springs, and therefore each cell contains 6 torsional springs.  The
torsional energy is 
\begin{multline}
  E_{t}({\bf Q})
  :=
  \sum_{i,j=1}^{N_2} \frac{k_t}{2}\left[\left(\phi\left(\bb^1_{ij},\bb^2_{ij}\right)-2\pi/3\right)^2+\left(\phi\left(\bb^2_{ij},\bb^3_{ij}\right)-2\pi/3\right)^2+\left(\phi\left(\bb^1_{ij},\bb^3_{ij}\right)-2\pi/3\right)^2\right.\\ \left. 
 + \left(\phi\left(-\bb^3_{i-1j},-\bb^1_{ij}\right)-2\pi/3\right)^2+\left(\phi\left(-\bb^2_{ij-1},-\bb^1_{ij}\right)-2\pi/3\right)^2+\left(\phi\left(-\bb^3_{i-1j},-\bb^2_{ij-1}\right)-2\pi/3\right)^2\right], \label{Et}
\end{multline}
where $k_t$ is the torsional spring constant and $\phi({\bf a},{\bf c})$ is the angle between the vectors ${\bf a}$ and ${\bf c}$. Right of Figure~\ref{fig:bonds} shows the six angles corresponding to the cell $i,j$.

Assuming that admissible
in-plane deformations of $\hat{\mathcal A}_2$ are small, we can use
the approximation  
$$\phi = \arccos(x) \approx \arccos(-1/2) + \frac{1}{\sin(\arccos(-1/2))}(x+1/2) = \frac{2\pi}{3} + \frac{2}{\sqrt{3}}\left(\cos(\phi) + \frac{1}{2} \right)$$
to rewrite $E_{t}$ as
\begin{multline}
  E_{t}({\bf Q})
  =
  \sum_{i,j=1}^{N_2} \frac{k_t}{2} \frac{4}{3} \left[\left(\frac{\bb^1_{ij}\cdot\bb^2_{ij}}{\|\bb^1_{ij}\|\|\bb^2_{ij}\|}+\frac{1}{2}\right)^2 +\left(\frac{\bb^2_{ij}\cdot\bb^3_{ij}}{\|\bb^2_{ij}\|\|\bb^3_{ij}\|}+\frac{1}{2}\right)^2+\left(\frac{\bb^1_{ij}\cdot\bb^3_{ij}}{\|\bb^1_{ij}\|\|\bb^3_{ij}\|}+\frac{1}{2}\right)^2\right.\\ \left. +\left(\frac{\bb^3_{i-1j}\cdot\bb^1_{ij}}{\|\bb^3_{i-1j}\|\|\bb^1_{ij}\|}+\frac{1}{2}\right)^2+\left(\frac{\bb^2_{ij-1}\cdot\bb^1_{ij}}{\|\bb^2_{ij-1}\|\|\bb^1_{ij}\|}+\frac{1}{2}\right)^2+\left(\frac{\bb^3_{i-1j}\cdot\bb^2_{ij-1}}{\|\bb^3_{i-1j}\|\|\bb^2_{ij-1}\|}+\frac{1}{2}\right)^2\right].
\label{ede2.6}
\end{multline}

\begin{figure}
\centering
\includegraphics[scale=0.35] {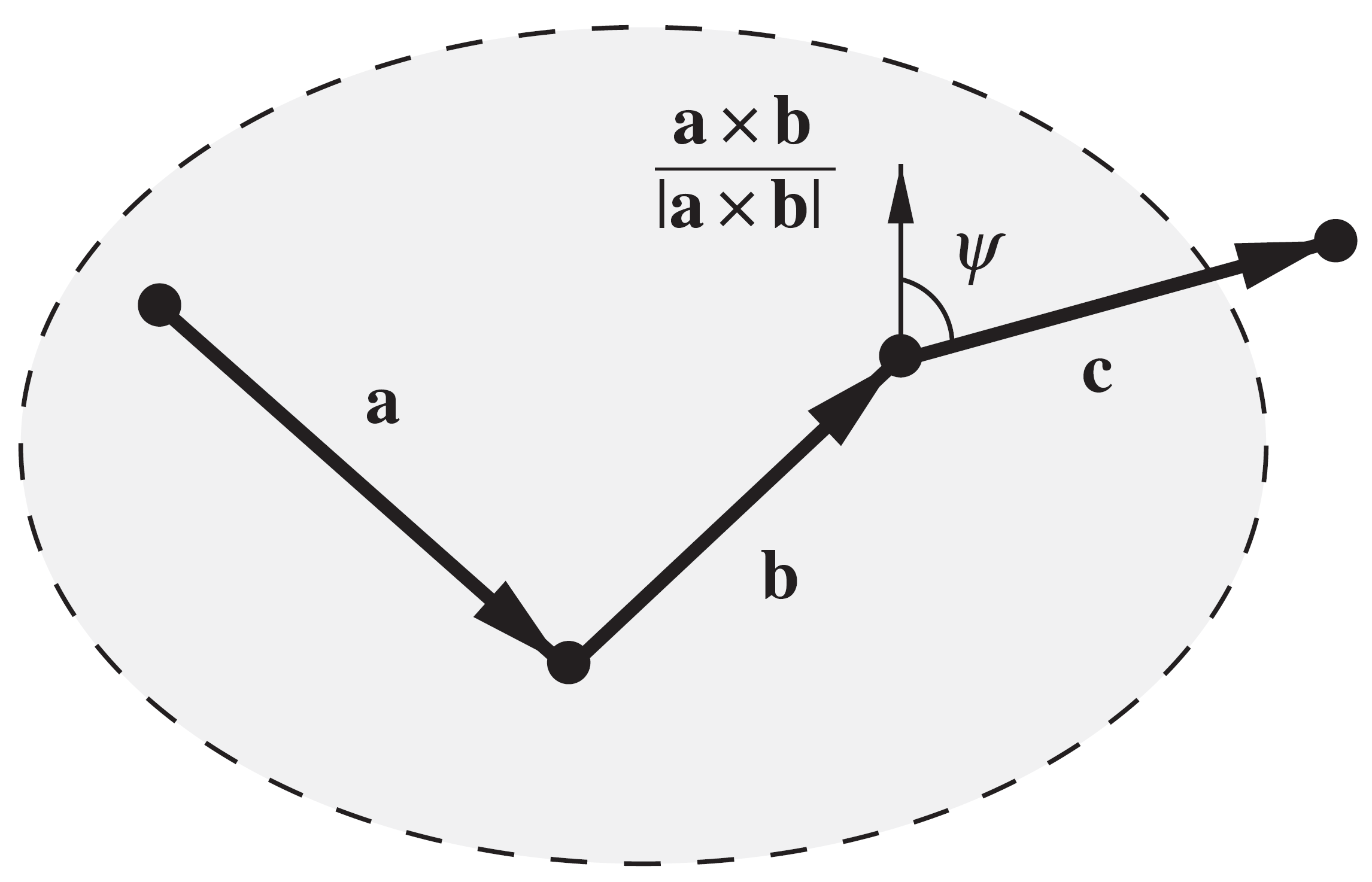}
  \caption{A dihedral spring connecting vectors ${\bf a}$, ${\bf b}$, and ${\bf c}$. The spring energy is minimized when $\psi= \pi/2$.}
  \label{fig:dih}
\end{figure}

Certain out-of-plane deformations are not penalized by the extensional
and torsional energies.  For example, there is no energy cost for
folding along a direction parallel to the sides of the domain $D$.
To penalize for such deformations, we introduce the dihedral energy by
assuming that a dihedral spring connects every triplet of adjacent
bonds.  This spring energy is minimized when the third bond lies in
the plane formed by the first two bonds.  See Figure~\ref{fig:dih}.
We assume that the energy of a dihedral spring is
\[e_d({\bf
  a,b,c}):=\frac{k_d}{2}\cos^2\psi=\frac{k_d}{2}\frac{{\left(({\bf
      a}\times{\bf b})\cdot{\bf c}\right)}^2}{{\|{\bf a}\times{\bf
      b}\|}^2{\|{\bf c}\|}^2},
\]
where $k_d$ is the dihedral spring constant and $\psi$ is the dihedral angle defined as in Figure~\ref{fig:dih}.
For each $i,j=1,\ldots,N_2$, each bond $\bb^p_{ij}$, $p=1,2,3$ is the `middle' bond in 4 different triplets of adjacent
bonds.  See Figure~\ref{fig:bonds}.

%
%Note that the actual number of the dihedral
%springs connected to this and all other atoms is larger due to
%periodicity of the lattice.  
%
Hence the total dihedral energy is
\begin{align}
  E_{d}({\bf Q})
  :=
  \sum_{i,j=1}^{N_2} \frac{k_d}{2} &
  \left\{
    \frac{{\left((\bb^1_{ij}\times\bb^2_{ij})\cdot\bb^2_{ij-1}\right)}^2}{{\|\bb^1_{ij}\times\bb^2_{ij}\|}^2{\|\bb^2_{ij-1}\|}^2}+
    \frac{{\left((\bb^1_{ij}\times\bb^2_{ij})\cdot\bb^3_{i-1j}\right)}^2}{{\|\bb^1_{ij}\times\bb^2_{ij}\|}^2{\|\bb^3_{i-1j}\|}^2} +
    \frac{{\left((\bb^1_{ij}\times\bb^3_{ij})\cdot\bb^2_{ij-1}\right)}^2}{{\|\bb^1_{ij}\times\bb^3_{ij}\|}^2{\|\bb^2_{ij-1}\|}^2} +\right. \nonumber\\
 &\quad
    \frac{{\left((\bb^1_{ij}\times\bb^3_{ij})\cdot\bb^3_{i-1j}\right)}^2}{{\|\bb^1_{ij}\times\bb^3_{ij}\|}^2{\|\bb^3_{i-1j}\|}^2} +
 \frac{{\left((\bb^2_{ij}\times\bb^3_{i-1j+1})\cdot\bb^1_{ij}\right)}^2}{{\|\bb^2_{ij}\times\bb^3_{i-1j+1}\|}^2{\|\bb^1_{ij}\|}^2}+
    \frac{{\left((\bb^2_{ij}\times\bb^3_{i-1j+1})\cdot\bb^3_{ij}\right)}^2}{{\|\bb^2_{ij}\times\bb^3_{i-1j+1}\|}^2{\|\bb^3_{ij}\|}^2}  + \nonumber\\ 
& \quad
   \frac{{\left((\bb^2_{ij}\times\bb^1_{ij+1})\cdot\bb^1_{ij}\right)}^2}{{\|\bb^2_{ij}\times\bb^1_{ij+1}\|}^2{\|\bb^1_{ij}\|}^2}
    +\frac{{\left((\bb^2_{ij}\times\bb^1_{ij+1})\cdot\bb^3_{ij}\right)}^2}{{\|\bb^2_{ij}\times\bb^1_{ij+1}\|}^2{\|\bb^3_{ij}\|}^2}+
 \frac{{\left((\bb^3_{ij}\times\bb^2_{ij})\cdot\bb^1_{i+1j}\right)}^2}{{\|\bb^3_{ij}\times\bb^2_{ij}\|}^2{\|\bb^1_{i+1j}\|}^2} +\nonumber \\ 
& \left. \quad
    \frac{{\left((\bb^3_{ij}\times\bb^2_{ij})\cdot\bb^2_{i+1j-1}\right)}^2}{{\|\bb^3_{ij}\times\bb^2_{ij}\|}^2{\|\bb^2_{i+1j-1}\|}^2}
    +\frac{{\left((\bb^3_{ij}\times\bb^1_{ij})\cdot\bb^1_{i+1j}\right)}^2}{{\|\bb^3_{ij}\times\bb^1_{ij}\|}^2{\|\bb^1_{i+1j}\|}^2}
    +\frac{{\left((\bb^3_{ij}\times\bb^1_{ij})\cdot\bb^2_{i+1j-1}\right)}^2}{{\|\bb^3_{ij}\times\bb^1_{ij}\|}^2{\|\bb^2_{i+1j-1}\|}^2}
\right\}. \label{e_dih}
\end{align}

The bending between interatomic bonds is penalized by introducing
harmonic torsional springs and dihedral angles between the bonds.  
The expressions for the extensional and torsional springs,
respectively, show that the sum of the corresponding energy components
is minimized when the atoms on $\hat{\mathcal A}_2$ form a hexagonal
lattice with sides of length $h_2/\sqrt{3}$.

Finally, we consider 2 choices for the energy of the weak interaction between
$\hat{\mathcal A}_1$ and $\hat{\mathcal A}_2$.  Our first choice is to
define
\begin{equation}\label{eq:DiscreteE_w}
  E_w({\bf Q}) 
  =
  \omega \sum_{i,j=1}^{N_2} \sum_{k=1}^{2}
  \sum_{\ell,m=-\infty}^{\infty} \sum_{\tilde k=1}^{2}
  g_{{\rm LJ}}\left(\frac{\|\bq^k_{ij}-\bp^{\tilde k}_{\ell m}\|}{\sigma}\right),
\end{equation} 
where $g_{{\rm LJ}}$ is the classical Lennard-Jones 12-6 potential
\begin{equation}
g_{{\rm LJ}}(r) = r^{-12}-2r^{-6}.\label{ee1.0}
\end{equation} 
The parameters $\sigma$ and $\omega$ define the equilibrium
interatomic distance and the strength of the Lennard-Jones potential, respectively.
%For carbon–carbon interactions
%in a graphene bilayer, $\omega =2.39$ meV and $\sigma =3.41\angstrom$
%\cite{Kozio__2019}.
Note that the inner double sum in \eqref{eq:DiscreteE_w}
is taken over the entire rigid lattice to properly account for weak
interactions between the lattices.

The Lennard-Jones potential fails to adequately account for the
registry dependence in the interaction between bilayers.  The
Kolmogorov-Crespi potential addresses this deficiency
\cite{PhysRevB.71.235415}. Therefore, the second choice we consider for the weak
interaction is
\begin{equation}\label{eq:DiscreteE_w_kc}
  E_w({\bf Q}) 
  =
  \omega \sum_{i,j=1}^{N_2} \sum_{k=1}^{2}
  \sum_{\ell,m=-\infty}^{\infty} \sum_{\tilde k=1}^{2}
  g_{{\rm KC}}
  \left(\frac{\|\bq^k_{ij}-\bp^{\tilde k}_{\ell m}\|}{\sigma},
        \frac{\hat{\rho}(\bq^k_{ij}-\bp^{\tilde k}_{\ell m})}{\delta}
  \right),
\end{equation} 
where $g_{{\rm KC}}$ is a version of the Kolmogorov-Crespi potential having the form 
\begin{equation}
  g_{{\rm KC}}(r,\rho)
  =
  {\rm e}^{-\tilde\lambda \left(r - 1\right)}\left[ \tilde C + 2f(\rho)\right]-r^{-6}
  \quad
  \text{with}
  \quad  
  f(\rho)
  =
  e^{-\rho^{2}}
  \left(
  \tilde C_0
  +
  \tilde C_2 \rho^2
  +
  \tilde C_4 \rho^4
  \right).
  \label{ee2}
\end{equation} 
In \eqref{eq:DiscreteE_w_kc}, $\hat{\rho}(\mathbf{a})$ is the length
of the projection of
$\mathbf{a}$ onto the $xy$-plane, which is the plane containing the
fixed lattice $\hat{\mathcal  A}_1$.  For $g_{{\rm KC}}$, $\omega$ can be associated
with the strength of the potential.  The parameters $\sigma$ and
$\delta$ are lengths;  $\sigma$ is related to the equilbrium spacing
between the layers. The constants $\tilde{\lambda}$, $\tilde{C}$, $\tilde{C}_{0}$,
$\tilde{C}_{2}$, and $\tilde{C}_{4}$ are dimensionless parameters
\cite{PhysRevB.71.235415}.   
%For
%carbon–carbon interactions in a graphene bilayer, $\omega=10.238$ meV,
%and $\sigma=3.34$\angstrom$, \tilde C_0=15.71/\omega$ meV, $\tilde
%C_2=12.29/\omega$ meV, $\tilde C_4=4.933/\omega$ meV, $C=3.030$ meV,
%$\delta=0.578\angstrom$, and $\lambda = 3.629 \angstrom^{-1}$\cite{}. 

\section{Continuum Model} \label{sec: continuum model}

First, we briefly describe the approach we take to derive the
continuum model.  We assume that the atoms on the deformable lattice
$\hat{\mathcal A}_2$ are embedded in a smooth surface $\mathcal
A_2\subset\mathbb R^3$ and we describe this surface parametrically in
terms of the displacement field.  Nondimensionalizing the discrete
problem introduces a small geometric parameter $\varepsilon=\sigma/L$,
equal to the ratio of the equilibrium distance of the weak interaction
to the length of the side of the domain $D$.  Evaluating the
displacements at atomic positions, substituting these into the
expression \eqref{eq:entot} for the discrete energy, expanding the
result in terms of $\varepsilon$, and converting summation into
integration, leads to an expansion in terms of $\varepsilon$ for the
continuum energy, written as a functional of the displacement field.

We identify the leading-order terms in this expansion, up to the order
at which contributions from the extensional, torsional, and dihedral
springs, as well the van der Waals interactions are included.  The
resulting continuum energy is of Ginzburg-Landau type and contains
terms of different powers in $\varepsilon$.  The minimizers of the
continuum energy typically exhibit bulk regions of registry, separated
by thin walls where the gradient of the displacement field is
large. Thus, within the walls, the contributions from higher-order
terms generally cannot be neglected. We choose to cut off the
expansion that leads to the continuum energy at the order when all
components of the displacement contribute to the energy density inside
the walls at leading order. Finally, in the next section, we present
the results of simulations confirming that the behavior of minimizers
of the continuum energy match that of minimizers of the discrete
energy.

We assume that the deformed configuration of a periodic
cell of $\hat{\mathcal A}_2$ is embedded in a sufficiently smooth
surface
$\mathcal A_2
  =
  \left\{
  \left({\bf x}+{\bf u}({\bf x}), \sigma + v({\bf x})\right)
  \colon
  ({\bf x}, \sigma) \in \mathcal A^{0}_2
  \right\},$
where $({\bf u}({\bf x}), v({\bf x}))$ % =(u_1(x_1,x_2),u_2(x_1,x_2),v(x_1,x_2))$
is the displacement of the point $({\bf x},\sigma)$
on $\mathcal A^{0}_2$.
%Hence, the deformed
%surface $\mathcal A_2$ is given by
%%
%\begin{equation}
%  \left\{({\bf x}+{\bf u}({\bf x}), \sigma + v({\bf x}))=(x_1+u_1(x_1,x_2),x_2+u_2(x_1,x_2),\sigma+v(x_1,x_2))\colon (x_1,x_2,\sigma) \in \mathcal A^{0}_2\right\}. 
%  \label{ee12}
%\end{equation}
%%
%
Next, we assume $\sigma<<L$, that is, the length scale associated with the
equilibrium spacing between the lattices is much smaller than the
lateral extent of the system.  Then, we define $\varepsilon =
\sigma/L$ and introduce the rescalings
\begin{equation}
   \boldsymbol\chi = \frac{\bf x}{L},\ \ 
  \boldsymbol\xi = \frac{\bf u}{\varepsilon L},\ \ 
  \eta = \frac{v}{\varepsilon L}, \ \
  \mathcal E=\frac{\varepsilon E}{\omega}.
  \label{ee1}
\end{equation}
We define the nondimensional parameters
\begin{equation}
  \delta_{1} = \frac{h_{1}}{\sigma},\ \ 
  \delta_{2} = \frac{h_{2}}{\sigma},\ \ 
  \gamma_s=\frac{6\sqrt{3}k_s}{\omega\delta_2^2}, \ \
  \gamma_t=\frac{64\sqrt{3}k_t}{\omega\delta_2^2}, \ \
  \gamma_d=\frac{\sqrt{3}k_d}{4\omega}.
  \label{ee1.1}
\end{equation}
The numerical coefficients in the definitions of $\gamma_s$,
$\gamma_t$, and $\gamma_d$ are explained later.
The scaling for the displacements are appropriate for small
deformations considered here. With a slight abuse of notation, we now 
set
\begin{equation}
  \mathcal A^{0}_2
  =
  \left\{
  \left(\boldsymbol\chi,\varepsilon\right)
  \colon
  %\boldsymbol\chi = A_2{\bf y}
  %\mbox{ with }  {\bf y} \in [0,1]^2
  \boldsymbol\chi \in D_{1}
  \right\},
  \quad
  \mathcal A_2
%  &=
%  \left\{(\chi_1+\varepsilon \xi_1(\boldsymbol\chi),\chi_2+\varepsilon \xi_2(\boldsymbol\chi),\varepsilon+\varepsilon \eta(\boldsymbol\chi))\colon \boldsymbol\chi=(\chi_1, \chi_2)=A_2\boldsymbol\psi \mbox{ with } \boldsymbol\psi \in[0,1]^2\right\} \nonumber \\
  =
  \left\{\left(\boldsymbol\chi+\varepsilon
         {\boldsymbol\xi}(\boldsymbol\chi),\varepsilon+\varepsilon
         \eta(\boldsymbol\chi)\right)
         \colon
         (\boldsymbol\chi,\varepsilon)\in\mathcal A^{0}_2
         \right\},
  \label{ee13}
\end{equation}
where $D_1$ is the rescaled parallelogram $D$ now of side 1.
We assume that $\delta_i=\mathcal{O}(1),\ i=1,2$, that is, the lattice parameters for $\hat{\mathcal A}_1$ and $\hat{\mathcal A}_2$ are comparable to the distance
between $\hat{\mathcal A}_1$ and $\hat{\mathcal A}_2$ (and hence both are much smaller than the lateral extent of the
system).  Furthermore, in order to observe the registry effects on a macroscale,
we assume that
\begin{equation}
  \alpha:=\frac{\delta_{1}-\delta_{2}}{\varepsilon\delta_2}=\mathcal{O}(1)
  \label{ee9},
\end{equation}
so that the mismatch between the equilibrium lattice parameters of $\hat{\mathcal A}_1$ and $\hat{\mathcal A}_2$ is small.

In the rescaled coordinates, the atoms on $\mathcal{A}_{2}^0$ are
located at the points $\br_{ij}^k=(\boldsymbol\chi^k_{ij},\eps)$,
where
\begin{equation}
\boldsymbol\chi^k_{ij}
=
\eps\delta_2\left((i + k/3) {\bf a}^2_1+(j+k/3){\bf a}^2_2\right)
\label{ee102}
\end{equation}
for $i,j=1,\ldots,N_2$ and
$k=1,2$, are obtained by dividing ${\bf r}^k_{ij}$ by $L$ in
\eqref{eq:R}. Atom $k$ in the $i,j$ cell is then displaced to the point
\begin{equation}
\bq^k_{ij}=(\boldsymbol\chi^k_{ij}+\eps\boldsymbol\xi(\boldsymbol\chi^k_{ij}),\eps+\eps\eta(\boldsymbol\chi^k_{ij})).
\label{qij}
\end{equation}
Note that here and in what follows we continue to use the notation ${\bf q}^k_{ij}$, ${\bf r}^k_{ij}$, and ${\bf b}_{ij}^p$, but now to denote the corresponding nondimensional quantities.

\subsection{Elastic Energy Contribution} \label{elencon}

The developments in this section closely follow those in
\cite{espanol2018discrete}. Using that $\varepsilon$ is small, we
Taylor expand the rescaled versions 
of the components \eqref{eq:DiscreteE_s}, \eqref{ede2.6},
and \eqref{e_dih} of the discrete elastic energy in $\varepsilon$
about $\boldsymbol\xi(\boldsymbol\chi^2_{ij})$.  After truncating the
expansions, we end up with approximate energies as a function of
$\boldsymbol\xi$ and $\eta$ evaluated at $\boldsymbol\chi^2_{ij}$, where
$i,j=1,\ldots,N_2$.  The details of the derivations that led to these
truncated expansions and the definitions of
$\mathbf{v}_{1}$, $\mathbf{v}_{2}$, and $\mathbf{v}_{3}$
are given in the Supplementary Material~\ref{appe1}.  This yields

\begin{align}
\label{stretchs}
  \mathcal E_s[\boldsymbol\xi,\eta]
  \sim
  \sum_{i,j=1}^{N_2} \frac{9k_s \varepsilon^{3}}{2\omega}  &
  \left[ \left(\bold{v}_1\cdot \nabla \boldsymbol \xi \bold{v}_1 + \frac{\varepsilon}{2} \bold{v}_1 \cdot  (\nabla \eta \otimes \nabla \eta) \bold{v}_1 \right)^2 \right. \nonumber \\
 & +  \left(\bold{v}_2\cdot \nabla \boldsymbol \xi \bold{v}_2 + \frac{\varepsilon}{2} \bold{v}_2 \cdot  (\nabla \eta \otimes \nabla \eta) \bold{v}_2 \right)^2 \nonumber \\
   &\left. +\left(\bold{v}_3\cdot \nabla \boldsymbol \xi \bold{v}_3 + \frac{\varepsilon}{2} \bold{v}_3 \cdot  (\nabla \eta \otimes \nabla \eta) \bold{v}_3 \right)^2 \right],
\end{align}
\begin{align}
\label{torsional}
\mathcal E_t[\boldsymbol\xi,\eta] 
\sim
\sum_{i,j=1}^{N_2} \frac{48 k_t}{\omega}
\varepsilon^3 & \left[ \left(\frac{\bold{v}_1 \cdot \nabla \xi \bold{v}_2 + \bold{v}_2\cdot \nabla \xi \bold{v}_1}{2} + \frac{\varepsilon}{2} \left( \bold{v}_1\cdot (\nabla \eta \otimes \nabla \eta) \bold{v}_2 \right)\right)^2 \right. \nonumber \\
&+\left(\frac{\bold{v}_2 \cdot \nabla \xi \bold{v}_3 + \bold{v}_3\cdot \nabla \xi \bold{v}_2}{2} + \frac{\varepsilon}{2} \left( \bold{v}_2\cdot (\nabla \eta \otimes \nabla \eta) \bold{v}_3 \right)\right)^2 \nonumber \\
& \left.  + \left(\frac{\bold{v}_1 \cdot \nabla \xi \bold{v}_3 + \bold{v}_3\cdot \nabla \xi \bold{v}_1}{2} + \frac{\varepsilon}{2} \left( \bold{v}_1\cdot (\nabla \eta \otimes \nabla \eta) \bold{v}_3 \right)\right)^2
\right],
\end{align}
and 
\begin{equation}
   \mathcal E_d[\boldsymbol\xi,\eta]
  \sim
  \sum_{i,j=1}^{N_2} \frac{3k_d\delta_{2}^{2}\varepsilon^{5}}{8\omega}
  \left[
     7\eta_{,11}^2+16\eta_{,12}^2 - 2\eta_{,11}\eta_{,22} + 7\eta_{,22}^2
    \right].  
  \label{dihes}
\end{equation}
As we did in \cite{espanol2018discrete}, we neglect some third-order
terms in $\varepsilon$, in particular, the terms that contain
second-order derivatives in $\boldsymbol\xi$ or are cubic in
derivatives of $\boldsymbol\xi$.  Further, we include some quartic
terms in the derivative of $\eta$, that allow us to complete squares
in \eqref{stretchs} and \eqref{torsional}.  We showed in
\cite{espanol2018discrete} that including/deleting these higher
order terms from the truncated energy gives minimizers with the
structure close to that of the minimizers of the discrete energy as
$\varepsilon\to0$.

We now observe that $\mathcal A_2^0$ has area $\sqrt{3}/2$
in nondimensional coordinates and that the spacing between the atoms
is $\eps\delta_2/\sqrt{3}\ll1$. Hence, the number of atoms on
$\mathcal A_2^0$ is $\sim\frac1{\eps^2}$.
These observations justify replacing the sums in \eqref{stretchs},
\eqref{torsional}, and \eqref{dihes} with integrals to define the
continuum elastic energies
\begin{equation}
{\small
  \mathcal F_s^\eps[\boldsymbol\xi,\eta]
  =:
  \frac{\gamma_s\eps}{2}\int_{D_1}\left[
   {\left(\bold{v}_1\cdot \nabla \boldsymbol \xi \bold{v}_1 + \frac{\varepsilon}{2}|\nabla \eta \cdot \bold{v}_1|^{2}\right)}^2+{\left(\bold{v}_2 \cdot \nabla \boldsymbol \xi \bold{v}_2 + \frac{\varepsilon}{2}|\nabla \eta \cdot \bold{v}_2|^{2}\right)}^2+ {\left(\bold{v}_3\cdot \nabla \boldsymbol \xi \bold{v}_3 + \frac{\varepsilon}{2}|\nabla \eta \cdot \bold{v}_3|^{2}\right)}^2  \right]\,d{\boldsymbol\chi}, \label{ee103}}
\end{equation}
\begin{align}
\label{tor_cont}
\mathcal F_t^\eps[\boldsymbol\xi,\eta]&=:\gamma_t\eps\int_{D_1} \left[ \left(\frac{\bold{v}_1 \cdot \nabla \boldsymbol\xi \bold{v}_2 + \bold{v}_2\cdot \nabla \boldsymbol\xi \bold{v}_1}{2} + \frac{\varepsilon}{2} \left( \bold{v}_1\cdot (\nabla \eta \otimes \nabla \eta) \bold{v}_2 \right)\right)^2 \right. \nonumber \\
& \qquad \qquad \qquad   +\left(\frac{\bold{v}_2 \cdot \nabla \boldsymbol\xi \bold{v}_3 + \bold{v}_3\cdot \nabla \boldsymbol\xi \bold{v}_2}{2} + \frac{\varepsilon}{2} \left( \bold{v}_2\cdot (\nabla \eta \otimes \nabla \eta) \bold{v}_3 \right)\right)^2 \nonumber \\
& \qquad \qquad  \qquad \left.  + \left(\frac{\bold{v}_1 \cdot \nabla \boldsymbol\xi \bold{v}_3 + \bold{v}_3\cdot \nabla \boldsymbol\xi \bold{v}_1}{2} + \frac{\varepsilon}{2} \left( \bold{v}_1\cdot (\nabla \eta \otimes \nabla \eta) \bold{v}_3 \right)\right)^2
\right]\,d{\boldsymbol\chi},
\end{align}
and
\begin{equation}
\label{bend_cont}
\mathcal F_d^\eps[\boldsymbol\xi,\eta]=:\gamma_d\eps^3\int_{D_1} \left[
    7\eta_{,11}^2+16\eta_{,12}^2 - 2\eta_{,11}\eta_{,22} + 7\eta_{,22}^2
    \right]\,d{\boldsymbol\chi}.
\end{equation}

\subsection{Van der Waals Energy Contribution} \label{vaencon}

We derive the continuum versions of \eqref{eq:DiscreteE_w} and \eqref{eq:DiscreteE_w_kc}, 
%the contribution to the energy from the weak interactions between the
%deformable and the rigid lattice.
which we shall see have the form
\begin{equation}
  \mathcal F^\eps_w[\boldsymbol\xi,\eta]
  =
  \frac{1}{\varepsilon}\int_{D_1} G\left(\boldsymbol\chi,\boldsymbol\xi,\eta\right)\,d\boldsymbol\chi.
  \label{vdwfin}
\end{equation}
An important feature of our approach is that, to define the function $G$,
we develop an expression for the local mismatch between the two
lattices in their reference configurations.  This expression depends
on $\boldsymbol\chi$ and on the relative rotation and lattice
parameter mismatch.

\begin{figure}[htb]
\centering
    \includegraphics[width=.6\linewidth]{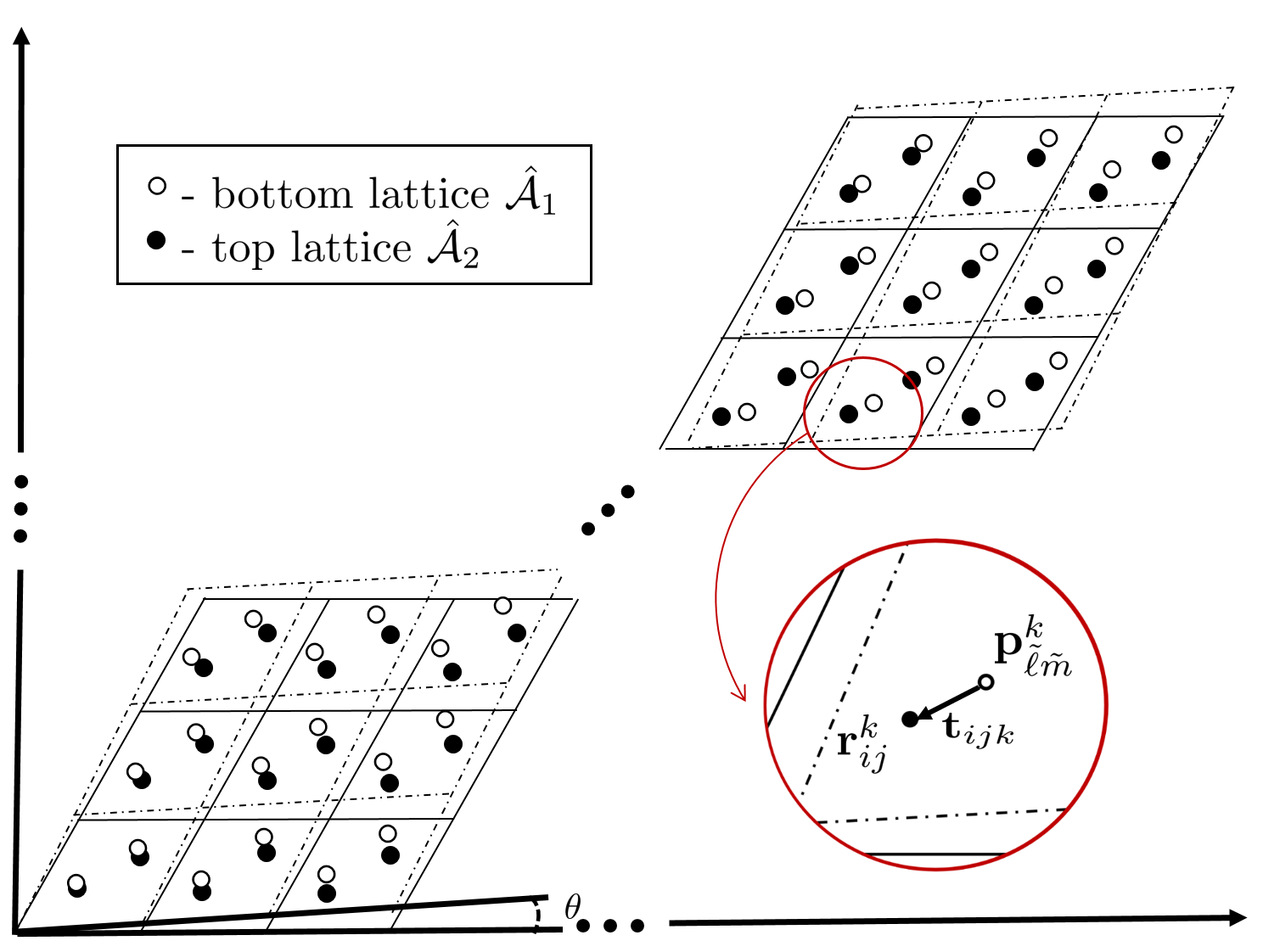}
  \caption{The reference configuration of the system of two hexagonal lattices ${\hat{\mathcal A}}_1$ and ${\hat{\mathcal A}}_2$, mismatched by an angle $\theta$, in nondimensional coordinates. Only two patches of the lattice structure are shown: near and away from the origin. The region of two atoms in the reference configurations, $\br^k_{ij}$ and $\bp^{k}_{\tilde \ell  \tilde m}$, is zoomed in to show the corresponding local horizontal mismatched $\mathbf{t}_{ijk}$ between them.}
  \label{f11}
\end{figure}

The Lennard-Jones potential is considered first.
Starting with the inner double sum on the right-hand side of
\eqref{eq:DiscreteE_w}, we observe that the total interaction energy corresponding to the atom $ijk$ on $\hat{\mathcal A}_2$ with all the atoms on
the rigid lattice $\hat{\mathcal A}_1$ is
\begin{equation}
  \sum_{\ell,m=-\infty}^{\infty} \sum_{\tilde k=1}^{2} g_{\rm LJ}\left(\frac{\|\bq^k_{ij}-\bp^{\tilde k}_{\ell m}\|}{\varepsilon}\right).
  \label{ee3}
\end{equation}
Next, we define $\mathbf{t}_{ijk}$, the local horizontal mismatch
between the lattices $\hat{\mathcal{A}}_{1}$ and
$\hat{\mathcal{A}}_{2}$ in the reference configuration as measured at
atom $ijk$ on $\hat{\mathcal{A}}_{2}$.
To determine $\mathbf{t}_{ijk}$, we project the vector ${\bf r}^k_{ij}$
onto the plane of $\hat{\mathcal{A}}_{1}$.  The projection falls
inside one of the unit cells of $\hat{\mathcal{A}}_{1}$.  Let $\tilde
\ell,\tilde m$ be the indices of that unit cell.  We define
$\mathbf{t}_{ijk}$ as the vector from the atom
${\bf p}_{\tilde \ell  \tilde m}^k$ on $\hat{\mathcal{A}}_{1}$
to the endpoint of the projection of
${\bf  r}_{ij}^k$ onto the plane of $\hat{\mathcal{A}}_{1}$.  See Figure~\ref{f11}.

We can now write 
{\small 
\begin{align}
\left\|\bq^k_{ij}-\bp^{\tilde k}_{\ell m}\right\| &= \left\|\left(\bq^k_{ij}- \br^k_{ij}\right) + \left(\br^k_{ij} - \bp^{k}_{\tilde \ell  \tilde m}\right) + \left(\bp^k_{\tilde \ell \tilde m} - \bp^{\tilde k}_{\ell m}\right)\right \| \nonumber \\
& =\left \| \left(\eps\boldsymbol\xi(\boldsymbol\chi^k_{ij}),\eps\eta(\boldsymbol\chi^k_{ij})\right)  + (\mathbf{t}_{ijk},\eps) + 
\left(\eps \delta_1 \left( \left(\tilde \ell - \ell + (k-\tilde k)/3\right)\ba^1_1 + \left(\tilde m - m + (k-\tilde k)/3\right)\ba^1_2 \right),0 \right) \right\| \nonumber \\
&=\left(\left\| \eps \delta_1 \left( \left(\tilde \ell - \ell + (k-\tilde k)/3\right)\ba^1_1 + \left(\tilde m - m + (k-\tilde k)/3\right)\ba^1_2 \right)+\mathbf{t}_{ijk}+\varepsilon\boldsymbol\xi\left(\boldsymbol\chi^k_{ij}\right)\right\|^2  + \left(\varepsilon+\varepsilon\eta\left(\boldsymbol\chi_{ij}^k\right)\right)^2\right)^{1/2}.
\label{eq:dijk}
\end{align}}

Recall from \eqref{ee9} that $\alpha\varepsilon$, the relative mismatch between
the lattice parameters, is $\mathcal{O}(\varepsilon)$.
In a similar way, we define 
\begin{equation}
\Theta:=\frac{\theta}{\varepsilon}=O(1),
\label{angdef}
\end{equation}
so that $\Theta\varepsilon$, the relative rotation between the
lattices, is $\mathcal{O}(\varepsilon)$.
Hence, we can linearize $\mathbf{t}_{ijk}$ in $\varepsilon$ to write
%so that the
%relative contributions to $\mathbf{t}_{ijk}$ from mismatch and
%misorientation can be computed separately and then added together,
%that is,
$\mathbf{t}_{ijk} = \mathbf{t}_{ijk}^t + \mathbf{t}_{ijk}^r$, 
where $\mathbf{t}_{ijk}^t$ is the relative mismatch when $\theta=0$
and
$\mathbf{t}_{ijk}^r$ is the relative mismatch when $\alpha=0$.
We next derive expressions for $\mathbf{t}_{ijk}^t$ and
$\mathbf{t}_{ijk}^r$.  Recall that we assume that 
the left bottom corner of the 0,0 cell of $\hat{\mathcal{A}}_{2}$ lies directly above the left bottom corner of the 0,0 cell of 
$\hat{\mathcal{A}}_{1}$.

To write down $\mathbf{t}_{ijk}^t$, we assume that $\theta=0$ and
hence ${\ba}_i^1={\ba}_i^2$ for $i=1,2$.   We then have
\begin{align}
  \mathbf{t}_{ijk}^t &
  =
  \varepsilon\delta_2\left((i+k/3){\ba}_1^2+(j+k/3){\ba}_2^2\right)-\varepsilon\delta_1\left((\tilde \ell + k/3){\ba}_1^1+(\tilde m+k/3){\ba}_2^1\right)\nonumber \\ 
 & = \left((i+k/3)(\delta_{2} - \delta_{1}) \varepsilon + i_1 \delta_{1} \varepsilon \right) \ba^2_1 + \left ((j+k/3)(\delta_{2}-\delta_1) \varepsilon + j_1\delta_{1}\varepsilon \right) \ba^2_2\nonumber\\
 & =  \left((i+k/3)\left(\frac{\delta_{2} - \delta_{1}}{\delta_2\varepsilon}\right) \delta_2\varepsilon^2 + i_1 \delta_{1} \varepsilon \right) \ba^2_1 + \left ((j+k/3)\left(\frac{\delta_{2} - \delta_{1}}{\delta_2\varepsilon}\right) \delta_2\varepsilon^2 + j_1 \delta_{1}\varepsilon \right) \ba^2_2\nonumber\\
 & =  \alpha \eps  \left(\delta_2\eps(i+k/3)\right) \ba^2_1 + \alpha \eps \left (\delta_2 \eps (j+k/3)\right) \ba^2_2 + \delta_{1} \eps i_1\ba^2_1 + \delta_{1}\eps j_1\ba^2_2%,\nonumber\\
 = \alpha \eps  \boldsymbol\chi^k_{ij} + \delta_1 \eps \left( i_1  \ba^2_1 + j_1 \ba^2_2\right),
  \label{ee4}
\end{align}
where $i_1 = i-\tilde \ell$ and $j_1 = j-\tilde m$, and where the
final equal sign uses \eqref{ee102}.  
%Because 
%%
%\begin{equation}
% \alpha \eps \left(\delta_2\eps(i+k/3)\right) \ba^2_1 + \alpha \eps \left (\delta_2 \eps (j+k/3)\right) \ba^2_2=\alpha \eps\boldsymbol\chi^k_{ij}, \label{ee5}
%\end{equation}
%%
%the continuum approximation of $\mathbf{t}_{ijk}^t$ is 
%\begin{equation}
%  \mathbf{t}_{ijk}^t
% =
% \alpha \eps  \boldsymbol\chi^k_{ij} + \delta_1 \eps \left( i_1  \ba^2_1 + j_1 \ba^2_2\right).
%\end{equation}
Note that we could instead write 
\begin{equation}
  \mathbf{t}_{ijk}^t
 =
 \alpha \eps  \boldsymbol\chi^k_{ij} + \delta_1 \eps \left( i_1  \ba^1_1 + j_1 \ba^1_2\right).
  \label{eekt}
\end{equation}

To write down $\mathbf{t}_{ijk}^r$, we assume $\delta_1=\delta_2$, and in this case, we have 
\begin{align}
  \mathbf{t}_{ijk}^r
  &=
  \varepsilon\delta_1\left((i+k/3){\ba}_1^2+(j+k/3){\ba}_2^2\right)-\varepsilon\delta_1\left((\tilde \ell + k/3){\ba}_1^1+(\tilde m+k/3){\ba}_2^1\right)\nonumber \\ 
  &=\varepsilon\delta_1\left((i+k/3)\left({\ba}_1^2-{\ba}_1^1\right)+(j+k/3)\left({\ba}_2^2-{\ba}_2^1\right)\right)+\varepsilon\delta_1\left(i_2{\ba}_1^1+j_2{\ba}_2^1\right),
  \label{ee4.1}
\end{align}
where $i_2=i-\tilde \ell$ and $j_2=j-\tilde m$.  Recalling that the lattice
$\hat{\mathcal{A}}_{1}$ is rotated with respect to
$\hat{\mathcal{A}}_{2}$ by the angle $\theta=\varepsilon\Theta$, we see that
%rotation matrix 
%\[R(\theta)=
%\left(
%\begin{array}{ccc}
% \cos{\theta} & -\sin{\theta}   \\
% \sin{\theta} & \cos{\theta}
%\end{array}
%\right)
%=\left(\begin{array}{ccc}
% \cos{\varepsilon\Theta} & -\sin{\varepsilon\Theta}   \\
% \sin{\varepsilon\Theta} & \cos{\varepsilon\Theta}
%\end{array}
%\right)
%\sim\left(\begin{array}{ccc}
% 1 & -\varepsilon\Theta  \\
% \varepsilon\Theta & 1
%\end{array}
%\right)
%\]
%with respect to the basis $\left\{{\ba}_1^2,{\ba}_2^2\right\}$, where
%we use \eqref{angdef}.  It follows that
%
\[
  {\ba}_i^2-{\ba}_i^1
  =
  {\ba}_i^2-R(\theta)\,{\ba}_i^2
  =
  \left(
  \begin{array}{ccc}
    1-\cos{\varepsilon\Theta} & \sin{\varepsilon\Theta}   \\
    -\sin{\varepsilon\Theta} & 1-\cos{\varepsilon\Theta}
  \end{array}
  \right)  
  {\ba}_i^2
  \sim
  \varepsilon\Theta
  \left(\begin{array}{ccc}
    0 & 1  \\
    -1 & 0
  \end{array}\right){\ba}_i^2
\quad
\text{for $i=1,2$.}
\]
%
%and
%\[{\ba}_2^2-{\ba}_2^1={\ba}_2^2-R(\theta)\,{\ba}_2^2\sim\varepsilon\Theta \left(\begin{array}{ccc}
% 0 & -1  \\
% 1 & 0
%\end{array}\right){\ba}_2^2.\]
Inserting these expressions into \eqref{ee4.1} yields
\begin{equation}
\mathbf{t}_{ijk}^r\sim\varepsilon\Theta(\boldsymbol\chi_{ij}^k)^\perp+\varepsilon\delta_1\left(i_2{\ba}_1^1+j_2{\ba}_2^1\right),
\label{ekr}
\end{equation}
where $(\boldsymbol\chi_{ij}^k)^\perp= \left(\begin{array}{ccc}
 0 & 1  \\
 -1 & 0
\end{array}\right)\boldsymbol\chi_{ij}^k$.

Combining \eqref{eekt} and \eqref{ekr} gives
\begin{equation}
  \mathbf{t}_{ijk}
 =
 \mathbf{t}_{ijk}^t+\mathbf{t}_{ijk}^r
 \sim \varepsilon\alpha\boldsymbol\chi^k_{ij}+\varepsilon\Theta(\boldsymbol\chi_{ij}^k)^\perp+\varepsilon\delta_1\left(i_1\ba^1_1 + j_1 \ba^1_2+  i_2{\ba}_1^1+j_2{\ba}_2^1  \right).
  \label{eek}
\end{equation}
Now we substitute \eqref{eek} into \eqref{eq:dijk}.  Also, we use that
$i_1=i_2=i-\tilde \ell$, $j_1=j_2=j-\tilde m$ and that the basis of
$\hat{\mathcal{A}}_1$ is a small perturbation of the basis of
$\hat{\mathcal{A}}_2$, which follows from the smallness of $\theta$.
Lastly, we use that $\delta_1$ and $\delta_2$ are close.
From \eqref{eq:dijk}, we thereby obtain
\begin{align}
  \left\|\bq^k_{ij}-\bp^{\tilde k}_{\ell m}\right\|
  &\sim
  \eps \left(\left\|  \delta_1 \left( \left(\tilde \ell - \ell +
  (k-\tilde k)/3\right)\ba^1_1 + \left(\tilde m - m + (k-\tilde
  k)/3\right)\ba^1_2  + 2 i_1\ba^1_1 + 2 j_1 \ba^1_2 \right)
  \right. \right. \nonumber \\
  & \left. \left. \qquad
  +\,\alpha\boldsymbol\chi^k_{ij}+\Theta(\boldsymbol\chi_{ij}^k)^\perp
  +\boldsymbol\xi\left(\boldsymbol\chi^k_{ij}\right)\right\|^{2}  
 + \left(1+\eta\left(\boldsymbol\chi^k_{ij}\right)\right)^{2}\right)^{\frac12} \nonumber \\
 & \sim
 \eps \left(\left\|  \delta_1 \left( \left(2 i - \ell - \tilde \ell +
 (k-\tilde k)/3\right)\ba^1_1 + \left(2 j - m - \tilde m + (k-\tilde
 k)/3\right)\ba^1_2  \right) \right. \right. \nonumber \\
 & \left. \left. \qquad
 +\,\alpha\boldsymbol\chi^k_{ij}+\Theta(\boldsymbol\chi_{ij}^k)^\perp
 +\boldsymbol\xi\left(\boldsymbol\chi^k_{ij}\right)\right\|^{2}  
 + \left(1+\eta\left(\boldsymbol\chi^k_{ij}\right)\right)^{2}\right)^{\frac12}\nonumber\\
 & \sim
 \eps
 \left(
 \left\|  \delta_2 \left( \left(2 i - \ell - \tilde \ell +
 (k-\tilde k)/3\right)\ba^2_1 + \left(2j - m - \tilde m+ (k-\tilde k)/3\right)\ba^2_2 \right)
 \right.\right.\nonumber \\ 
 & \left. \left. \qquad
 +\,\alpha\boldsymbol\chi^k_{ij}+\Theta(\boldsymbol\chi_{ij}^k)^\perp
 +\boldsymbol\xi\left(\boldsymbol\chi^k_{ij}\right)\right\|^{2}
 +
 \left(1+\eta\left(\boldsymbol\chi^k_{ij}\right)\right)^{2}\right)^{\frac12} \nonumber\\
 &= 
 \eps
 \left(
 \left\|  \delta_2 \left( \left(\ell +
 (k-\tilde k)/3\right)\ba^2_1
 +
 \left(m + (k-\tilde k)/3\right)\ba^2_2 \right)
 \right.\right.\nonumber \\ 
 & \left. \left. \qquad
 +\,\alpha\boldsymbol\chi^k_{ij}+\Theta(\boldsymbol\chi_{ij}^k)^\perp
 +\boldsymbol\xi\left(\boldsymbol\chi^k_{ij}\right)\right\|^{2}
 +
 \left(1+\eta\left(\boldsymbol\chi^k_{ij}\right)\right)^{2}\right)^{\frac12},
\end{align}
where in the last line
we changed the indices $\ell \to 2i - \ell -\tilde \ell$ and $m \to 2j - m -\tilde m$.
Therefore, we can write
\begin{equation}
  g_{\rm LJ}\left(\frac{\|\bq^k_{ij}-\bp^{\tilde k}_{\ell
      m}\|}{\varepsilon}\right)
  =
  g_{\rm{LJ}}
  \left(
  \sqrt{
    \rho
    \left(
    \alpha\boldsymbol\chi^k_{ij}
    +
    \Theta(\boldsymbol\chi_{ij}^k)^\perp
    +
    \boldsymbol\xi\left(\boldsymbol\chi^k_{ij}\right)
    \right)^{2}
    +
    \left(
    1 + \eta\left(\boldsymbol\chi^k_{ij}\right)
    \right)^{2}
  }
  \right), 
  \label{ee100}
\end{equation}
where
$$\rho(\bp) = \left\|\delta_2(\ell + (k-\tilde
k)/3)\,{\ba}_1^2+\delta_2 (m + (k-\tilde k)/3)\,{\ba}_2^2+{\bf
  p}\right\|$$ 
for every ${\bf p}\in\mathbb R^2$.

%Returning to \eqref{ee3}, we see that the function $G$  that gives a
%continuum description of the van der Waals energy arising from the
%local lattice mismatch is defined by 

To complete the definition of $G$ for the Lennard-Jones potential,
we now set
\begin{equation}
  G(\boldsymbol\chi,\boldsymbol\xi,\eta)
  :=
  \mathcal
  G_{\rm LJ}\left(\alpha\boldsymbol\chi+\Theta\boldsymbol\chi^\perp+\boldsymbol\xi,\eta\right),  
  \label{gcont}
\end{equation}
where 
\begin{equation}
\mathcal G_{\rm{LJ}}({\bf p},t):= \sum_ {k=1}^{2} \sum_{\ell,m=-\infty}^{\infty} \sum_{\tilde k=1}^{2}
        g_{\rm{LJ}}\left(
          \sqrt{\rho(\bp)^{2}
  + \left(1+t\right)^{2}}
          \right)
          \label{calg0}
\end{equation}
for $t>-1$.
%or 
%\begin{equation}
%  \mathcal G_{\rm{KC}}({\bf p},t)
%  :=
%  \sum_ {k=1}^{2} \sum_{\ell,m=-\infty}^{\infty} \sum_{\tilde k=1}^{2}
%  g_{\rm{KC}}
%  \left(
%   \sqrt{ \rho(\bp)^{2} + \left(1+t\right)^{2}},
%   \frac{\varepsilon L}{\delta}\rho(\bp)
%  \right)
%  \label{calg1}
%\end{equation}
%with 
%$$\rho(\bp) = \left\|\delta_2(\ell + (k-\tilde
%k)/3)\,{\ba}_1^2+\delta_2 (m + (k-\tilde k)/3)\,{\ba}_2^2+{\bf
%  p}\right\|$$ 
%for every ${\bf p}\in\mathbb R^2$ and $t>-1$.
%

For the Kolmogorov-Crespi potential, we note that
$\rho
\left(
\alpha\boldsymbol\chi^k_{ij}
+
\Theta(\boldsymbol\chi_{ij}^k)^\perp
+
\boldsymbol\xi\left(\boldsymbol\chi^k_{ij}\right)
\right)$
is the length of
the projection of
$\bq^k_{ij}-\bp^{\tilde k}_{\ell m}$
onto the $xy$-plane, which is the plane
containing the fixed lattice $\hat{\mathcal A}_1$.
In this case, we set
$$
G(\boldsymbol\chi,\boldsymbol\xi,\eta)
:=
\mathcal
G_{\rm KC}\left(\alpha\boldsymbol\chi+\Theta\boldsymbol\chi^\perp+\boldsymbol\xi,\eta\right),
$$
where
\begin{equation}
  \mathcal G_{\rm{KC}}({\bf p},t)
  :=
  \sum_ {k=1}^{2} \sum_{\ell,m=-\infty}^{\infty} \sum_{\tilde k=1}^{2}
  g_{\rm{KC}}\left(
  \sqrt{\rho(\bp)^{2}
  +
  \left(1+t\right)^{2}},
  \frac{\varepsilon\rho(\bp)}{\delta/L}
  \right).
  \label{ee101}
\end{equation}

Finally, the nondimensional version of either \eqref{eq:DiscreteE_w}
or
\eqref{eq:DiscreteE_w_kc}
takes the form
\begin{align}\label{eq:DiscreteE_wND}
  {\mathcal E}_w({\bf Q}) 
  =
  \eps\sum_{i,j=1}^{N_2} \sum_{k=1}^{2} \sum_{\ell,m=-\infty}^{\infty}
  \sum_{\tilde k=1}^{2}  g%\left(\bq^k_{ij},\bp^{\tilde k}_{\ell m}\right)
  \sim
  \frac{1}{\eps}\sum_{i,j=1}^{N_2}
  G(\boldsymbol\chi,\boldsymbol\xi,\eta)\eps^2
  \sim
  \frac{1}{\varepsilon}\int_{D_1}
  G\left(\boldsymbol\chi,\boldsymbol\xi,\eta\right)\,d\boldsymbol\chi
  =:
  \mathcal F^\eps_w[\boldsymbol\xi,\eta],
\end{align}  
where $g$ is either $g_{\rm LJ}$ or $g_{\rm KC}$ with the appropriate arguments
and $G$ is defined using either $\mathcal G_{\rm{LJ}}$ or
$\mathcal G_{\rm{KC}}$.  
This establishes \eqref{vdwfin}.

\subsection{Continuum Energy} \label{encon}

Combining \eqref{ee103}--\eqref{bend_cont} and  \eqref{eq:DiscreteE_wND}
yields a continuum energy functional that can be written as
\begin{align}
	\mathcal F^\eps[\boldsymbol\xi,\eta] &:= 
        \frac{\eps}{2}\int_{D_1}f\left(D\left(V\nabla\boldsymbol\xi V^T \right)+\frac{\eps}{2}V\left(\nabla\eta\otimes \nabla\eta\right)V^T\right)\,d{\boldsymbol\chi} \nonumber\\
       & \quad\quad +
       \gamma_d\eps^3\int_{D_1} \left[
    7\eta_{,11}^2+16\eta_{,12}^2 - 2\eta_{,11}\eta_{,22} + 7\eta_{,22}^2
    \right]\,d{\boldsymbol\chi}
        +
        \frac{1}{\varepsilon}\int_{D_1} G\left(\boldsymbol\chi,\boldsymbol\xi,\eta\right)\,d\boldsymbol\chi.
        \label{ee6}
\end{align}
Here $D(A)=(A+A^T)/2$ is the symmetric part of $A$ for any $A\in
M^{3\times 3}$,
$V$ is the $3\times 2$ matrix whose $i$th row is $\mathbf{v}_{i}$
and
\[f\left(M\right)=\gamma_s\left(m_{11}^2+m_{22}^2 + m_{33}^2\right)+ \gamma_t \left(m_{12}^2 + m_{21}^2  + m_{13}^2 + m_{31}^2 + m_{23}^2 + m_{32}^2\right)\]
for any $M=(m_{ij})\in M^{3\times 3}_{sym}$.
%$$M=\left(
%\begin{array}{ccc}
%m_{11}   &  m_{12} & m_{13} \\
%m_{12}   &  m_{22} & m_{23}\\
%m_{13} &  m_{23} & m_{33}\\
%\end{array}
%\right)
%\in M^{3\times 3}_{sym}
%\quad \mbox{ and } \quad V= \left(
%\begin{array}{c}
% \bold{v}_1\\ 
% \bold{v}_2\\
% \bold{v}_3\\
%\end{array}
%\right)\in M^{3\times 2}
%.$$

Note that the elastic contribution to the energy \eqref{ee6} is like that of the F\"oppl--von K\'arm\'an theory. The corresponding variational problem is of Ginzburg-Landau type, where the minimizers are determined via a competition between the elastic energy and the potential energy, which has multiple wells associated with the low-energy commensurate regions. The system is forced to reside in these wells, with the sharp transition between the wells being smoothed out due to the penalty imposed by the elastic energy.  Consequently, we expect the minimizers of \eqref{ee6} to develop walls of characteristic width $\eps$.

Now let $B(\nabla\boldsymbol\xi,\nabla\eta) =
D\left(V\nabla\boldsymbol\xi
V^T\right)+\frac{\eps}{2}V\nabla\eta\otimes\nabla\eta V^T$
and
\begin{equation*}
  \hat M
  =
  \left( 
  \begin{array}{ccc}
    \gamma_s B_{11}   & \gamma_t B_{12} & \gamma_t B_{13} \\
    \gamma_t B_{21}   & \gamma_s B_{22} & \gamma_t B_{23}\\
    \gamma_t B_{31} &  \gamma_t B_{32} & \gamma_s B_{33}\\
  \end{array}
  \right),
\end{equation*}
where $B_{ij}$ are the entries in the matrix
$B(\nabla\boldsymbol\xi,\nabla\eta)$.  Let
${\bf b}_1(\nabla\boldsymbol\xi,\nabla\eta)$
and
${\bf b}_2(\nabla\boldsymbol\xi,\nabla\eta)$
be the first two columns of $V^T\hat M V$. 
It is shown in Supplementary Material~\ref{supp_EL} that the Euler-Lagrange equations for the
functional $\mathcal F^\eps$ are 
\begin{equation}
\hspace{-1mm}\left\{
\begin{aligned}
&	-\eps\,\mathrm{div}\left[{\bf b}_1(\nabla\boldsymbol\xi,\nabla\eta)
\right]
        +
        \frac1\varepsilon G_{\xi_1}(\boldsymbol\chi,\boldsymbol\xi,\eta)
        =0, \\
&	-\eps\,\mathrm{div}\left({\bf b}_2(\nabla\boldsymbol\xi,\nabla\eta)
\right]
        +
        \frac1\varepsilon G_{\xi_2}(\boldsymbol\chi,\boldsymbol\xi,\eta)
        =0,\\
&	14\eps^3\gamma_d\Delta^2\eta-\eps^2\,\mathrm{div}\left({\bf b}_1(\nabla\boldsymbol\xi,\nabla\eta)\cdot\nabla\eta,{\bf b}_2(\nabla\boldsymbol\xi,\nabla\eta)\cdot\nabla\eta
\right)
        +
        \frac1\varepsilon G_{\eta}(\boldsymbol\chi,\boldsymbol\xi,\eta)
        =0.         
\end{aligned}
\right.
\label{e17}
\end{equation}
Here the first two equations describe the force balance in the plane of the deformable lattice, while the last equation is the vertical force balance.

\section{Numerical Results} \label{s:results}

% figure counter at fr101

To validate our discrete-to-continuum procedure, in this section we
compare the predictions of the discrete and the continuum models.
We consider results for both the Lennard-Jones 12--6
potential~\eqref{ee1.0} and the Kolmogorov-Crespi
potential~\eqref{ee2}.  To solve the atomistic model numerically, we
use LAMMPS~\cite{PLIMPTON19951} to minimize the discrete
energy~\eqref{eq:entot}.  We use COMSOL~\cite{comsol} to numerically
solve the system of partial differential equations~\eqref{e17} derived
from the continuum model~\eqref{ee6}.  For the continuum simulations,
we use dissipation-dominated (gradient flow) dynamics to drive the
energy of the system toward a possibly local minimum.  The same task
was accomplished for the atomistic model by performing molecular
dynamics simulations at a sufficiently low temperature.

%Both sets of simulations were conducted assuming periodic boundary
%conditions in the plane of the rigid lattice with period $1$ in
%$\chi_1$- and $\chi_2$-directions.

\subsection{Periodic Boundary Conditions} \label{subsper}

\begin{figure}[htb]
\centering
  \includegraphics[width=1\linewidth]{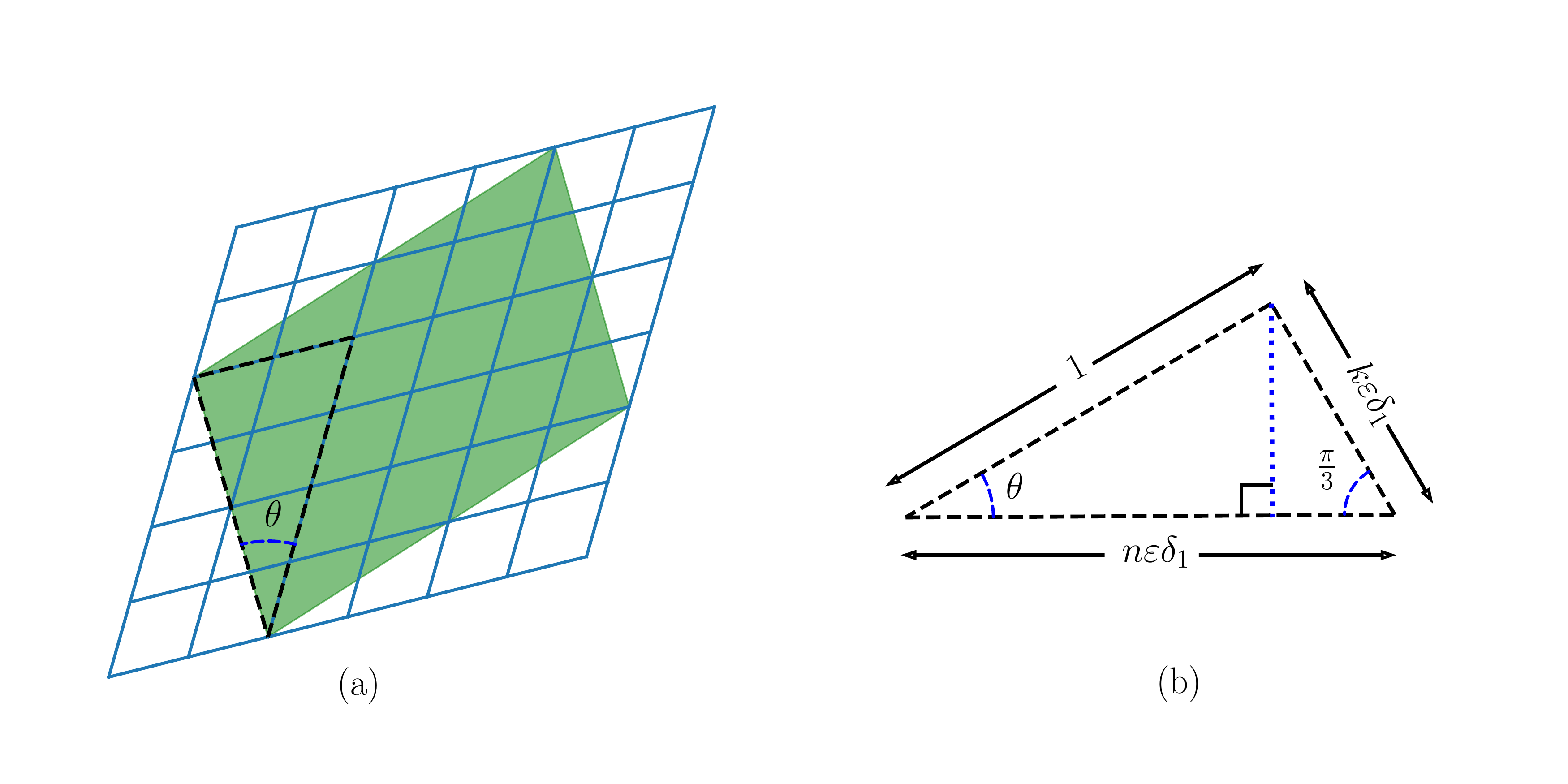}
  \caption{(a) A rotation of the reference configuration $\mathcal{A}_{2}^{0}$ (shaded) such that each
    corner sits above a corner of a unit cell in
    $\hat{\mathcal{A}}_{1}$.  (b)  Geometry of the dashed triangle in
    (a).  In the case depicted, $k=2$ and $n=4$.}  
  \label{fper}
\end{figure}

In this subsection we describe how the boundary conditions are
implemented in the simulations below.
Recall that we assume that $\hat{\mathcal{A}}_{2}$ is
infinite in extent and deforms periodically.
This requires that the surface $\mathcal{A}_{2}^{0}$ 
satisfies the following constraint.
If we pick two points identified on the edges of 
$\mathcal{A}_{2}^{0}$
and project the position vectors for these points onto $\hat{\mathcal{A}}_{1}$, then
$\hat{\mathcal{A}}_{1}$ is invariant when translated by the vector between the two
projected points.  See Figure~\ref{fper} (a).
This in turn places constraints on the combinations of rotation and
lattice mismatch that can be used to define the reference
configuration. 

As suggested by Figure~\ref{fper}(a), a sufficient condition for
periodicity is that each corner of the reference configuration
$\mathcal{A}_{2}^{0}$ sits above a corner of a unit cell in
$\hat{\mathcal{A}}_{1}$.
From the triangle in Figure~\ref{fper}(b), we derive the relations
\begin{equation}
  \cos\theta = n\varepsilon\delta_{1}-\frac{1}{2}k\varepsilon\delta_{1},
  \quad 
  \sin\theta = \frac{\sqrt{3}}{2}k\varepsilon\delta_{1},
  \quad
  \text{and}
  \quad
  \tan\theta = \frac{\frac{\sqrt{3}}{2}k}{n-\frac{k}{2}}
  \label{eq:cos:sin}
\end{equation}
among $\theta, \varepsilon\delta_{1}, n$, and $k$.

We use \eqref{eq:cos:sin} to generate parameter values ensuring that
periodicity is satisfied.
Recall that we denote the number of cells
in each side of the deformable lattice $\hat{\mathcal A}_2$ in $D_1$
by $N_2$, so that $\varepsilon\delta_2 N_2=1$.  We pick $N_{2}$ and
$\delta_{2}$ and set $\varepsilon=(\delta_{2}N_{2})^{-1}$.
Given $\varepsilon$, we pick 
$k, n\in\mathbb N$ and set
%$k\in\mathbb N$, and define the number of cells in
%one side of the rigid lattice $\hat {\mathcal A}_1$ as 
%
%\begin{equation}N_1=N_2-m\label{Nper},\end{equation} 
%the angle $\theta$ from
%
\begin{equation}
  \theta
  :=
  \mbox{arctan}\left(\frac{\frac{\sqrt{3}}{2}k}{n-\frac{k}{2}}\right)
  \mbox{ and }
  \delta_{1}
  :=
  \frac{2}{\sqrt{3}k\varepsilon}\sin \theta
  =
  \frac{1}{\eps\sqrt{\left(n-\frac{k}{2}\right)^2+\frac{3}{4}k^2}}.
  \label{thper}
\end{equation}
Note that the first condition in \eqref{eq:cos:sin} is also
satisfied.  Hence the reference configuration will satisfy the
`corner-to-corner' periodicity condition.  

Next we use \eqref{thper} to derive useful expressions for the parameters
$\alpha$ and $\Theta$ appearing in \eqref{gcont}.  We define $m$ by $n=N_{2}-m$.
%
%
%Also using that $d=\delta_1\varepsilon$ and 
%$$1 = \cos^2(\theta) + \sin^2(\theta)  = \left(n-\frac{k}{2}\right)^2d^2 + \frac{3}{4} k^2d^2,$$
%we get
%$$1  = \left(n-\frac{k}{2}\right)^2(\delta_1\varepsilon)^2 + \frac{3}{4}k^2(\delta_1\varepsilon)^2,$$
%and therefore the lattice constant for $\hat{\mathcal A}_1$ as
%\begin{equation}
%\delta_1=\frac{1}{\eps\sqrt{\left(n-\frac{k}{2}\right)^2+\frac{3}{4}k^2}}.
%\label{delper}
%\end{equation}
%These expressions are obtained by combining the equations in \eqref{eq:cos:sin} with $n=N_1$ and $d=\delta_1\varepsilon$.
%The parameters $N_1$, $\theta$, and $\delta_1$ as given by
%\eqref{Nper}-\eqref{delper} are used in the discrete simulations
%presented below. We are now in a position to determine the parameters
%$\alpha$ and $\Theta$ for the corresponding continuum simulations.
From \eqref{angdef} we have
\begin{align}
  \Theta&
  =
  \frac{\theta}{\varepsilon}
  =
  \frac{1}{\eps}\mbox{arctan}
  \left(\frac{\frac{\sqrt{3}}{2}k\varepsilon\delta_{2}}{n\varepsilon\delta_{2}-\frac{k}{2}\varepsilon\delta_{2}}\right)
%  =
%  \frac{1}{\eps}\mbox{arctan}\left(\frac{\sqrt{3}k}{2(N_2-m)-k}\right)
  =
  \frac{1}{\eps}\mbox{arctan}
  \left(\frac{\frac{\sqrt{3}}{2}k\delta_2\varepsilon}{1-\left(m+\frac{k}{2}\right)\delta_2\varepsilon}\right)
%  \nonumber \\ 
%  %
%  &
  \sim
  \frac{\sqrt{3}}{2}k\delta_2,
\end{align}
using that $\varepsilon\delta_2N_2=1$. Further, 
\begin{align}
  \alpha
  &=
  \frac{\delta_1-\delta_2}{\varepsilon\delta_2}
  =
%  \frac{1}{\eps\delta_2\sqrt{\left(N_1-\frac{k}{2}\right)^2+\frac{3}{4}k^2}}-1
%  =
  \frac{1}{\eps^{2}\delta_2\sqrt{\left(N_2-m-\frac{k}{2}\right)^2+\frac{3}{4}k^2}}-1
  =
  \frac{1}{\varepsilon\sqrt{\left(1-(m+\frac{k}{2})\eps\delta_2\right)^2+\frac{3}{4}(\eps\delta_2 k)^2}}-1
%  \nonumber\\
%  %
%  &
  \sim
  \left(m+\frac{k}{2}\right)\delta_2.
  \label{ae34}
\end{align}
%
%by using \eqref{Nper} and \eqref{delper} and that $\varepsilon\delta_2N_2=1$. It follows from \eqref{ae34} that the parameter $\alpha$, defined in \eqref{ee9}, is given by
%\begin{equation}
%\alpha\sim\left(m+\frac{k}{2}\right)\delta_2.
%\label{alper}
%\end{equation}

\subsection{Results for the Lennard-Jones Potential}

\begin{figure}[!h]
\centering
\begin{subfigure}{0.3\textwidth}
  %trim=left botm right top ** THESE ARE MARGINS **
  \includegraphics[width=\textwidth, clip, trim=1.25in 3in 1.25in 1.1in]
                  {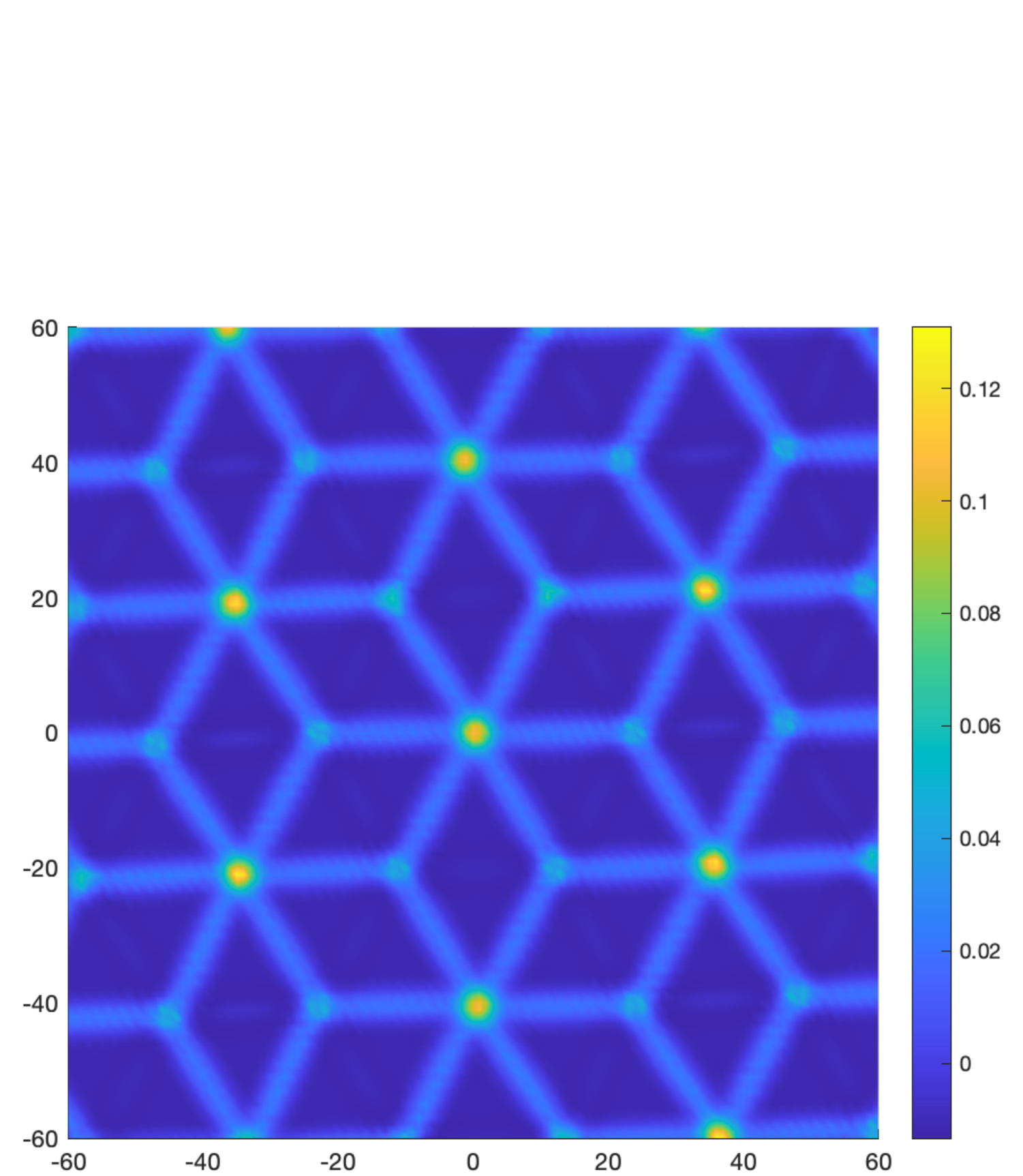}
  %\caption{eta\_LAMMPS.pdf.}
  \caption{$\eta$, Discrete Model.}
  \label{fr1}
\end{subfigure}
\hfill
\begin{subfigure}{0.3\textwidth}
  %trim=left botm right top ** THESE ARE MARGINS **
  \includegraphics[width=\textwidth, clip, trim=1.25in 3in 1.25in 1.1in]
                  {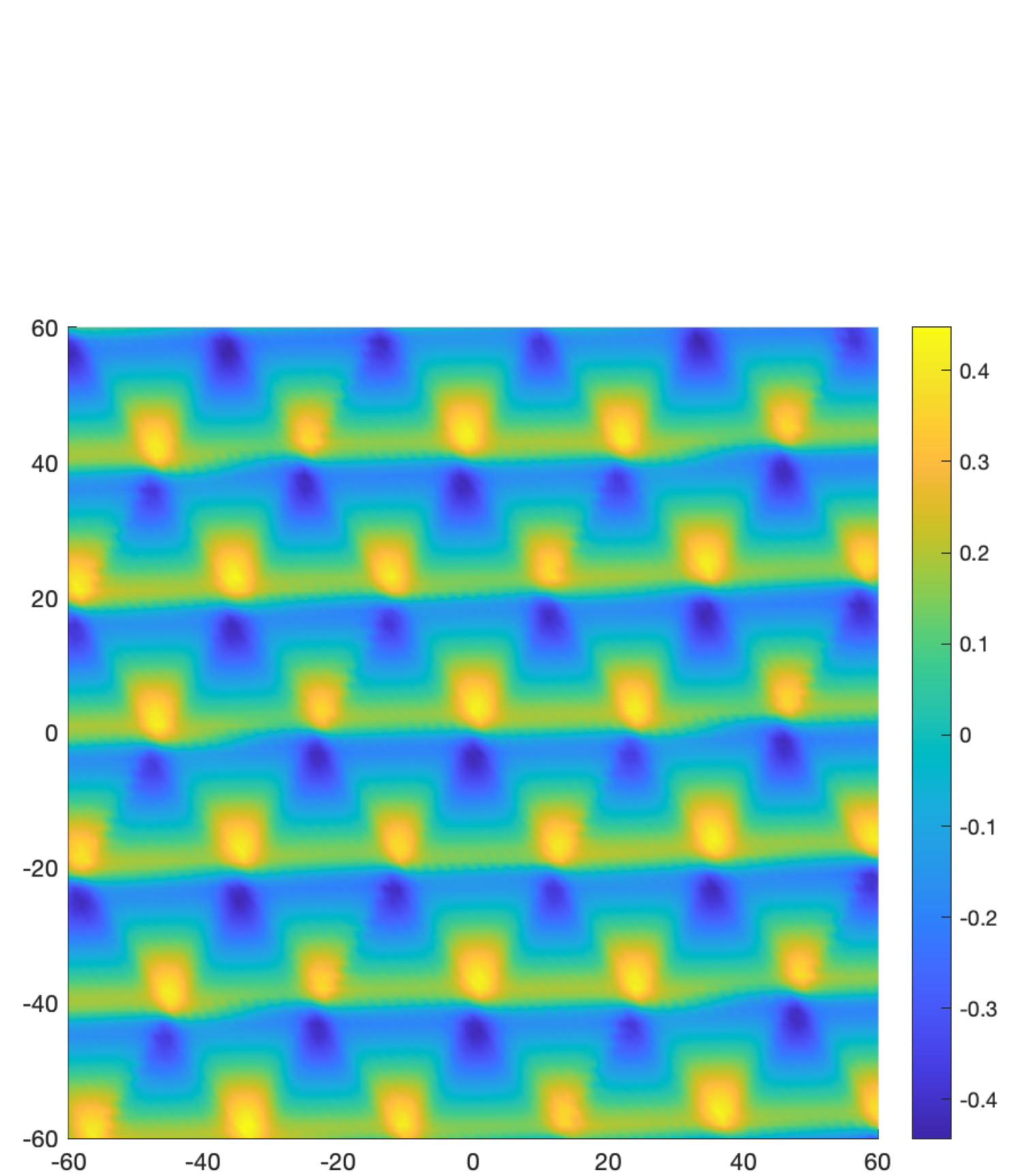}
  %\caption{xi1\_LAMMPS.pdf.}
  \caption{$\xi_{1}$, Discrete Model.}
  \label{fr5}
\end{subfigure}
\hfill
\begin{subfigure}{0.3\textwidth}
  %trim=left botm right top ** THESE ARE MARGINS **
  \includegraphics[width=\textwidth, clip, trim=1.25in 3in 1.25in 1.1in]
                  {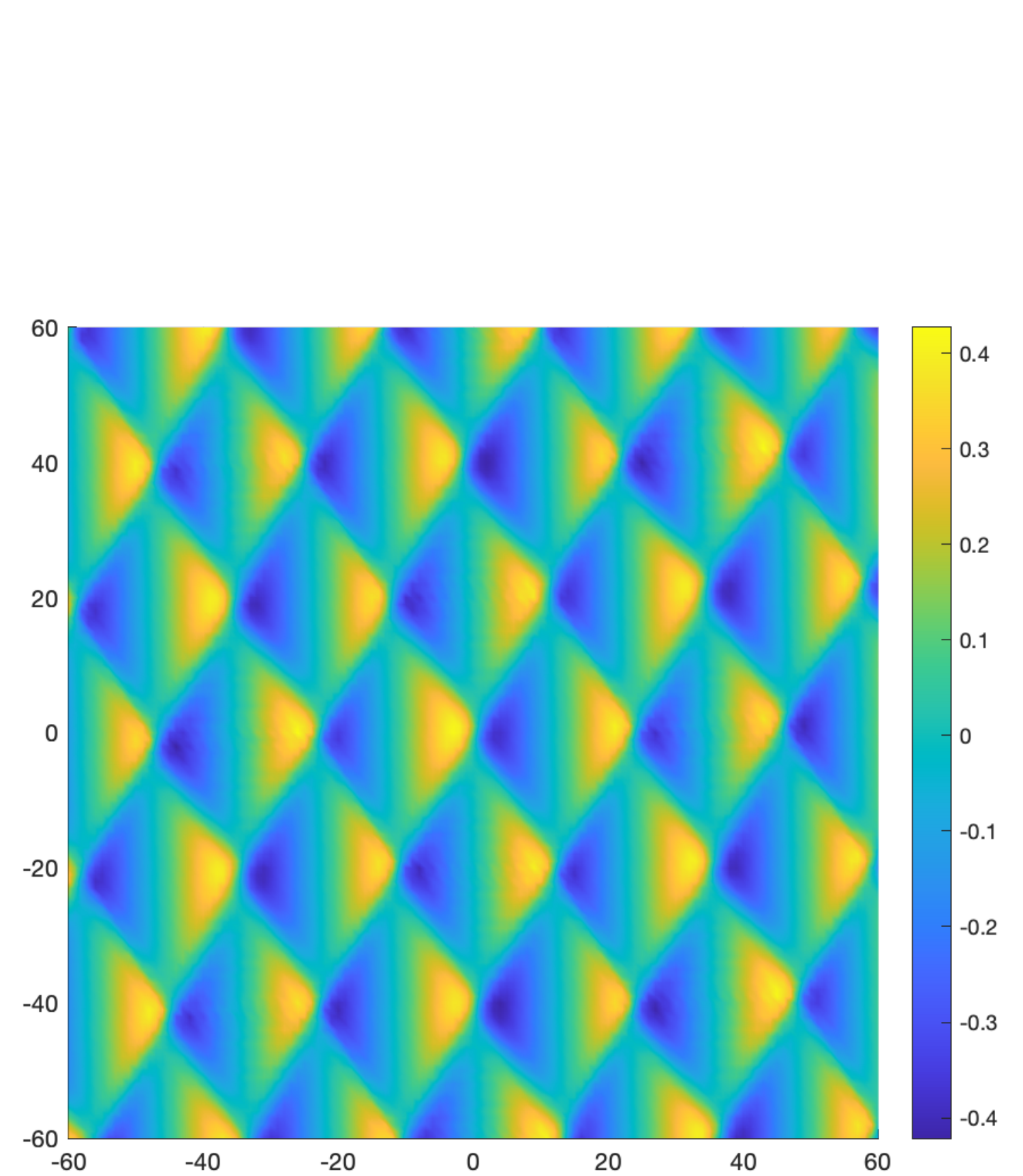}
  %\caption{xi2\_LAMMPS.pdf.}
  \caption{$\xi_{2}$, Discrete Model.}
  \label{fr7}
\end{subfigure}

\vspace*{-.5in}

\begin{subfigure}{0.3\textwidth}
  \includegraphics[width=\textwidth, clip, trim=1.25in 3in 1.25in 1.5in]
                  {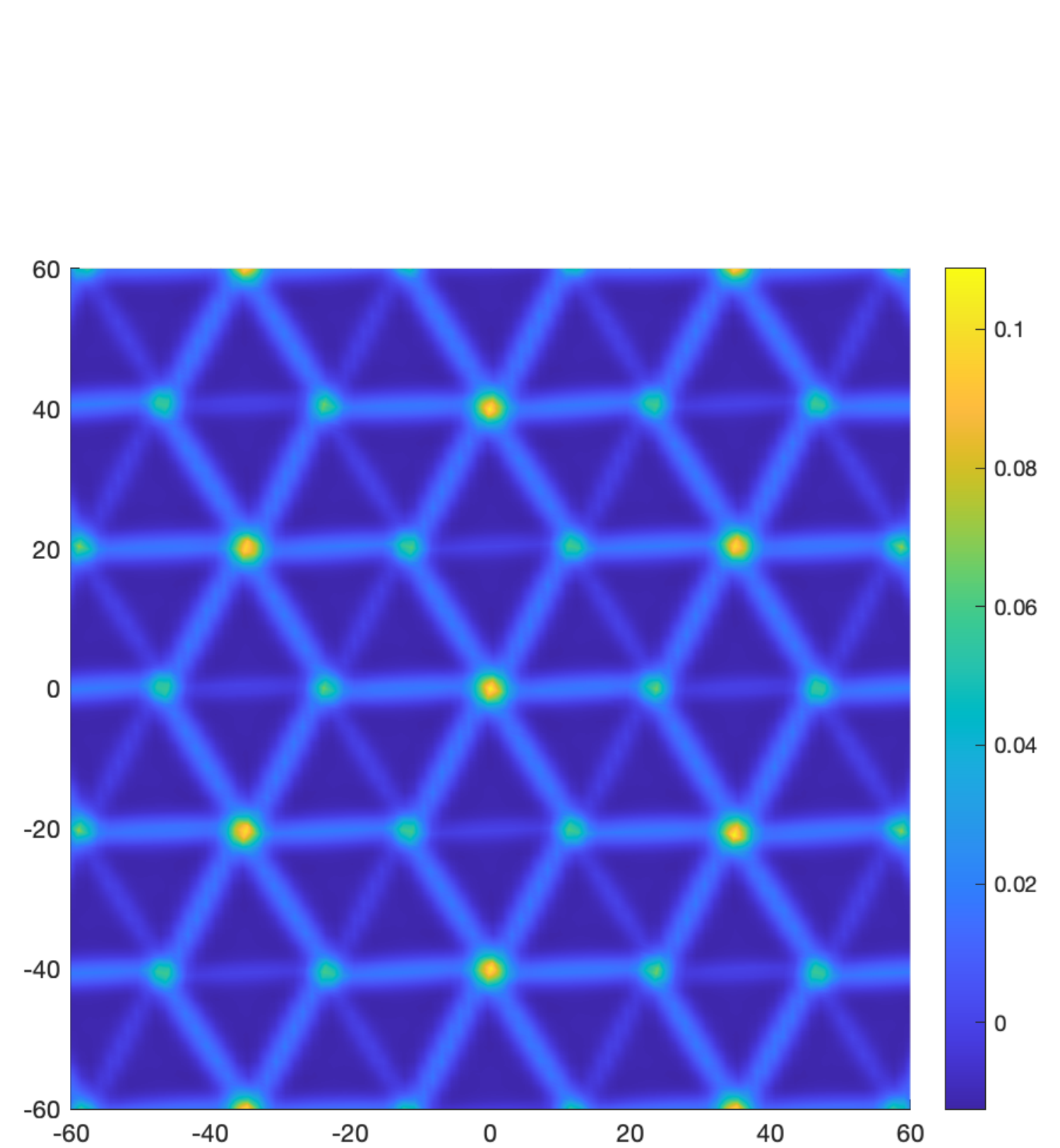}
  %\caption{eta\_COMSOL.pdf.}
  \caption{$\eta$, Continuum Model.}
  \label{fr2}
\end{subfigure}
\hfill
\begin{subfigure}{0.3\textwidth}
  \includegraphics[width=\textwidth, clip, trim=1.25in 3in 1.25in 1.5in]
                  {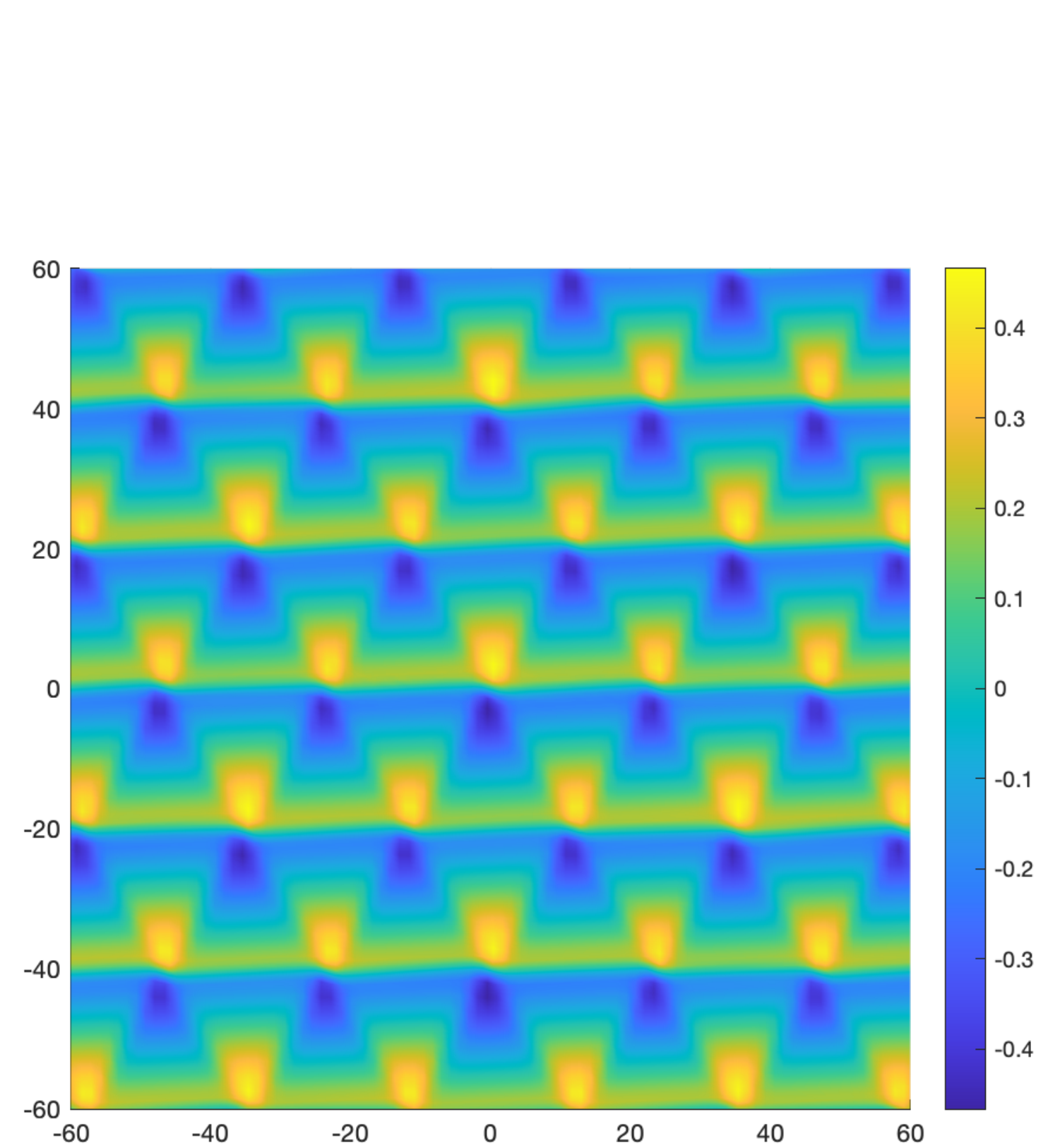}
  %\caption{xi1\_COMSOL.pdf.}
  \caption{$\xi_{1}$, Continuum Model.}
  \label{fr6}
\end{subfigure}
\hfill
\begin{subfigure}{0.3\textwidth}
  \includegraphics[width=\textwidth, clip, trim=1.25in 3in 1.25in 1.5in]
                  {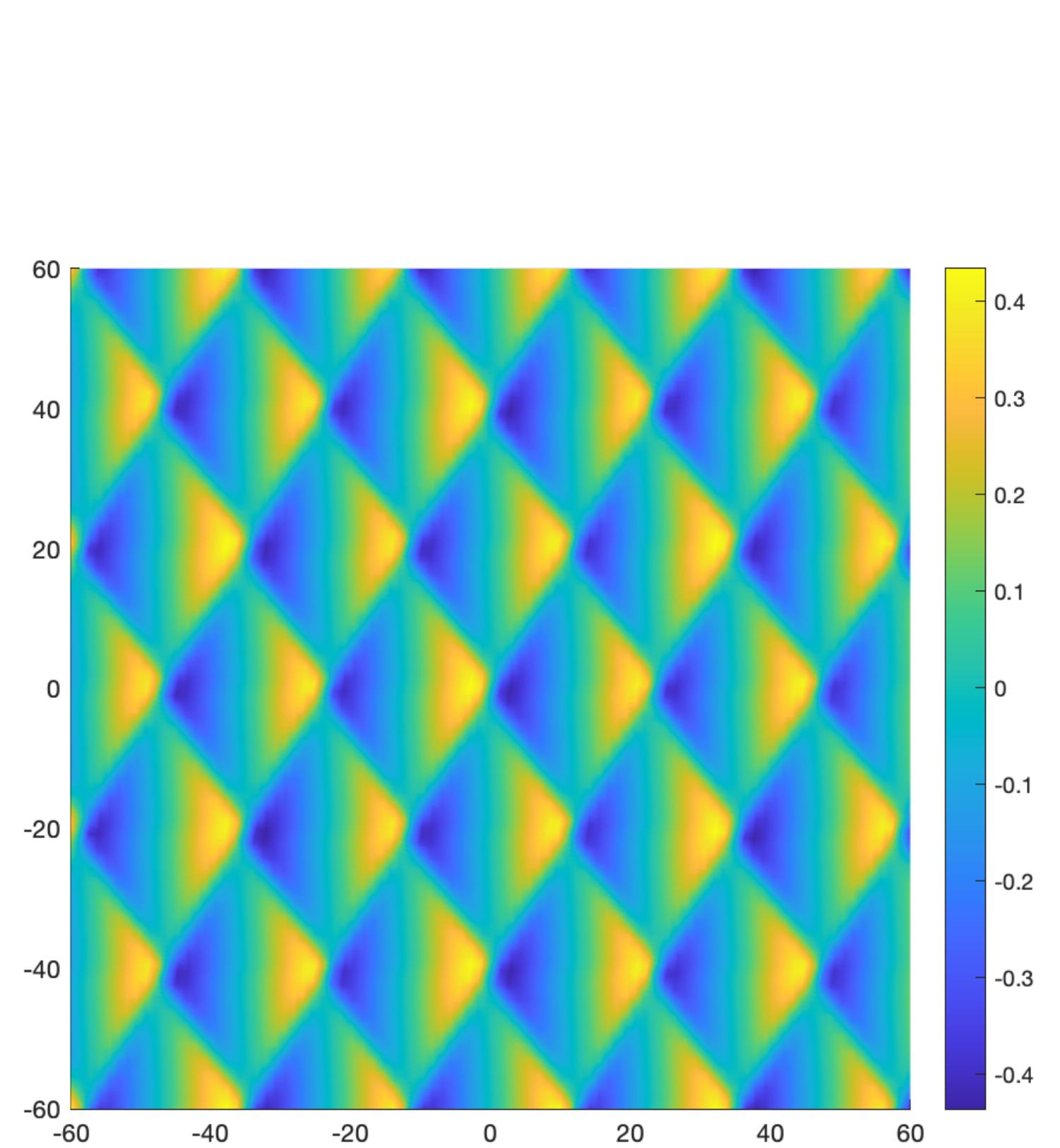}
  %\caption{xi2\_COMSOL.pdf.}
  \caption{$\xi_{2}$, Continuum Model.}
  \label{fr8}
\end{subfigure}

\caption{Simulation Results for Lennard-Jones Potential with
  $\varepsilon=0.0094$ and $\omega=0.5$.}
\label{fr19}
\end{figure}

\begin{figure}[!t]
\centering
\begin{subfigure}{0.3\textwidth}
  %trim=left botm right top ** THESE ARE MARGINS **
  \includegraphics[width=\textwidth, clip, trim=1.25in 3in 1.25in 3.in]
                  {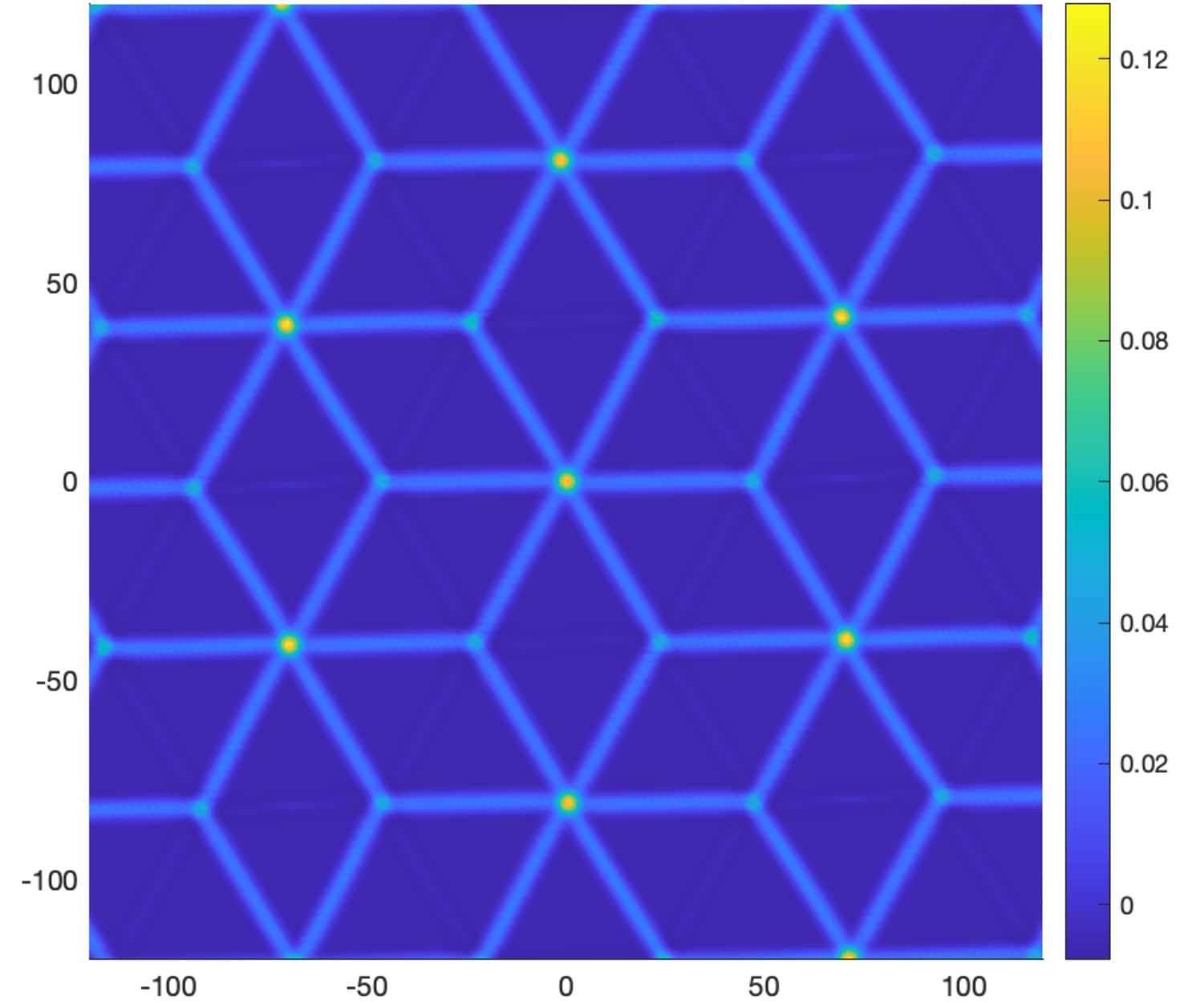}
  %\caption{eta\_LAMMPS\_vl.pdf.}
  \caption{$\eta$, Discrete Model.}
  \label{fr3}
\end{subfigure}
\hfill
\begin{subfigure}{0.3\textwidth}
  %trim=left botm right top ** THESE ARE MARGINS **
  \includegraphics[width=\textwidth, clip, trim=1.25in 3in 1.25in 3.4in]
                  {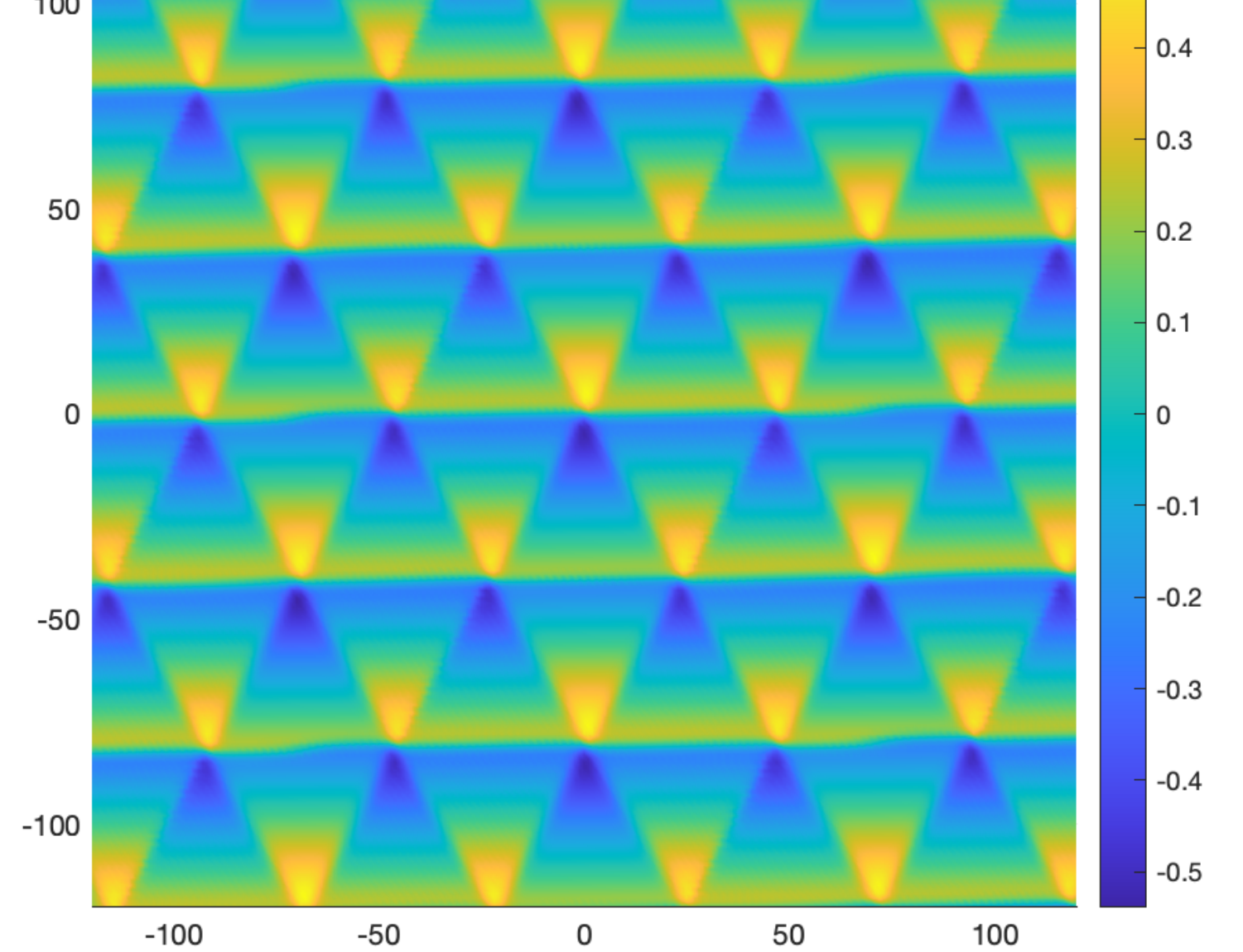}
  %\caption{xi1\_LAMMPS\_vl.pdf.}
  \caption{$\xi_{1}$, Discrete Model.}
  \label{fr9}
\end{subfigure}
\hfill
\begin{subfigure}{0.3\textwidth}
  %trim=left botm right top ** THESE ARE MARGINS **
  \includegraphics[width=\textwidth, clip, trim=1.25in 3in 1.25in 3.4in]
                  {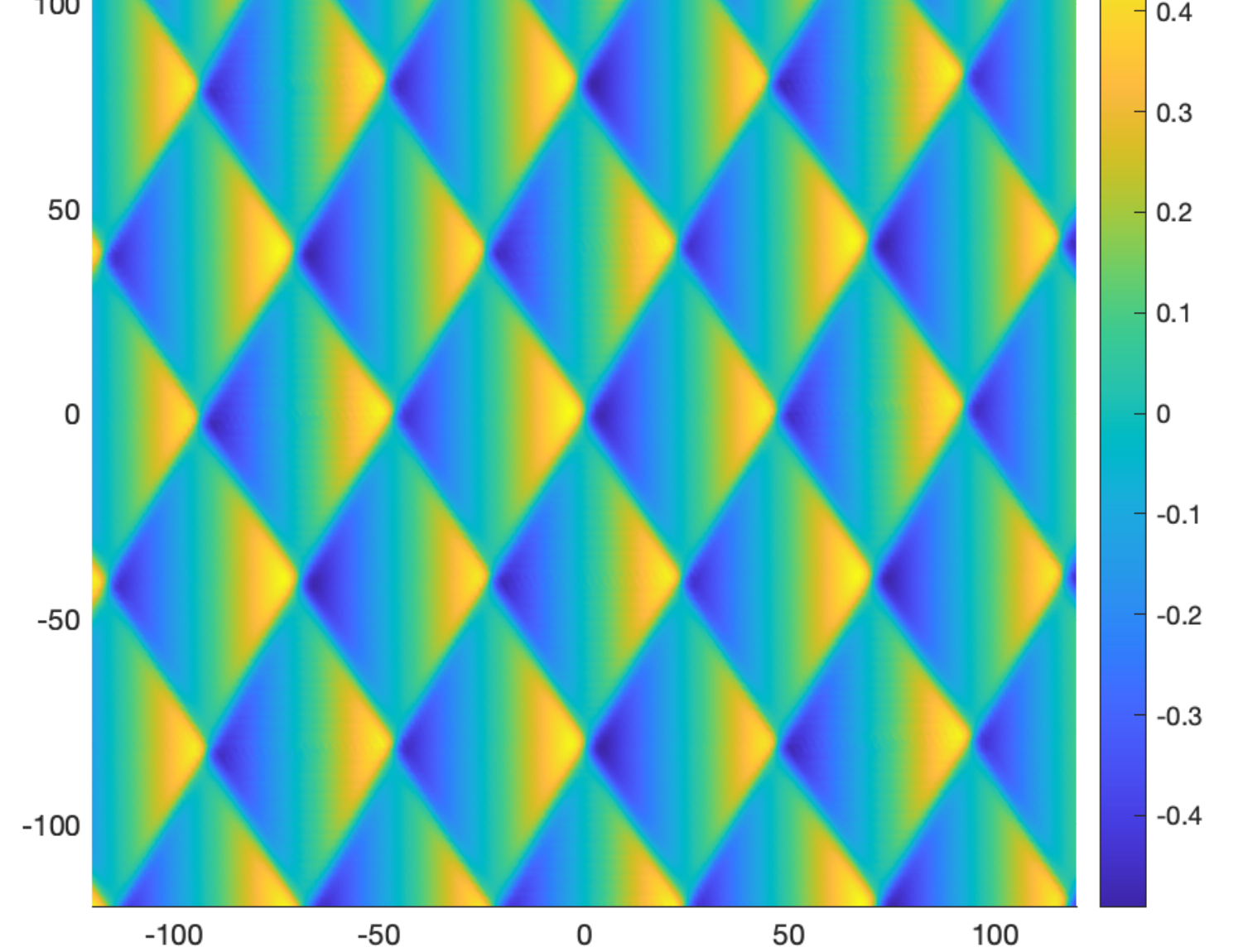}
  %\caption{xi2\_LAMMPS\_vl.pdf.}
  \caption{$\xi_{2}$, Discrete Model.}
  \label{fr11}
\end{subfigure}

\vspace*{-.5in}

\begin{subfigure}{0.3\textwidth}
  \includegraphics[width=\textwidth, clip, trim=1.25in 3in 1.25in 1.5in]
                  {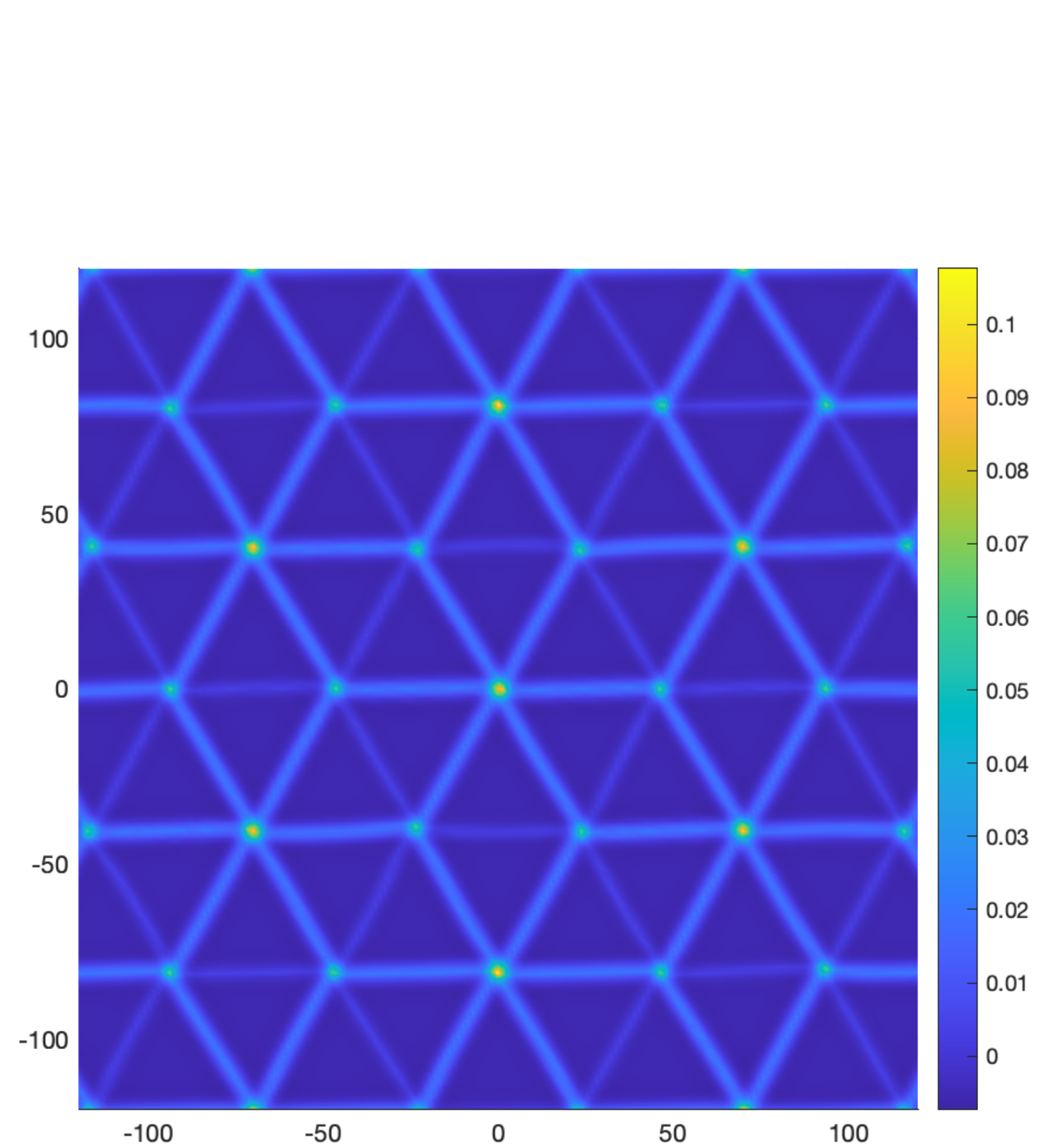}
  %\caption{eta\_COMSOL\_vl.pdf.}
  \caption{$\eta$, Continuum Model.}
  \label{fr4}
\end{subfigure}
\hfill
\begin{subfigure}{0.3\textwidth}
  \includegraphics[width=\textwidth, clip, trim=1.25in 3in 1.25in 1.5in]
                  {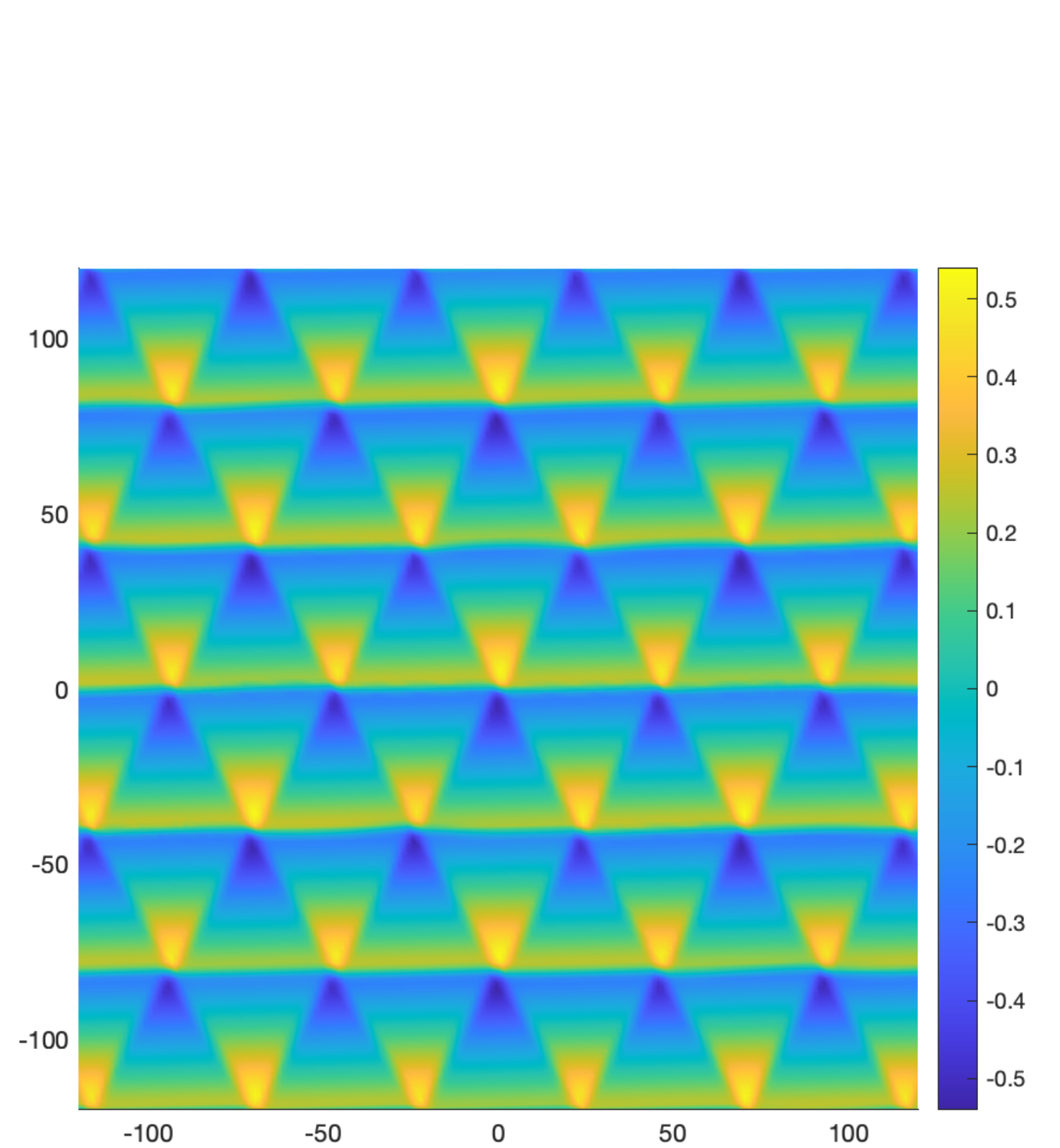}
  %\caption{xi1\_COMSOL\_vl.pdf.}
  \caption{$\xi_{1}$, Continuum Model.}
  \label{fr10}
\end{subfigure}
\hfill
\begin{subfigure}{0.3\textwidth}
  \includegraphics[width=\textwidth, clip, trim=1.25in 3in 1.25in 1.5in]
                  {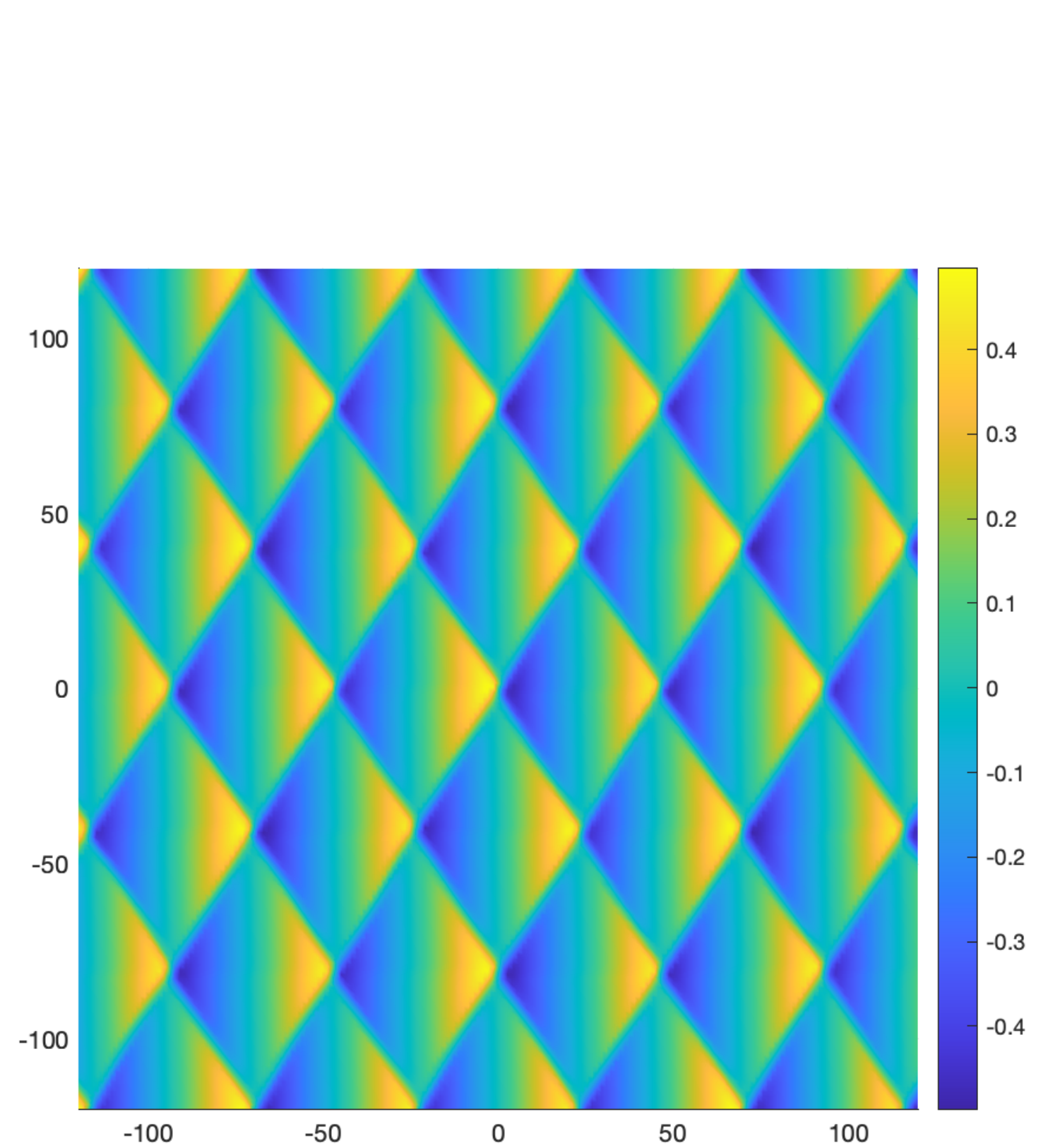}
  %\caption{xi2\_COMSOL\_vl.pdf.}
  \caption{$\xi_{2}$, Continuum Model.}
  \label{fr12}
\end{subfigure}

\caption{Simulation Results for Lennard-Jones Potential with
  $\varepsilon=0.0047$ and $\omega=0.5$.}
\label{fr22}
\end{figure}

For the Lennard-Jones potential, we start by comparing results for two
different values of $\varepsilon$.
Figure~\ref{fr19} shows results for $\varepsilon=0.0094$ and
Figure~\ref{fr22} shows results for $\varepsilon=0.0047$.
Each set of plots shows the out-of-plane displacement $\eta$ and the two
in-plane displacements $\xi_{1}$ and $\xi_{2}$ for both the discrete and
continuum models. 
The parameter values used in these simulations are given in
Tables~\ref{tt3}(a), (b).

In the plots shown in Figure~\ref{fr22}, the
lateral extent of the system is twice the lateral extent of the plots
in Figure~\ref{fr19}.  This difference reflects how
changes in $\varepsilon$ are implemented in the discrete simulations.
The continuum simulations are based on the partial differential equations
\eqref{e17},
in which $\varepsilon$ appears explicitly.  Hence we directly set the
value of $\varepsilon$.
However, $\varepsilon$ does not appear explicitly in the discrete
energy.  Instead, recalling that $\varepsilon=\sigma/L$ we run
simulations corresponding to different values of $\varepsilon$ by
changing the lateral size $L$ of the system, i.e., by making the fixed
and the deformable lattices smaller or larger.
%Figure~\ref{fr101} shows surface plots of $\xi_{1}$ and $\xi_{2}$ for
%the same parameters used in the simulation results shown in
%Figure~\ref{fr22}.  

%In Figures~\ref{fr19} and \ref{fr22}, we note that the values of the
%out-of-plane displacements are different, as indicated by the color
%bars to the right of each plot.
%%
%This occurs because ... the initial conditions are implemented
%differently in the discrete and continuum simulations ... .
%%
%Despite this difference, the ranges of the vertical displacements are
%approximately the same in Figures~\ref{fr1} and \ref{fr2}, about .14
%for the former and .12 in the latter.  The ranges in Figures~\ref{fr3}
%and \ref{fr4} are approximately the same.

%We note that the agreement between the predictions of the discrete and
%the continuum models improves as $\varepsilon$ gets smaller.  ... What
%is the evidence for this statement? ... We shall
%see that this is also true for the results predicted by the
%Kolmogorov-Crespi model.

The main observation from Figures~\ref{fr19} and \ref{fr22} is that
for both values of $\varepsilon$, we see good agreement between the
discrete and the continuum models.
We note that, for the out-of-plane displacement $\eta$, our
simulations predict spatial patterns that have been observed for
twisted graphene bilayers in many studies
\cite{van2015relaxation,van2014moire,PhysRevB.96.075311,jain2016structure,enaldiev2020stacking}.
The plots of $\eta$ in
Figures~\ref{fr1}, \ref{fr2}, \ref{fr3}, and \ref{fr4} 
exhibit hot spots, which are regions of relatively
large out-of-plane displacement localized about a point.  Neighboring
hot spots are connected by straight ridges or wrinkles, which are
regions of relatively large out-of-plane displacement localized about
the lines joining the hot spots.  These wrinkles form domain walls
between relatively large triangular commensurate regions.

In this same set of plots, we see that the hot spots occur at 2
different possible heights.  In this case, we refer to the wrinkles
emanating from the higher hot spots as \textit{primary wrinkles} and
the other wrinkles, which connect 2 lower hot spots, as
\textit{secondary wrinkles}.
We note that the secondary wrinkles are less discernible in the plots
from the discrete simulations compared to the plots from the continuum
simulations.  We can see this when comparing Figures~\ref{fr1} and
\ref{fr2} and when comparing Figures~\ref{fr3} and \ref{fr4}.  In
Figure~\ref{fr3}, it is difficult to discern any secondary wrinkles.
Another observation is that as $\varepsilon$ decreases, the hot
spots and the connecting wrinkles become more spatially concentrated,
as expected by the form of the energy for the continuum model.

%%%%%%%%%%%%%

Next we consider the effect of changing the value of the parameter
$\omega$, which appears in the Lennard-Jones potential and measures
the strength of the interaction, or the well-depth.
See~\eqref{eq:DiscreteE_w}.  Recall that we rescale the discrete energy
by $\omega$ when we derived the continuum model.  Decreasing
$\omega$ increases the value of the dimensionless elastic constants
$\gamma_{s}$, $\gamma_{t}$, and $\gamma_{d}$. We consider results for
two values of $\omega$. In Figure~\ref{fr22}, we see results for
$\omega=0.5$.  
%
%As noted above, the parameter values used in these simulations are also given in
%Tables~\ref{tt3}(a), (b) except that $L=280$, $\epsilon=0.004$, and 
%

\begin{figure}[!b]
\centering
\begin{subfigure}{0.3\textwidth}
  %trim=left botm right top ** THESE ARE MARGINS **
  \includegraphics[width=\textwidth, clip, trim=1.25in 3in 1.25in 3.4in]
                  {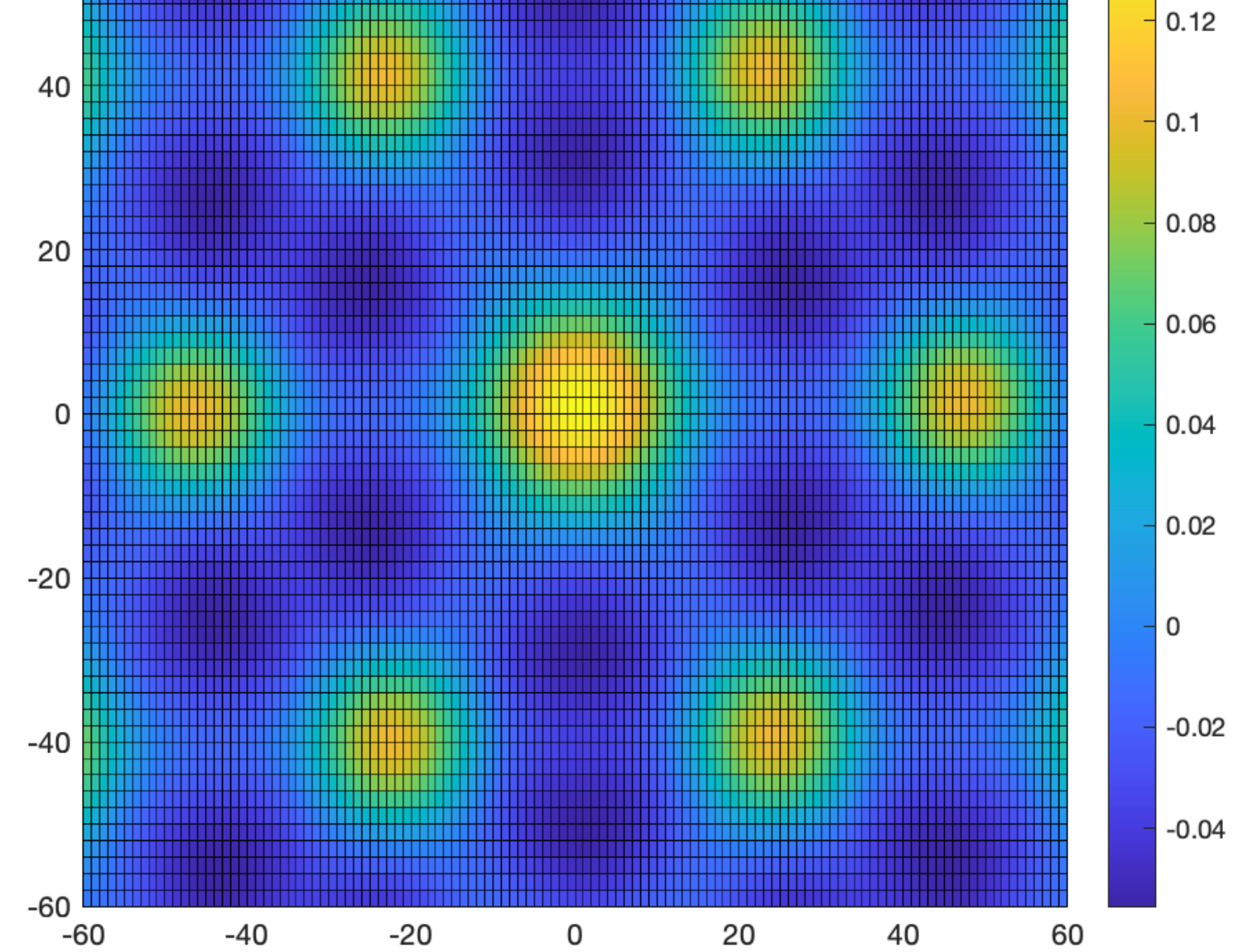}
  %\caption{eta\_LAMMPS\_vl.om.small.pdf.}
  \caption{$\eta$, Discrete Model.}
  \label{fr13}
\end{subfigure}
\hfill
\begin{subfigure}{0.3\textwidth}
  %trim=left botm right top ** THESE ARE MARGINS **
  \includegraphics[width=\textwidth, clip, trim=1.25in 3in 1.25in 3.4in]
                  {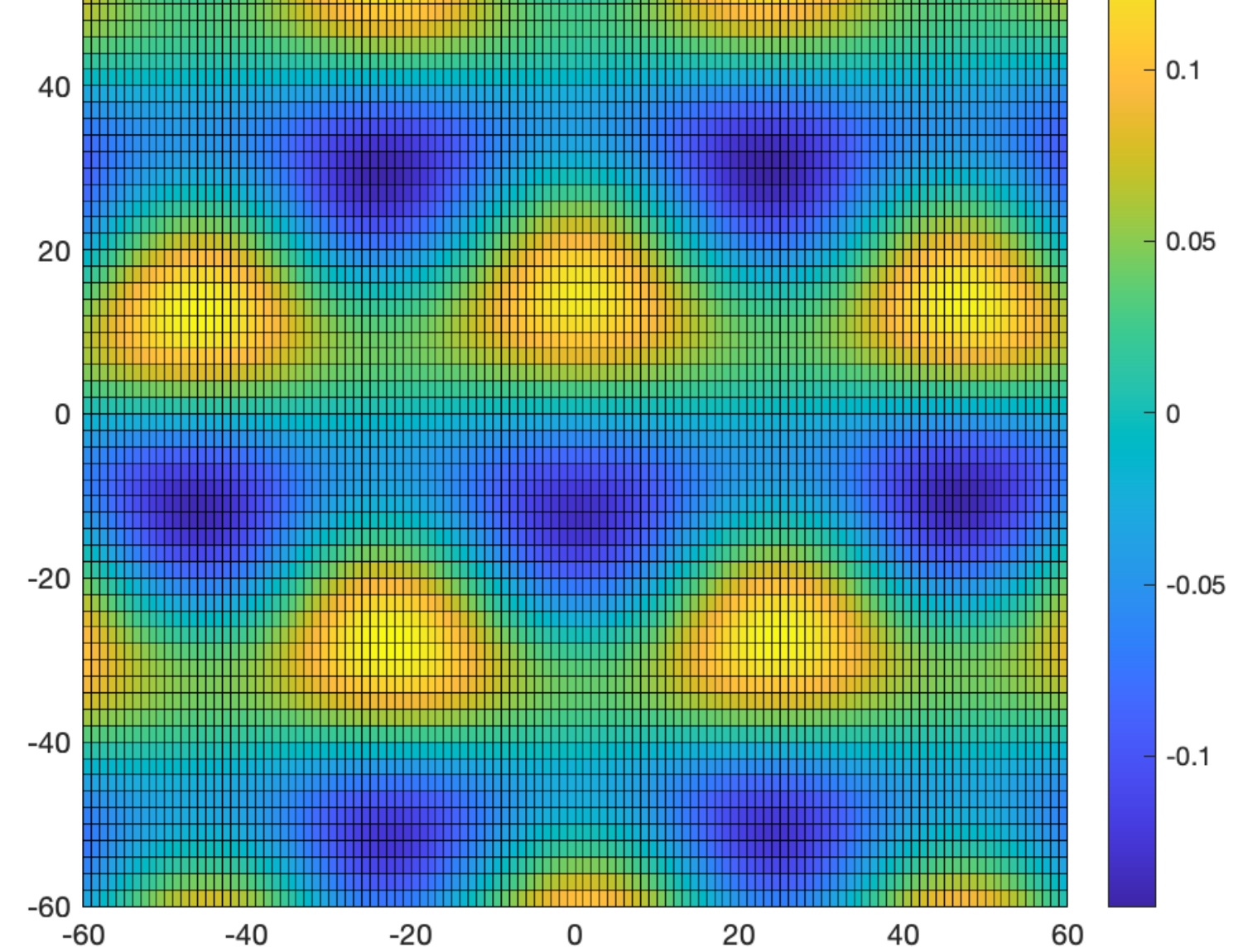}
  %\caption{xi1\_LAMMPS\_vl.om.small.pdf.}
  \caption{$\xi_{1}$, Discrete Model.}
  \label{fr15}
\end{subfigure}
\hfill
\begin{subfigure}{0.3\textwidth}
  %trim=left botm right top ** THESE ARE MARGINS **
  \includegraphics[width=\textwidth, clip, trim=1.25in 3in 1.25in 3.4in]
                  {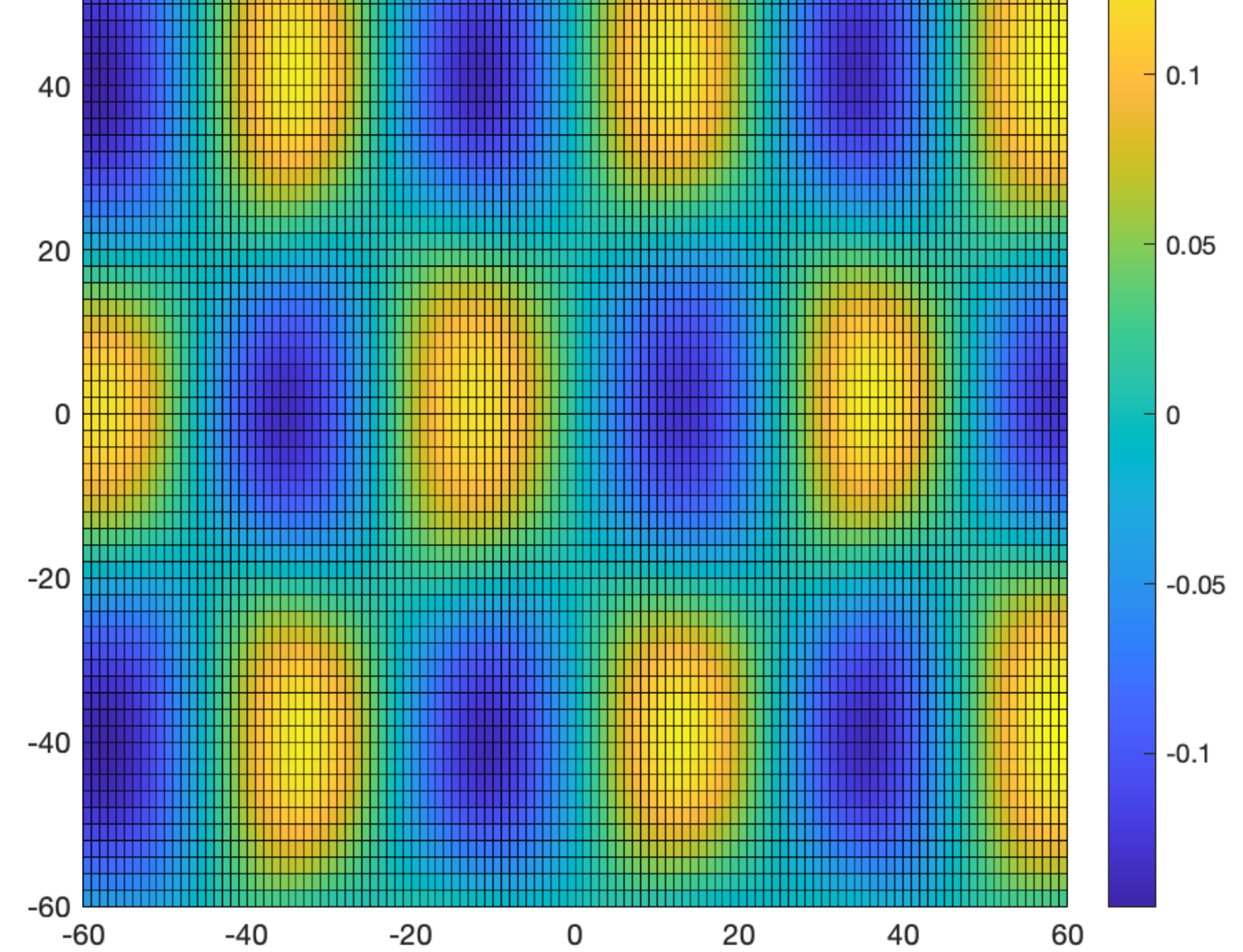}
  %\caption{xi2\_LAMMPS\_vl.om.small.pdf.}
  \caption{$\xi_{2}$, Discrete Model.}
  \label{fr17}
\end{subfigure}

\vspace*{-.5in}

\begin{subfigure}{0.3\textwidth}
  \includegraphics[width=\textwidth, clip, trim=1.25in 3in 1.25in 1.5in]
                  {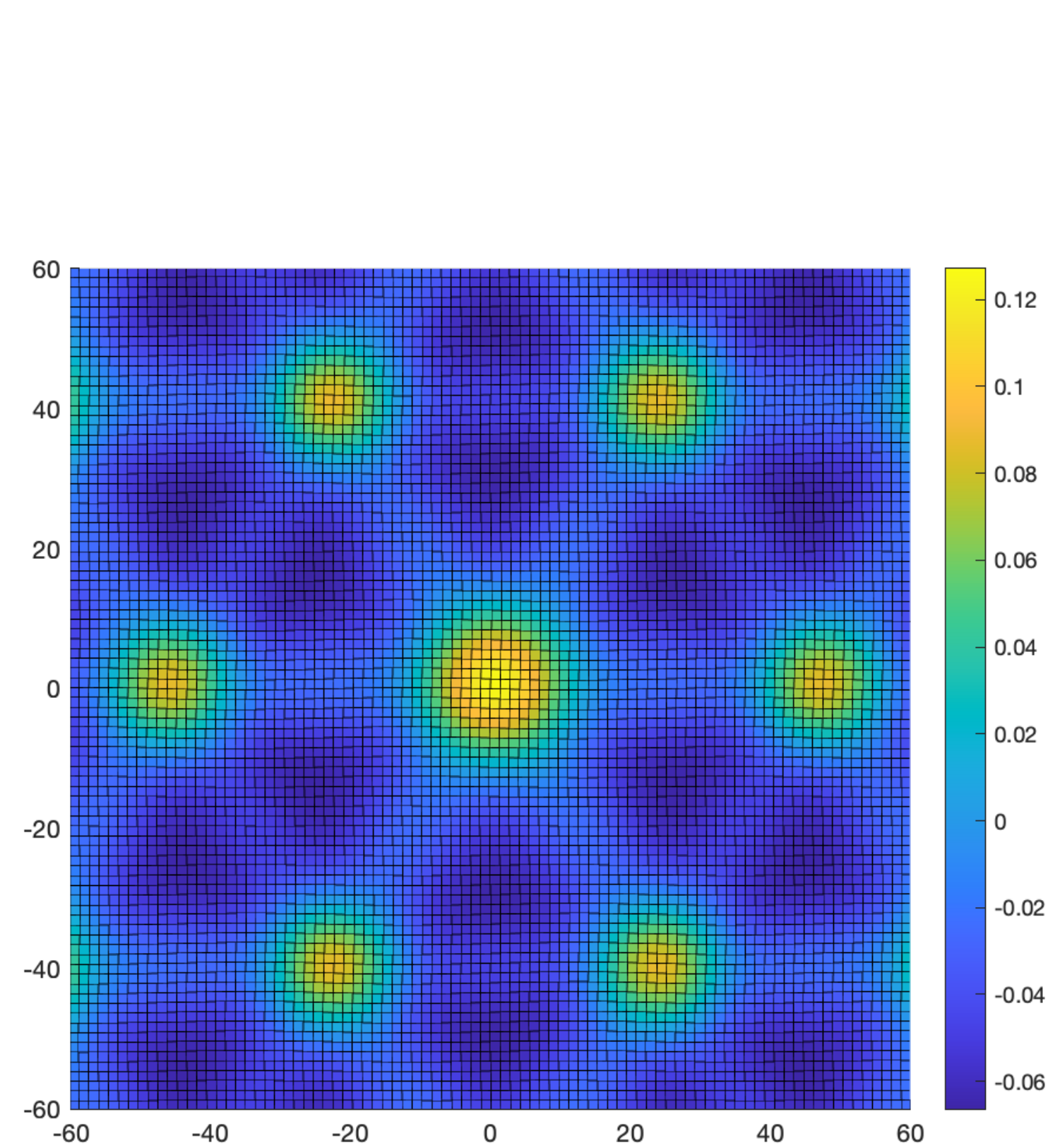}
  %\caption{eta\_COMSOL\_vl.om.small.pdf.}
  \caption{$\eta$, Continuum Model.}
  \label{fr14}
\end{subfigure}
\hfill
\begin{subfigure}{0.3\textwidth}
  \includegraphics[width=\textwidth, clip, trim=1.25in 3in 1.25in 1.5in]
                  {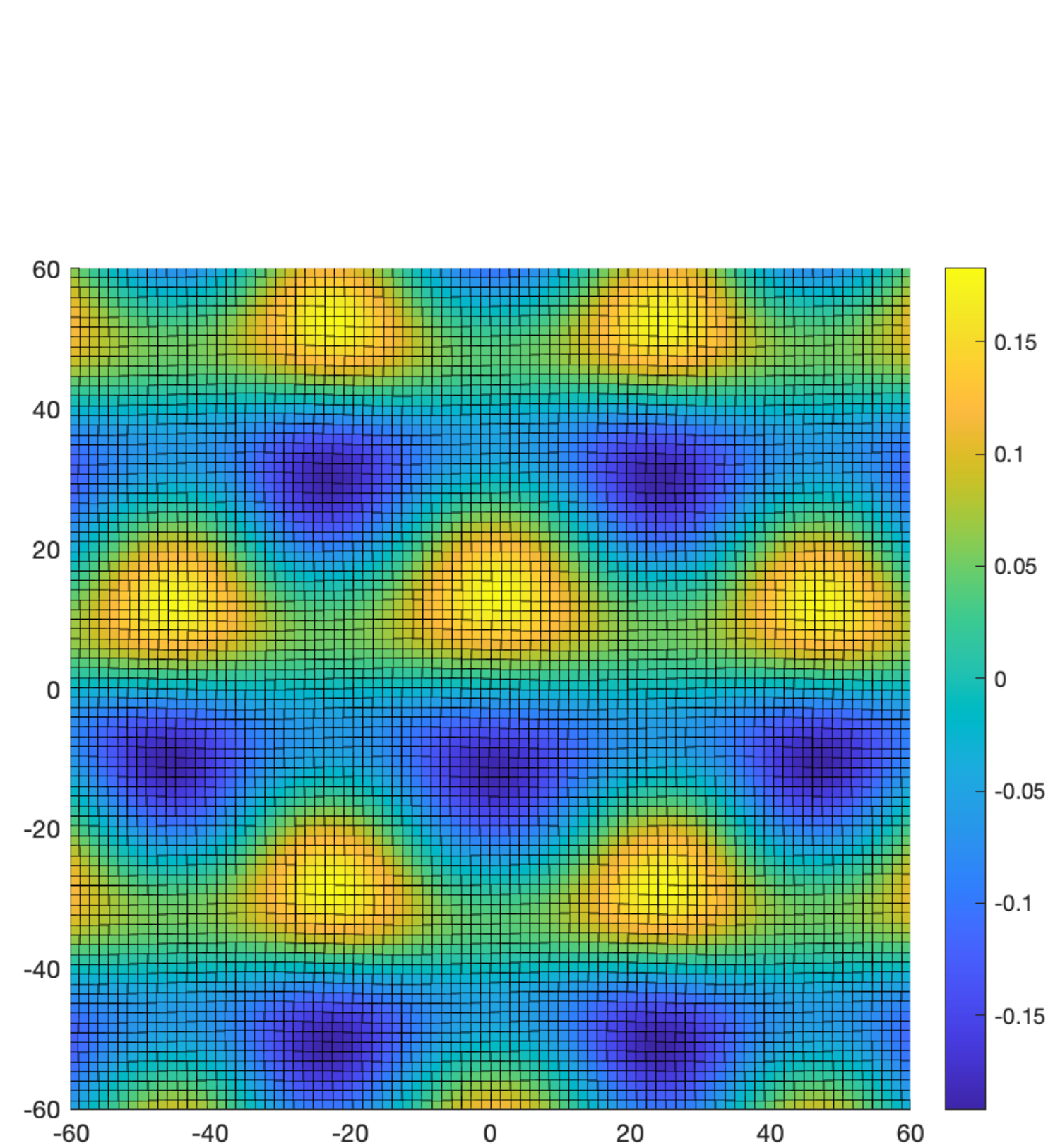}
  %\caption{xi1\_COMSOL\_vl.om.small.pdf.}
  \caption{$\xi_{1}$, Continuum Model.}
  \label{fr16}
\end{subfigure}
\hfill
\begin{subfigure}{0.3\textwidth}
  \includegraphics[width=\textwidth, clip, trim=1.25in 3in 1.25in 1.5in]
                  {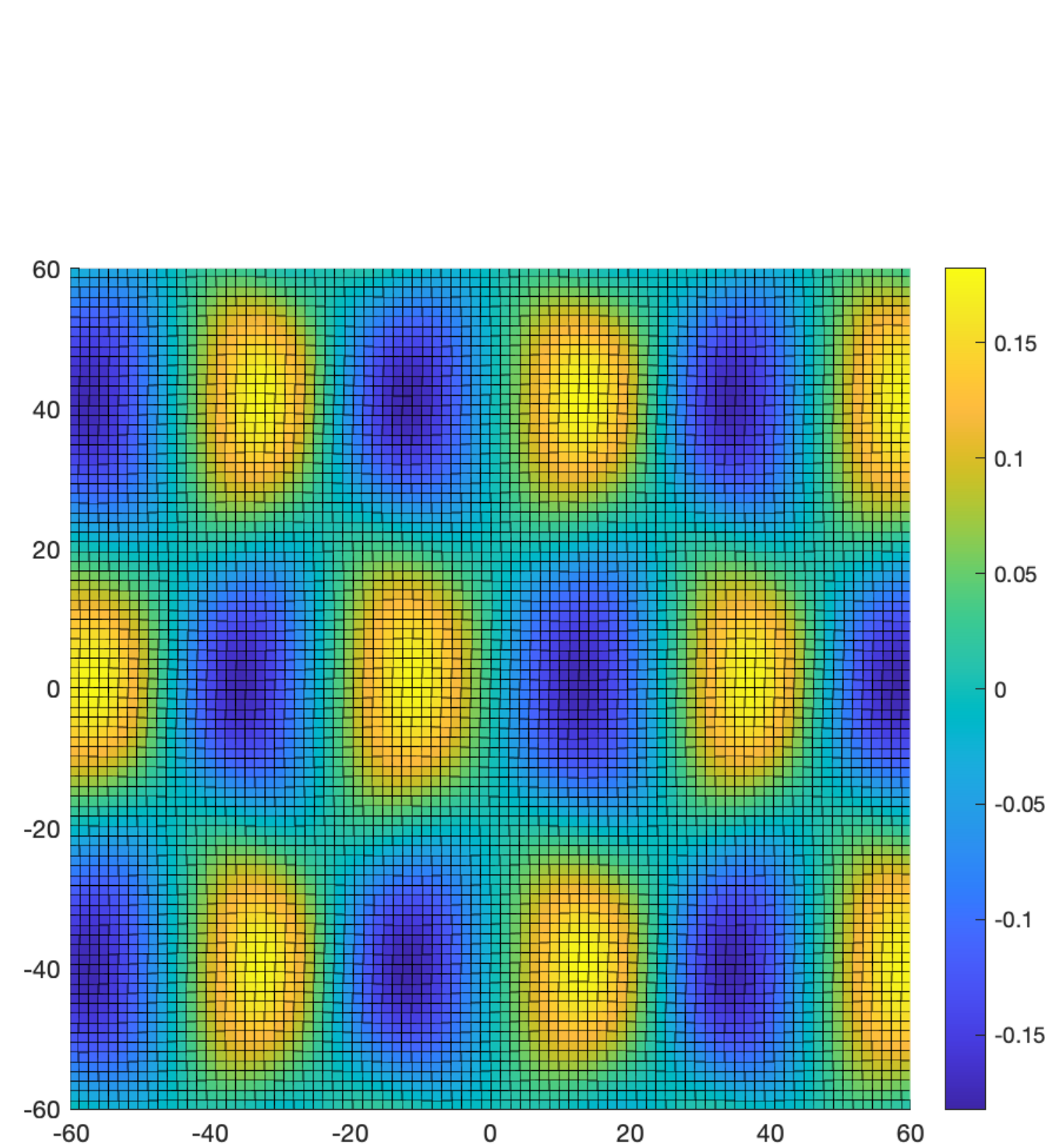}
  %\caption{xi2\_COMSOL\_vl.om.small.pdf.}
  \caption{$\xi_{2}$, Continuum Model.}
  \label{fr18}
\end{subfigure}

\caption{Simulation Results for Lennard-Jones Potential with
  $\varepsilon=0.0047$ and $\omega=0.0083$.}
\label{fr25}
\end{figure}

We now compare these plots to the results depicted in 
Figure~\ref{fr25}, for which the parameter values are the
same as those used for Figure~\ref{fr22} except that
$\omega=0.0083$ and that consequently the values of $\gamma_{s}$,
$\gamma_{t}$, and $\gamma_{d}$ are different. See Tables~\ref{tt3}(a), (b).
Note that the difference in lateral extent seen in these two sets of
plots is not because $\varepsilon$ is different.  Rather, in
Figure~\ref{fr25}, we show only a subset of the full domain in order
to better exhibit the details of each plot.
We see that for a smaller value of $\omega$, the hot spots and the wrinkles
become more spatially diffuse and the triangular commensurate regions
occupy a relatively smaller part of the lattice.
For smaller $\omega$, there is better agreement between the
results predicted by the discrete and the continuum models.  In
particular, we see good agreement between the heights of the
secondary wrinkles when comparing
Figure~\ref{fr13} to Figure~\ref{fr14}.
Decreasing the value of $\omega$ improves the match because it
increases the ratio of the elastic constants to the interaction
constant.  See \cite{espanol2017discrete} for further discussion of
this issue.

% Discussion of surface plots in Figure~\ref{fr101}

%%%%%%%%%%%%%%%%%%%%%%%%%%%%%%%%%%%%%%%%%%%%%%%%%%%%%%%%%%%%%%%%%%%%%%%%%%%%%%%
% Parameters

%% vertical
%\begin{table}[h!]
%\hfill
%  \begin{subtable}[h!]{.4\linewidth}
%    \label{tt1}
%    \begin{tabular}{c|l}
%      Parameter & Value \\ \hline\
%      $h_{1}$ & $3^{1/2}2^{1/6}$ \\
%      $h_{2}$ & $3^{1/2}2^{1/6}$ \\      
%      $L$ & $140$ \\
%      %$N_{2}$ & ?? \\
%      $\theta$ & $-2.76$ \\            
%%
%      $k_{s}$ & $25.2$ \\
%      $k_{t}$ & $1.5375$ \\
%      $k_{d}$ & $4.1$ \\
%%
%      $\sigma$ & $2^{1/6}$ \\
%      $\omega$ & $0.5$
%    \end{tabular}
%  \caption{Discrete.\hspace*{1.25in}}
%  \end{subtable}
%%
%\hfill
%%
%  \begin{subtable}[h!]{.4\linewidth}
%    \label{tt2}
%    \begin{tabular}{c|l}
%      Parameter & Value \\ \hline\
%      $\varepsilon$ & $0.008$\\
%      $\delta_{1}$ & $3^{1/2}$ \\
%      $\delta_{2}$ & $3^{1/2}$ \\
%%
%      $k$ & $4$ \\
%      $m$ & $-2$ \\
%      $\alpha$ & $0$ \\
%      $\Theta$ & $-6$ \\
%%
%      $\gamma_{s}$ & 174.58\\
%      $\gamma_{t}$ & 113.62\\
%      $\gamma_{d}$ & 3.55 
%    \end{tabular}
%    \caption{Continuum.\hspace*{1.25in}}
%  \end{subtable}
%  \hfill
%  \caption{Base-case Parameters Values for Simulations using
%    Lennard-Jones Potential.  \label{tt3}}
%\end{table}

% horizontal
\begin{table}[h!]
\hspace*{.12\linewidth}
  \begin{subtable}[h!]{.76\linewidth}
    \label{tt1}
    \begin{tabular}{l|c|c|c|c|c|c|c|c}
             &
      $h_{1}$, $h_{2}$ &      
      $L$    & 
      $\theta$ &   
      $k_{s}$  & 
      $k_{t}$  & 
      $k_{d}$  & 
      $\sigma$ & 
      $\omega$ \\  \hline
      Figures~\ref{fr19} &
      $3^{1/2}2^{1/6}$ &
      $120$   &          
      $-3.22$ &
      $25.2$   &         
      $1.5375$ &
      $4.1$    &        
      $2^{1/6}$ &
      $0.5$\\ 
      Figures~\ref{fr22}&%, \ref{fr101} &
      $3^{1/2}2^{1/6}$ &
      $240$   &          
      $-1.6$ &
      $25.2$   &         
      $1.5375$ &
      $4.1$    &        
      $2^{1/6}$ &
      $0.5$\\ 
      Figures~\ref{fr25} &
      $3^{1/2}2^{1/6}$ &
      $240$   &          
      $-1.6$ &
      $25.2$   &         
      $1.5375$ &
      $4.1$    &        
      $2^{1/6}$ &
      $0.008$
    \end{tabular}
  \caption{Parameters for the Discrete Model.  All lengths are in
    \r{A}.  $\theta$ is in degrees.  $k_{s}$, $k_{t}$, $k_{d}$, and $\omega$ are in eV.} %\hspace*{1.25in}
  \end{subtable}
\hspace*{.1\linewidth}
  \begin{subtable}[h!]{.8\linewidth}
    \label{tt2}
    \begin{tabular}{l|c|c|c|c|c|c|c|c|c}
                      &
      $\varepsilon$   &
      $\delta_{1}$, $\delta_{2}$     &
      $k$             &
      $m$             &
      $\alpha$        &
      $\Theta$        &
      $\gamma_{s}$    &
      $\gamma_{t}$    &
      $\gamma_{d}$     \\  \hline
   Figures~\ref{fr19} &
   $0.0094$  &
   $3^{1/2}$& 
   $4$      &   
   $-2$     &
   $0$      &     
   $-6$     &
   $174.58$   &
   $113.62$   &
   $3.55  $           \\ 
   Figures~\ref{fr22}&%, \ref{fr101} &
   $0.0047$  &
   $3^{1/2}$& 
   $4$      &   
   $-2$     &
   $0$      &     
   $-6$     &
   $174.58$   &
   $113.62$   &
   $3.55  $           \\ 
   Figures~\ref{fr25} &
   $0.0047$  &
   $3^{1/2}$& 
   $4$      &   
   $-2$     &
   $0$      &     
   $-6$     &
   $10474.79$  &
   $6817.35$   &
   $213.04$          
    \end{tabular}
    \caption{Dimensionless Parameters for the Continuum Model.
      $\Theta$ is in radians.} %\hspace*{1.25in}
  \end{subtable}
  \hfill
%  \vspace*{-.25in}
  \caption{Parameter Values for Simulations using
    LJ Potential.  \label{tt3}}
\end{table}

\vspace*{-.1in}

%%%%%%%%%%%%%%%%%%%%%%%%%%%%%%%%%%%%%%%%%%%%%%%%%%%%%%%%%%%%%%%%%%%%%%%%%%%%%%%

% eta

%% xi1
%\begin{figure}
%\centering
%\begin{subfigure}{0.4\textwidth}
%  %trim=left botm right top ** THESE ARE MARGINS **
%  \includegraphics[width=\textwidth, clip, trim=1.25in 1.75in 1.25in 1.5in]
%                  {./Figures-results/LJ-results/xi1_LAMMPS.pdf}
%  \caption{xi1\_LAMMPS.pdf.}
%  \label{fr5}
%\end{subfigure}
%\hfill
%\begin{subfigure}{0.4\textwidth}
%  \includegraphics[width=\textwidth, clip, trim=1.25in 1.75in 1.25in 1.5in]
%                  {./Figures-results/LJ-results/xi1_COMSOL.pdf}
%  \caption{xi1\_COMSOL.pdf.}
%  \label{fr6}
%\end{subfigure}
%
%\caption{xi1}
%\label{fr20}
%\end{figure}

%% xi2
%\begin{figure}
%\centering
%\begin{subfigure}{0.4\textwidth}
%  %trim=left botm right top ** THESE ARE MARGINS **
%  \includegraphics[width=\textwidth, clip, trim=1.25in 1.75in 1.25in 1.5in]
%                  {./Figures-results/LJ-results/xi2_LAMMPS.pdf}
%  \caption{xi2\_LAMMPS.pdf.}
%  \label{fr7}
%\end{subfigure}
%\hfill
%\begin{subfigure}{0.4\textwidth}
%  \includegraphics[width=\textwidth, clip, trim=1.25in 1.75in 1.25in 1.5in]
%                  {./Figures-results/LJ-results/xi2_COMSOL.pdf}
%  \caption{xi2\_COMSOL.pdf.}
%  \label{fr8}
%\end{subfigure}
%
%\caption{xi2}
%\label{fr21}
%\end{figure}

%%%%%%%%%%%%%%%%%%%%%%%%%%%%%%%%%%%%%%%%%%%%%%%%%

% eta_vl

vspace*{-.7in}

%% xi1_vl.om.small
%\begin{figure}
%\centering
%
%\caption{xi1 vl om small}
%\label{fr26}
%
%\end{figure}

%% xi2_vl.om.small
%\begin{figure}
%\centering
%
%\caption{xi2 vl om small}
%\label{fr27}
%
%\end{figure}

%\begin{figure}
%\centering
%
%\caption{}
%\label{fr94}
%
%\end{figure}

%\begin{figure}
%\centering
%  \includegraphics[width=\textwidth, clip, trim=1.25in 1.75in 1.25in 1.5in]
%                  {./Figures-results/LJ-results/surf_LAMMPS_vl.pdf}
%  \caption{surf\_LAMMPS\_vl.pdf.}
%  \label{fr95}
%\end{figure}
%
%
%\begin{figure}
%\centering
%  \includegraphics[width=\textwidth, clip, trim=1.25in 1.75in 1.25in 1.5in]
%                  {./Figures-results/LJ-results/surf_LAMMPS_vl.om.small.pdf}
%  \caption{surf\_LAMMPS\_vl.om.small.pdf.}
%  \label{fr96}
%\end{figure}

%\begin{figure}[h] 
%\includegraphics[width=1\linewidth]{./Figures-results/eta_COMSOL.pdf}
%\caption{eta\_COMSOL.pdf}\label{fr1}
%\end{figure}

\clearpage

%%%%%%%%%%%%%%%%%%%%%%%%%%%%%%%%%%%%%%%%%%%%%%%%%%%%%%%%%%%%%%%%%%%%%%%%%%%%%%%
%%%%%%%%%%%%%%%%%%%%%%%%%%%%%%%%%%%%%%%%%%%%%%%%%%%%%%%%%%%%%%%%%%%%%%%%%%%%%%% 

\subsection{Results for the Kolmogorov-Crespi Potential}

In this subsection, we compare the predictions of the discrete and
continuum models for the Kolmogorov-Crespi potential.  For the
discrete simulations, we generate results using the potential in
\eqref{ee2}, which we refer to as the Kolmogorov-Crespi-z potential.

\begin{figure}
\centering
\begin{subfigure}{0.3\textwidth}
  %trim=left botm right top ** THESE ARE MARGINS **
  \includegraphics[width=\textwidth, clip, trim=1.25in 3in 1.25in 3.4in]
                  {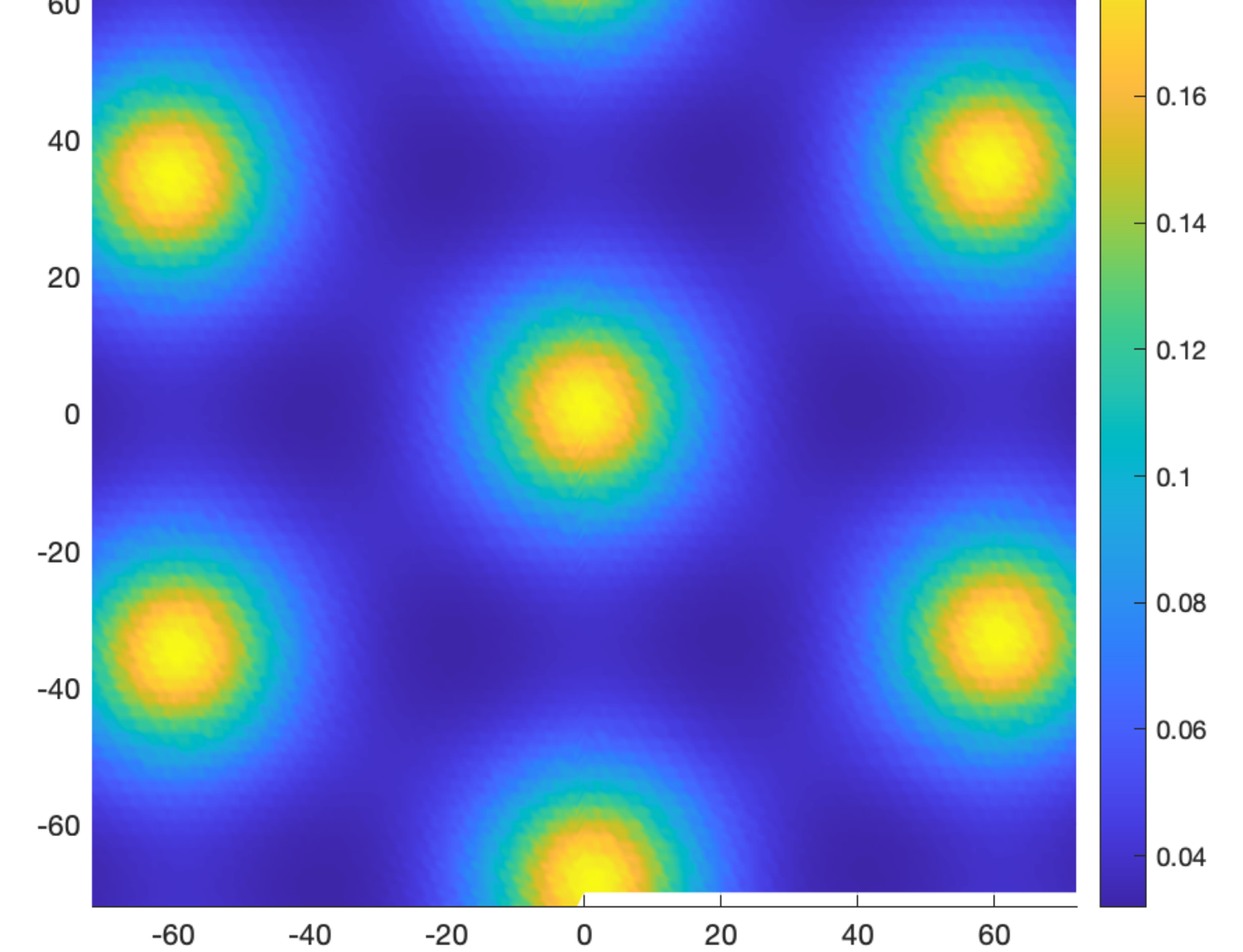}
  %\caption{eta\_LAMMPS\_half.pdf.}
  \caption{$\eta$, Discrete Model.}
  \label{fr40}
\end{subfigure}
\hfill
\begin{subfigure}{0.3\textwidth}
 %trim=left botm right top ** THESE ARE MARGINS **
  \includegraphics[width=\textwidth, clip, trim=1.25in 3in 1.25in 3.4in]
                  {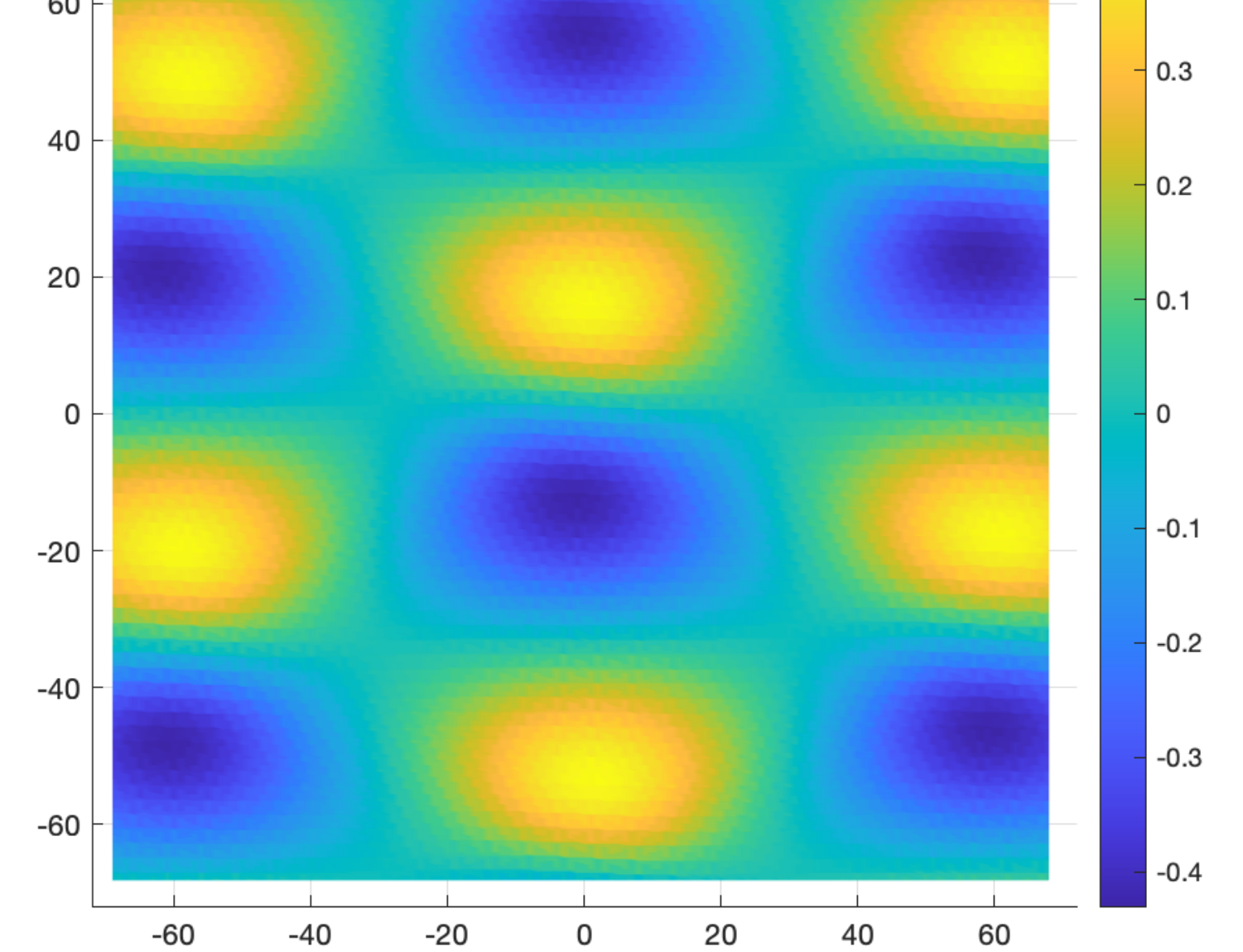}
  %\caption{xi1\_LAMMPS\_half.pdf.}
  \caption{$\xi_{1}$, Discrete Model.}
  \label{fr43}
\end{subfigure}
\hfill
\begin{subfigure}{0.3\textwidth}
 %trim=left botm right top ** THESE ARE MARGINS **
  \includegraphics[width=\textwidth, clip, trim=1.25in 3in 1.25in 3.4in]
                  {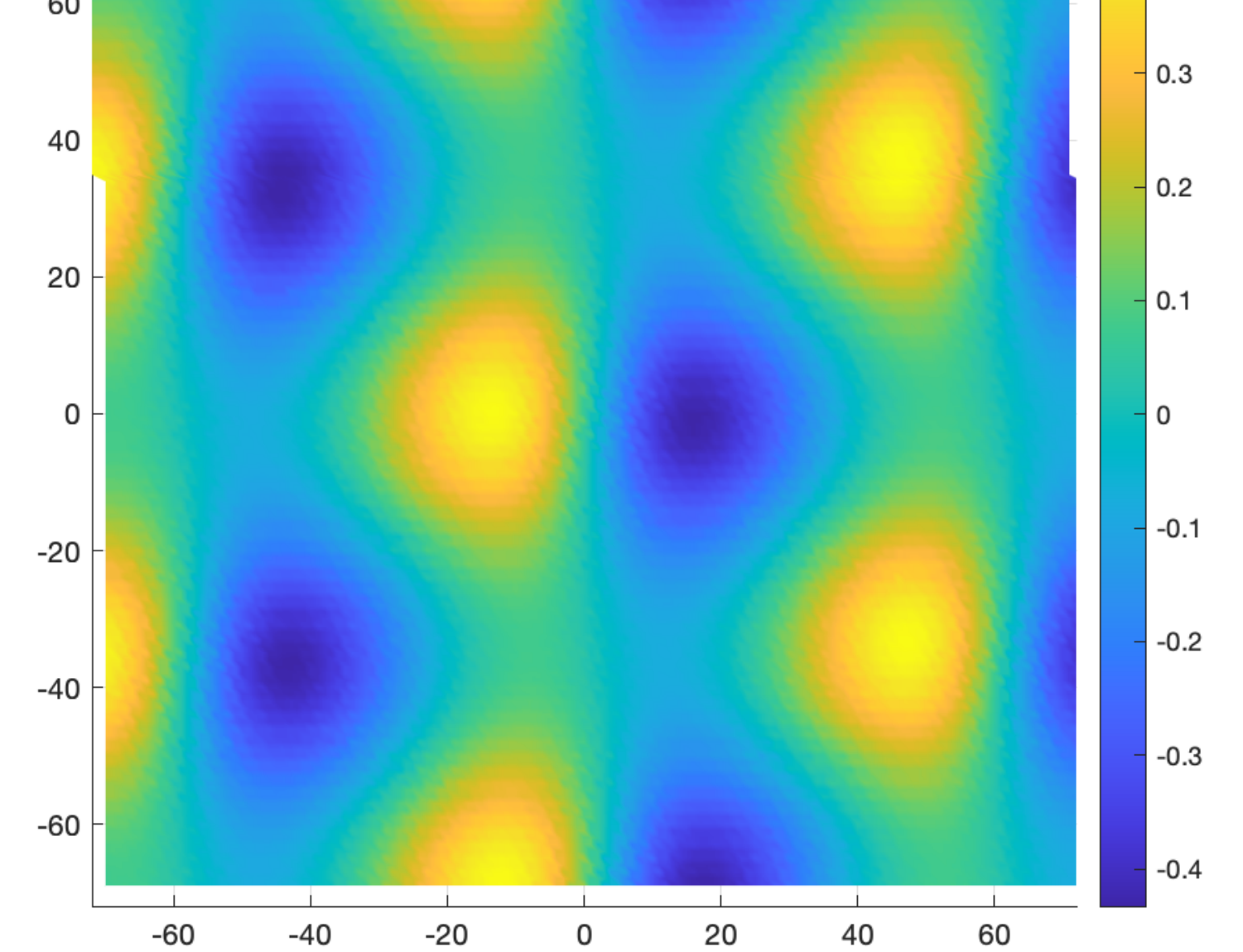}
  %\caption{xi2\_LAMMPS\_half.pdf.}
  \caption{$\xi_{2}$, Discrete Model.}
  \label{fr46}
\end{subfigure}

\vspace*{-.5in}

\begin{subfigure}{0.3\textwidth}
  \includegraphics[width=\textwidth, clip, trim=1.25in 3in 1.25in 1.5in]
                  {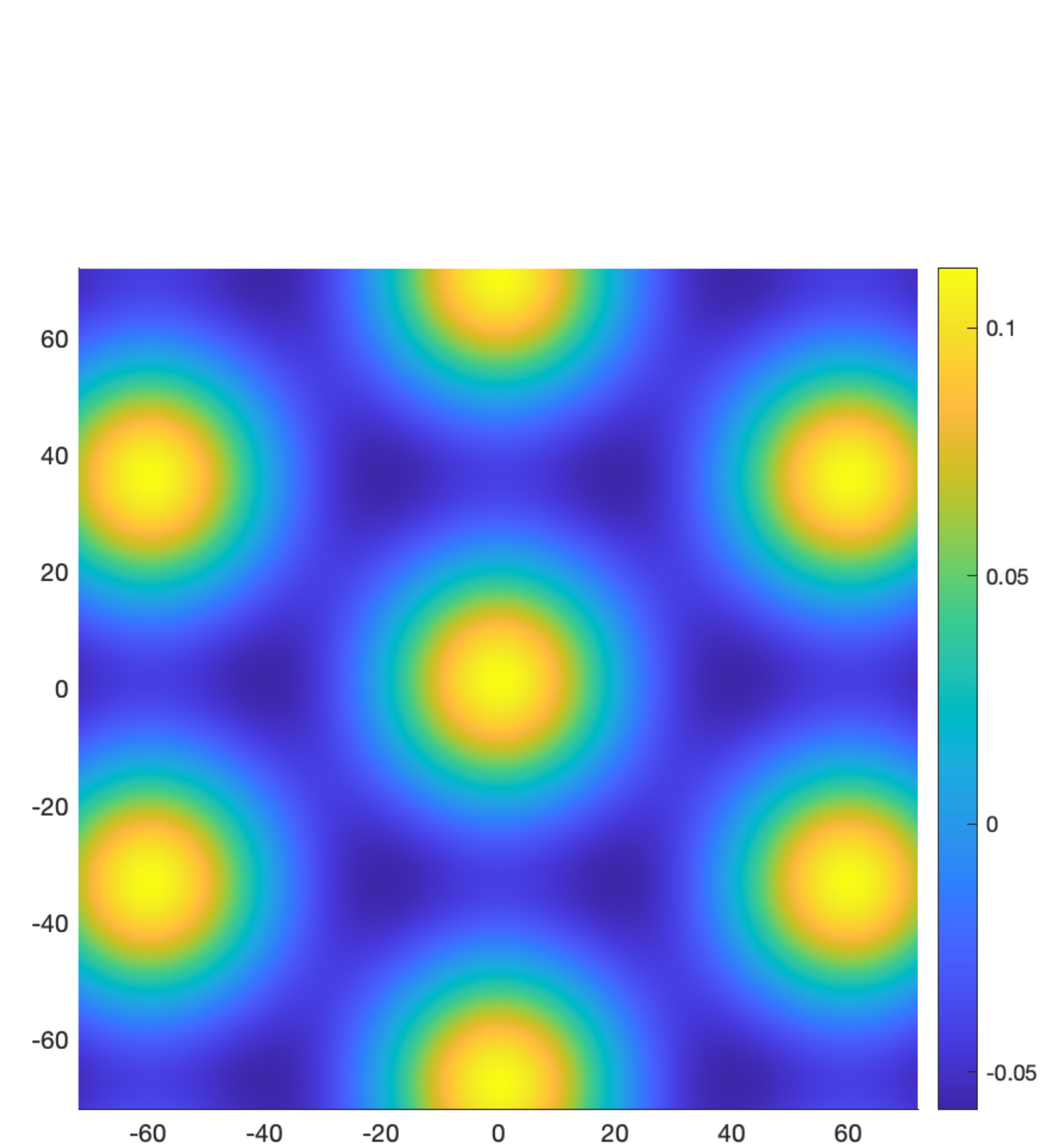}
  %\caption{eta\_COMSOLS\_half.pdf.}
  \caption{$\eta$, Continuum Model.}
  \label{fr41}
\end{subfigure}
\hfill
\begin{subfigure}{0.3\textwidth}
  \includegraphics[width=\textwidth, clip, trim=1.25in 3in 1.25in 1.5in]
                  {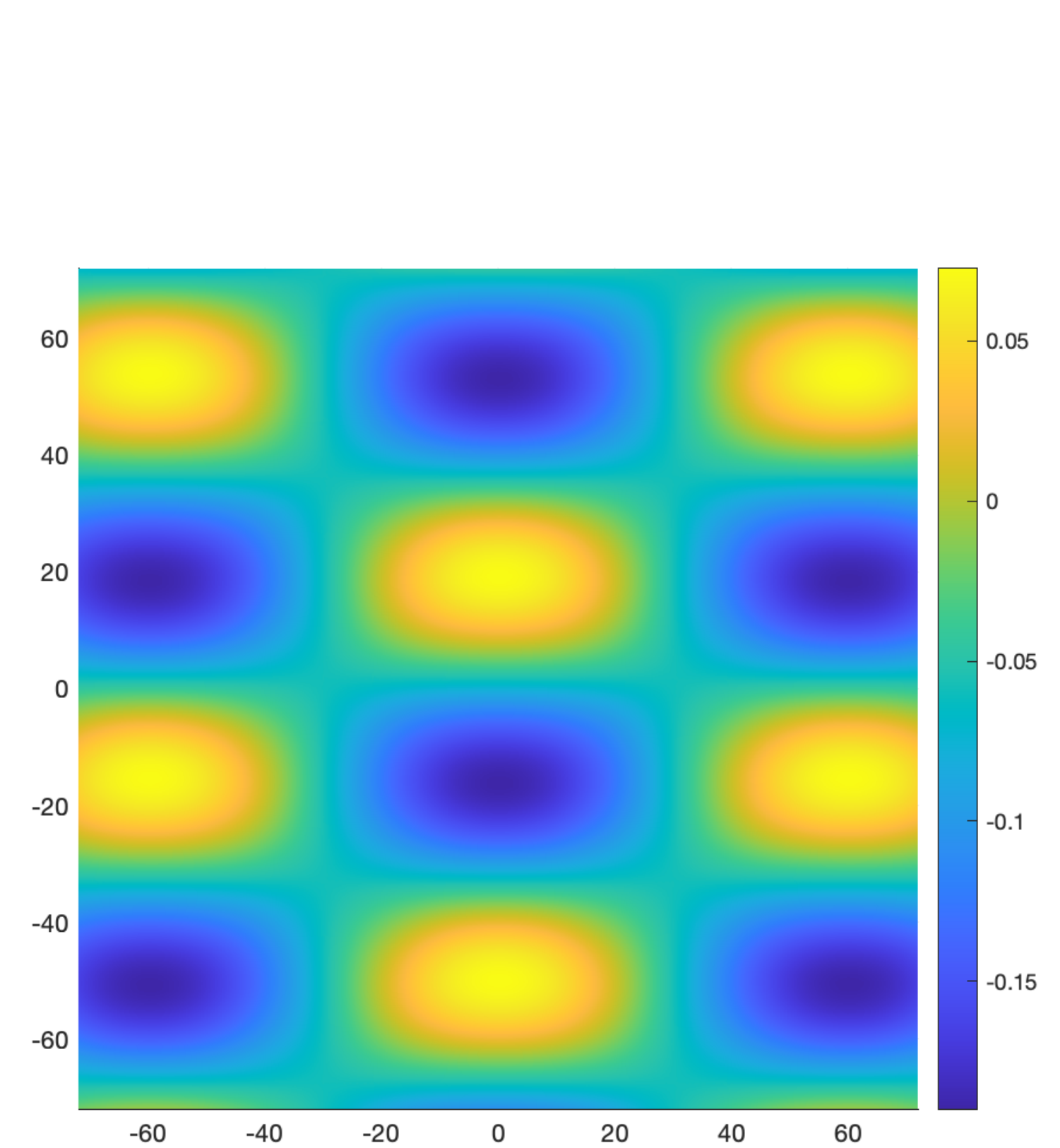}
  %\caption{xi1\_COMSOLS\_half.pdf.}
  \caption{$\xi_{1}$, Continuum Model.}
  \label{fr44}
\end{subfigure}
\hfill
\begin{subfigure}{0.3\textwidth}
  \includegraphics[width=\textwidth, clip, trim=1.25in 3in 1.25in 1.5in]
                  {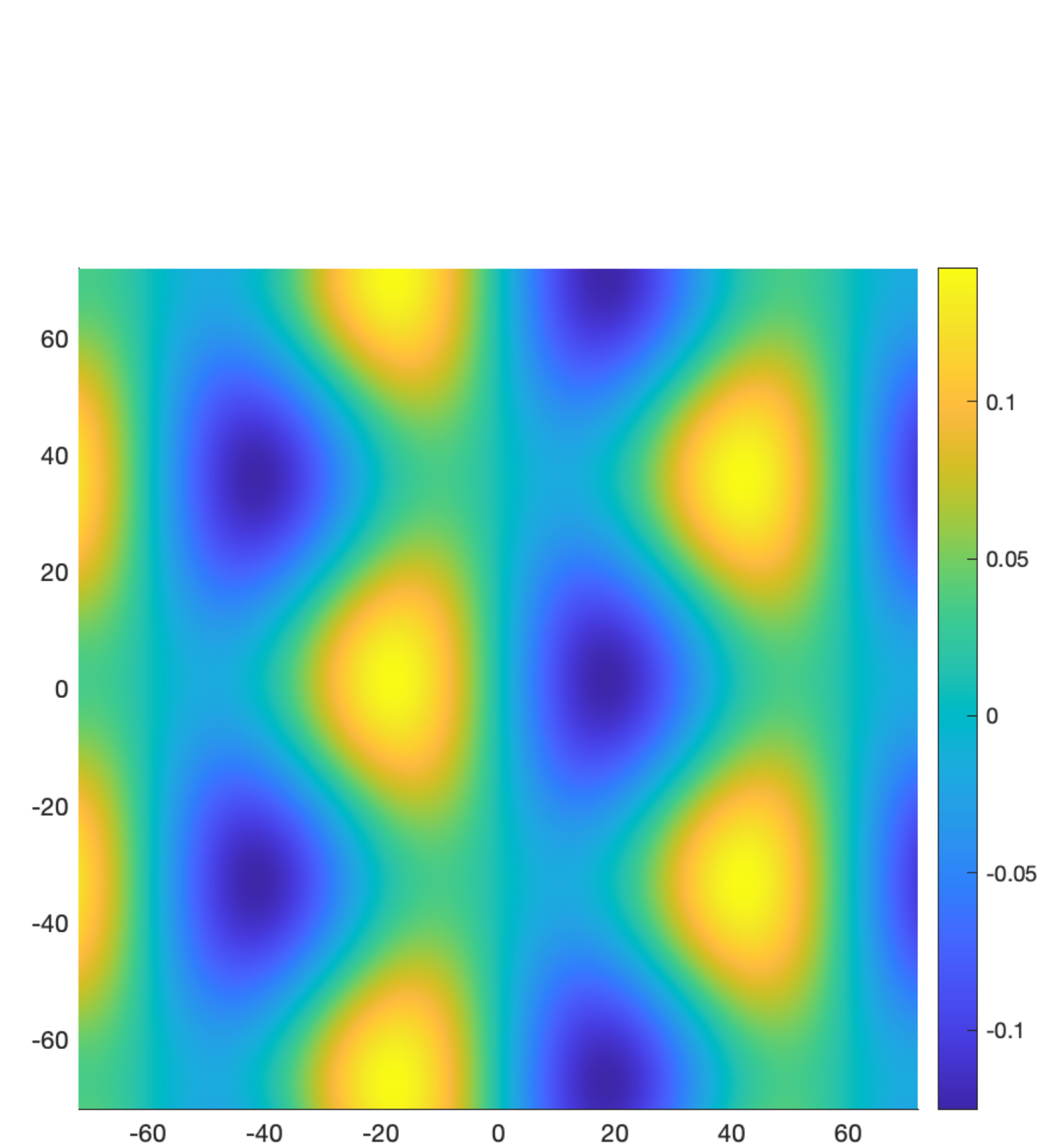}
  %\caption{xi2\_COMSOLS\_half.pdf.}
  \caption{$\xi_{2}$, Continuum Model.}
  \label{fr47}
\end{subfigure}

\caption{Simulation Results for Kolmogorov-Crespi Potential with
  $\varepsilon=0.024$.}
\label{fr42}
\end{figure}

\begin{figure}
\centering
\begin{subfigure}{0.3\textwidth}
  %trim=left botm right top ** THESE ARE MARGINS **
  \includegraphics[width=\textwidth, clip, trim=0in 0in 0in .4in]
                  {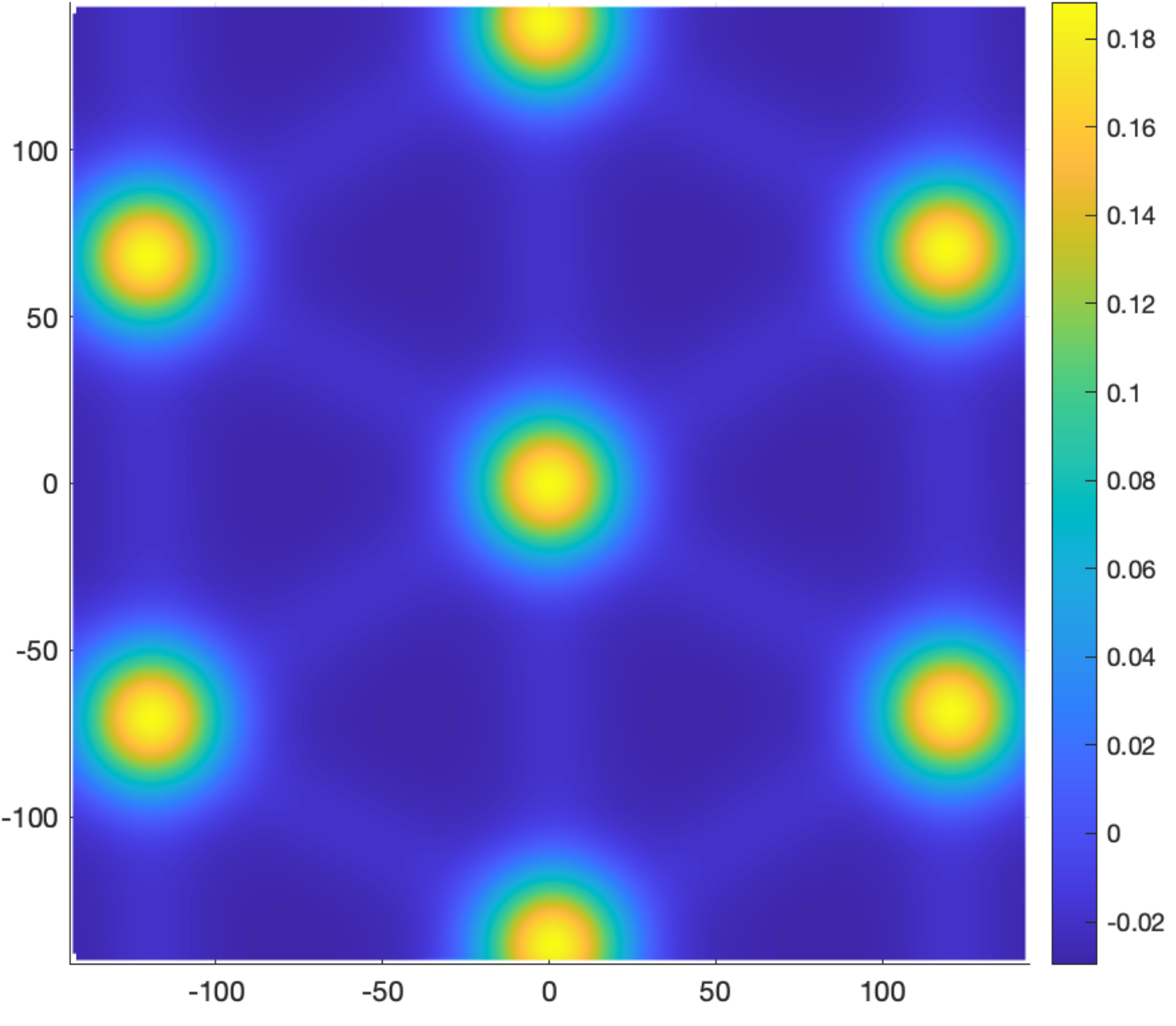}
  %\caption{eta\_LAMMPS\_full.pdf.}
  \caption{$\eta$, Discrete Model.}
  \label{fr49}
\end{subfigure}
\hfill
\begin{subfigure}{0.3\textwidth}
  %trim=left botm right top ** THESE ARE MARGINS **
  \includegraphics[width=\textwidth, clip, trim=1.25in 3in 1.25in 3.4in]
                  {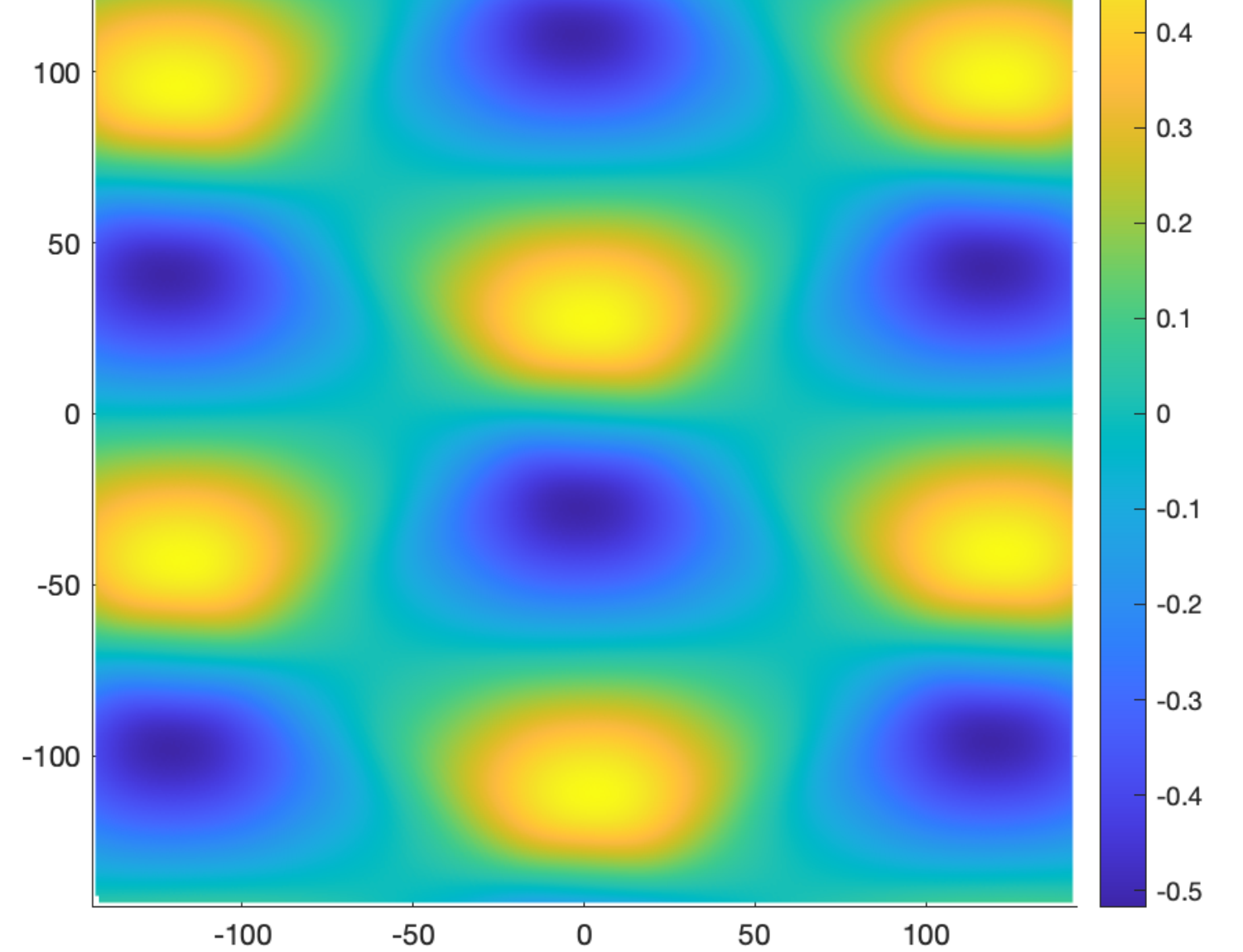}
  %\caption{xi1\_LAMMPS\_full.pdf.}
  \caption{$\xi_{1}$, Discrete Model.}
  \label{fr52}
\end{subfigure}
\hfill
\begin{subfigure}{0.3\textwidth}
  %trim=left botm right top ** THESE ARE MARGINS **
  \includegraphics[width=\textwidth, clip, trim=1.25in 3in 1.25in 3.4in]
                  {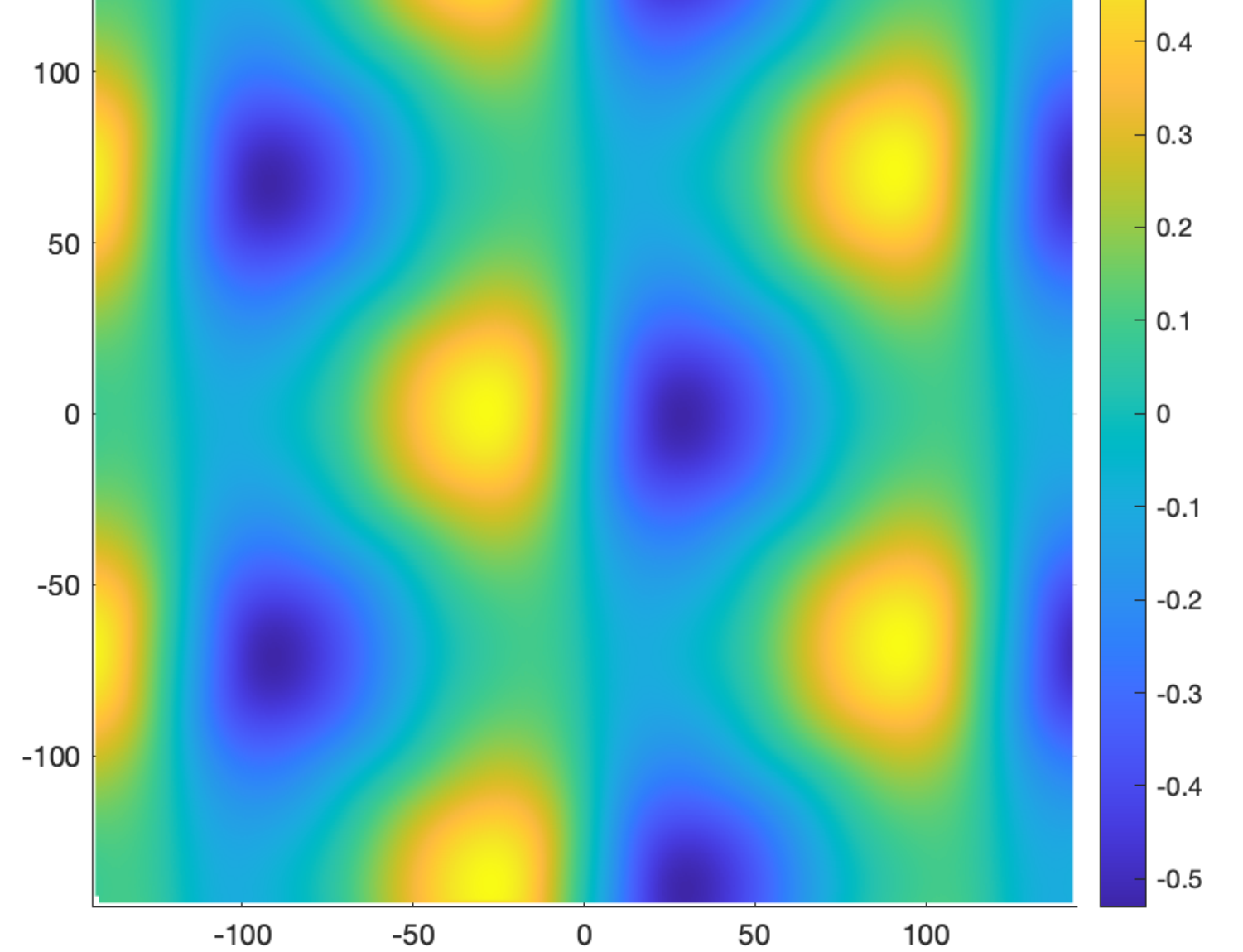}
  %\caption{xi2\_LAMMPS\_full.pdf.}
  \caption{$\xi_{2}$, Discrete Model.}
  \label{fr55}
\end{subfigure}

\vspace*{-.5in}

\begin{subfigure}{0.3\textwidth}
  \includegraphics[width=\textwidth, clip, trim=1.25in 3in 1.25in 1.5in]
                  {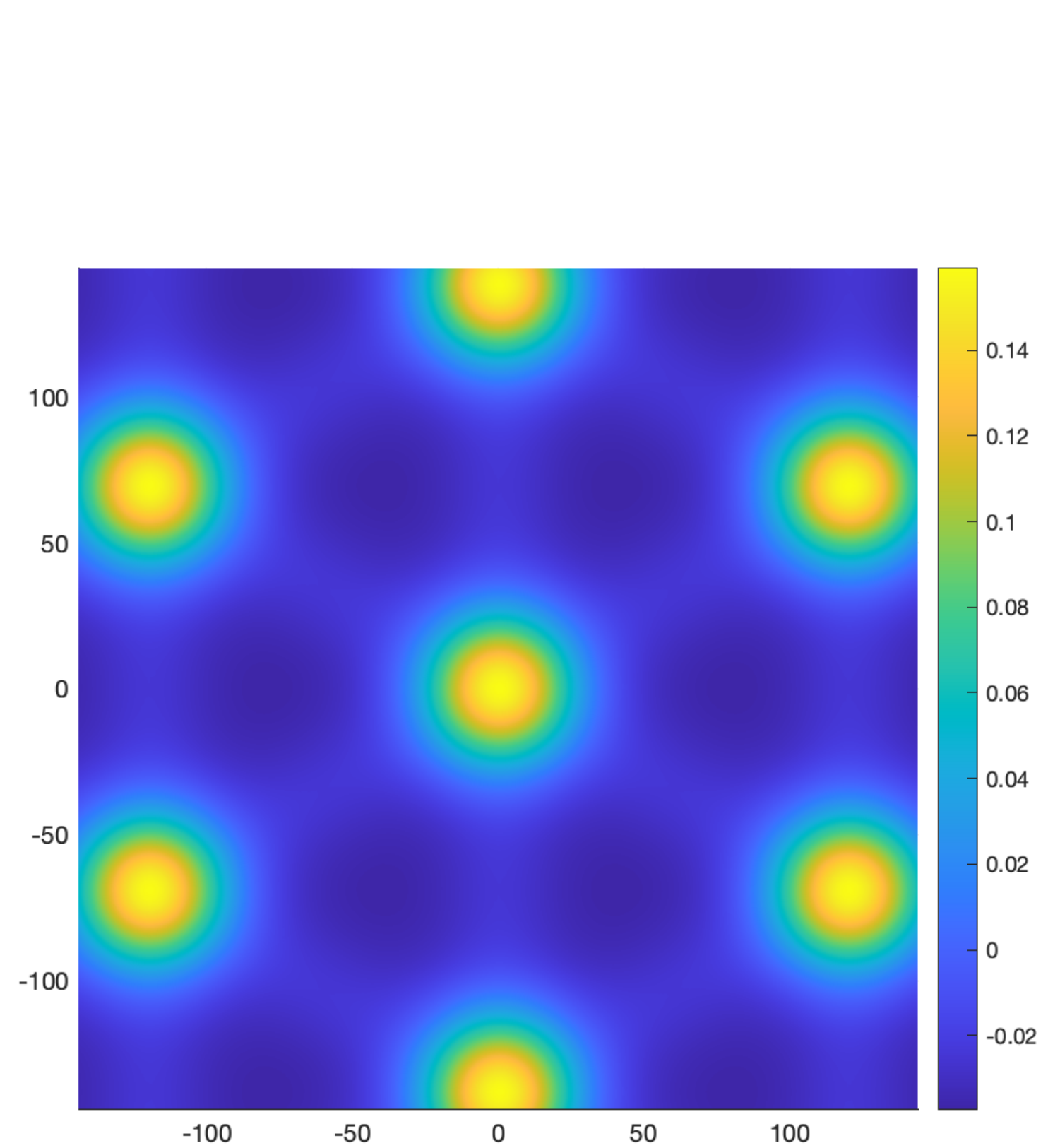}
  %\caption{eta\_COMSOLS\_full.pdf.}
  \caption{$\eta$, Continuum Model.}
  \label{fr50}
\end{subfigure}
\hfill
\begin{subfigure}{0.3\textwidth}
  \includegraphics[width=\textwidth, clip, trim=1.25in 3in 1.25in 1.5in]
                  {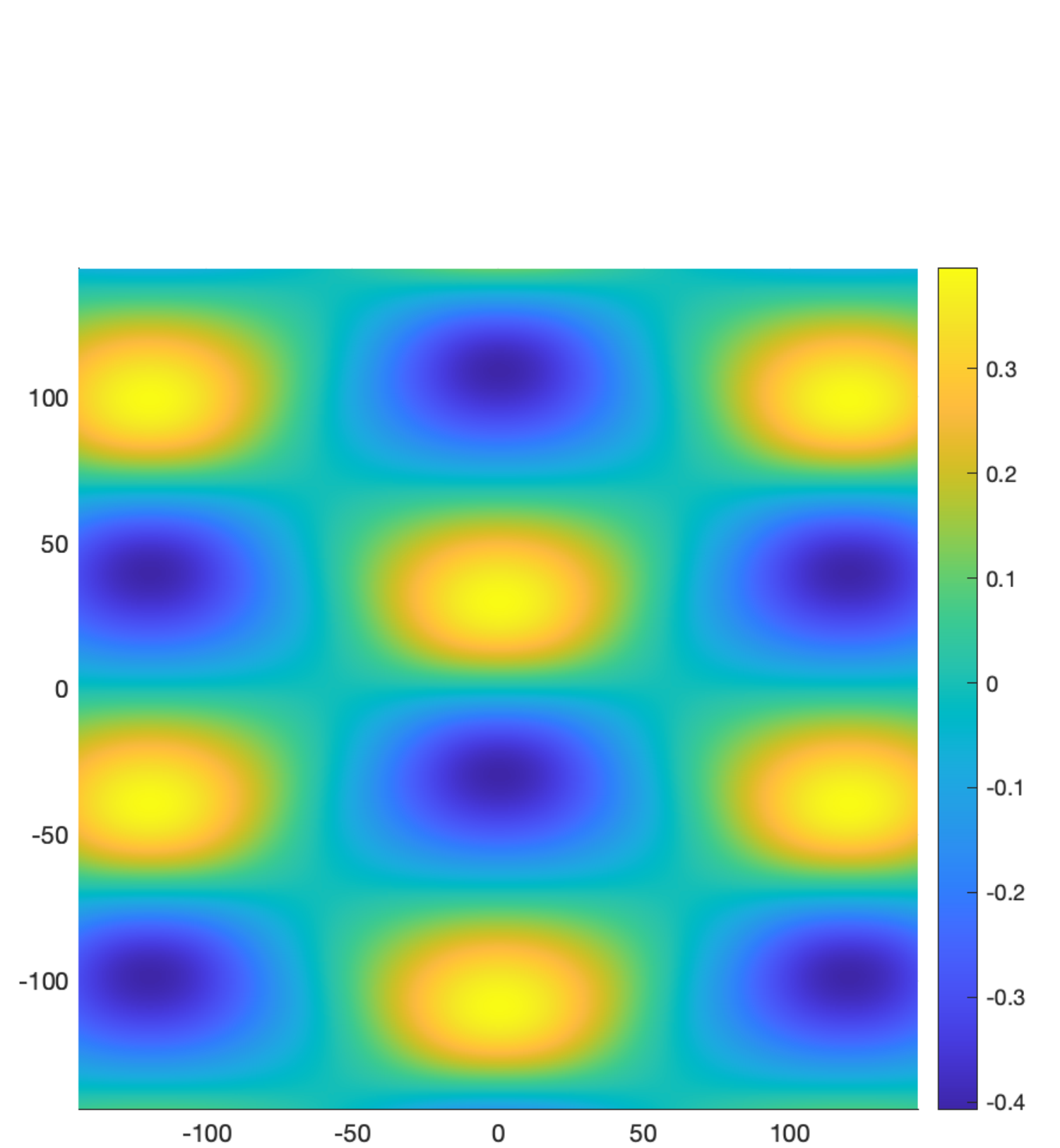}
  %\caption{xi1\_COMSOLS\_full.pdf.}
  \caption{$\xi_{1}$, Continuum Model.}
  \label{fr53}
\end{subfigure}
\hfill
\begin{subfigure}{0.3\textwidth}
  \includegraphics[width=\textwidth, clip, trim=1.25in 3in 1.25in 1.5in]
                  {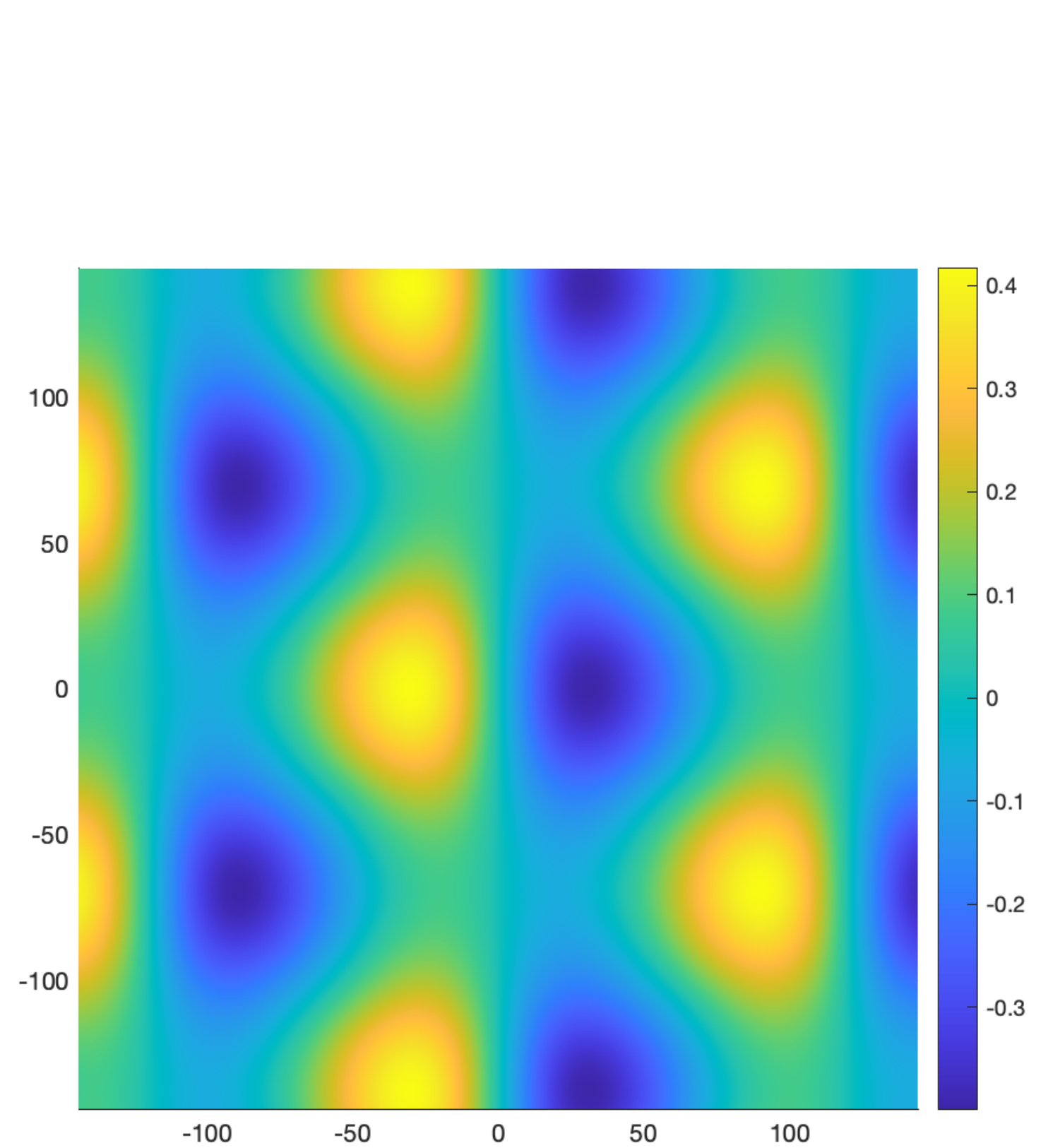}
  %\caption{xi2\_COMSOLS\_full.pdf.}
  \caption{$\xi_{2}$, Continuum Model.}
  \label{fr56}
\end{subfigure}

\caption{Simulation Results for Kolmogorov-Crespi Potential with
  $\varepsilon=0.012$.}
\label{fr51}
\end{figure}

% Smaller eps, better correspondance.
To demonstrate that the match between the discrete and the continuum
improves as $\varepsilon$ gets smaller, we consider results for two
different values of $\varepsilon$.
Figure~\ref{fr42} shows results for $\varepsilon=0.024$.
Figure~\ref{fr51} shows results for $\varepsilon=0.012$.
The other parameter values used in these simulations are listed in
Tables~\ref{tt6}(a), (b), and (c). 
As in the previous subsection, the main point is the good match we see
between the predictions of the discrete and continuum models.  
Also,
as we saw above, here we see by comparing Figures~\ref{fr40} and
\ref{fr49} and by comparing Figures~\ref{fr41} and \ref{fr50} that
reducing $\varepsilon$ spatially concentrates the hotspots and the
wrinkles.

\begin{table}[h!]
\hspace*{.12\linewidth}
  \begin{subtable}[h!]{.76\linewidth}
    \label{tt4}
    \begin{tabular}{l|c|c|c|c|c|c|c|c|c}
      &
      $h_{1}$, $h_{2}$ &      
      $L$    & 
      $\theta$ &   
      $k_{s}$  & 
      $k_{t}$  & 
      $k_{d}$  & 
      $\sigma$ & 
      $\omega$ & 
      $\delta$ \\  \hline
   Figure~\ref{fr42} &
   $2.46$  &
   $140$   &
   $-1.74$ &
    $3.23$ & 
    $0.75$ &
    $12.0$ &  
   $3.34$ &
   $0.01$ &
   $0.58$       \\
   Figure~\ref{fr51} &
   $2.46$  &
   $280$   &
   $-0.87$ &
    $3.23$ & 
    $0.75$ &
    $12.0$ &  
   $3.34$ &
   $0.01$ &
   $0.58$         

    \end{tabular}
  \caption{Parameters for the Discrete Model.  All lengths are in
    \r{A}.  $\theta$ is in degrees.  $k_{s}$, $k_{t}$, $k_{d}$, and $\omega$ are in eV.} %\hspace*{1.25in}
  \end{subtable}
\hspace*{.12\linewidth}
  \begin{subtable}[h!]{.76\linewidth}
    \label{tt5}
    \begin{tabular}{l|c|c|c|c|c|c|c|c|c}
                      &
      $\varepsilon$   &
      $\delta_{1}$, $\delta_{2}$  &
      $k$             &
      $m$             &
      $\alpha$        &
      $\Theta$        &
      $\gamma_{s}$    &
      $\gamma_{t}$    &
      $\gamma_{d}$     \\  \hline
   Figure~\ref{fr42} &
   $0.024$      &
   $0.74$     &
   $2$        &
   $-1$       &
   $0$        &
   $-1.28$    &
   $123.66$   &
   $306.64$   &
   $10.39$  \\
   Figure~\ref{fr51} &
   $0.012$      &
   $0.74$     &
   $2$        &
   $-1$       &
   $0$        &
   $-1.28$    &
   $123.66$   &
   $306.64$   &
   $10.39$ 
    \end{tabular}
    \caption{Dimensionless Parameters for the Continuum Model.
      $\Theta$ is in radians.} %\hspace*{1.25in}
  \end{subtable}

\hspace*{.34\linewidth}
  \begin{subtable}[h!]{.32\linewidth}
    \label{tt7}
    \begin{tabular}{c|c|c|c|c}
      $\tilde{\lambda}$ & 
      $\tilde{C}$       & 
      $\tilde{C}_{0}$    &
      $\tilde{C}_{2}$    &
      $\tilde{C}_{4}$    \\ \hline
  $12.12$ & 
  $0.30$  & 
  $1.53$  &
  $1.20$  &
  $0.48$   
    \end{tabular}
    \caption{Dimensionless Parameters for the KC Potential.}  %\hspace*{.5in}
%    \caption{Dimensionless Parameters for the Kolmogorov-Crespi Potential.}  %\hspace*{.5in}
  \end{subtable}  

  \caption{Parameter Values for Simulations using
    KC Potential.  \label{tt6}}
%    Kolmogorov-Crespi Potential.  \label{tt6}}
\end{table}

% eta half

%\vspace*{-.7in}

%% xi1 half
%\begin{figure}
%\centering
%
%\caption{xi1 half}
%\label{fr45}
%\end{figure}

%% xi2 half
%\begin{figure}
%\centering
%
%\caption{xi2 half}
%\label{fr48}
%\end{figure}

% eta full

%% xi1 full
%\begin{figure}
%\centering
%
%\caption{xi1 full}
%\label{fr54}
%\end{figure}

%% xi2 full
%\begin{figure}
%\centering
%
%\hfill
%
%\caption{xi2 full}
%\label{fr57}
%\end{figure}

\clearpage

%%%%%%%%%%%%%%%%%%%%%%%%%%%%%%%%%%%%%%%%%%%%%%%%%%%%%%%%%%%%%%%%%%%%%%%%%%%%%%%
%%%%%%%%%%%%%%%%%%%%%%%%%%%%%%%%%%%%%%%%%%%%%%%%%%%%%%%%%%%%%%%%%%%%%%%%%%%%%%% 

\subsection{Results for two versions of the Kolmogorov-Crespi potential}

Here we compare results based on the full Kolmogorov-Crespi potential
\cite{PhysRevB.71.235415} to results based on the 
Kolmogorov-Crespi-z potential \eqref{ee2}.  We compare the
predictions of the discrete model using both these potentials to the
predictions of the continuum model, which is derived using the
Kolmogorov-Crespi-z potential.  Note that we do not have a version of
the continuum model derived from the full Kolmogorov-Crespi
potential.  The goal of these comparisons is to explain why in some
cases we fail to see a good match between the predictions of the
discrete and the continuum models.

We present a set of results that indicates how the size of the elastic
constants affects the smoothness of the deformed configurations.  We
consider two cases, one case in which the elastic constants are
relatively small and one case in which the elastic constants are
relatively large.
The parameter values used in the simulations are listed in
Tables~\ref{tt13}~(a), (b) and Table~\ref{tt6}~(c).

\begin{figure}
\centering
\begin{subfigure}{0.3\textwidth}
  %trim=left botm right top ** THESE ARE MARGINS **
  \includegraphics[width=\textwidth, clip, trim=1.25in 3in 1.25in 1.5in]
                  {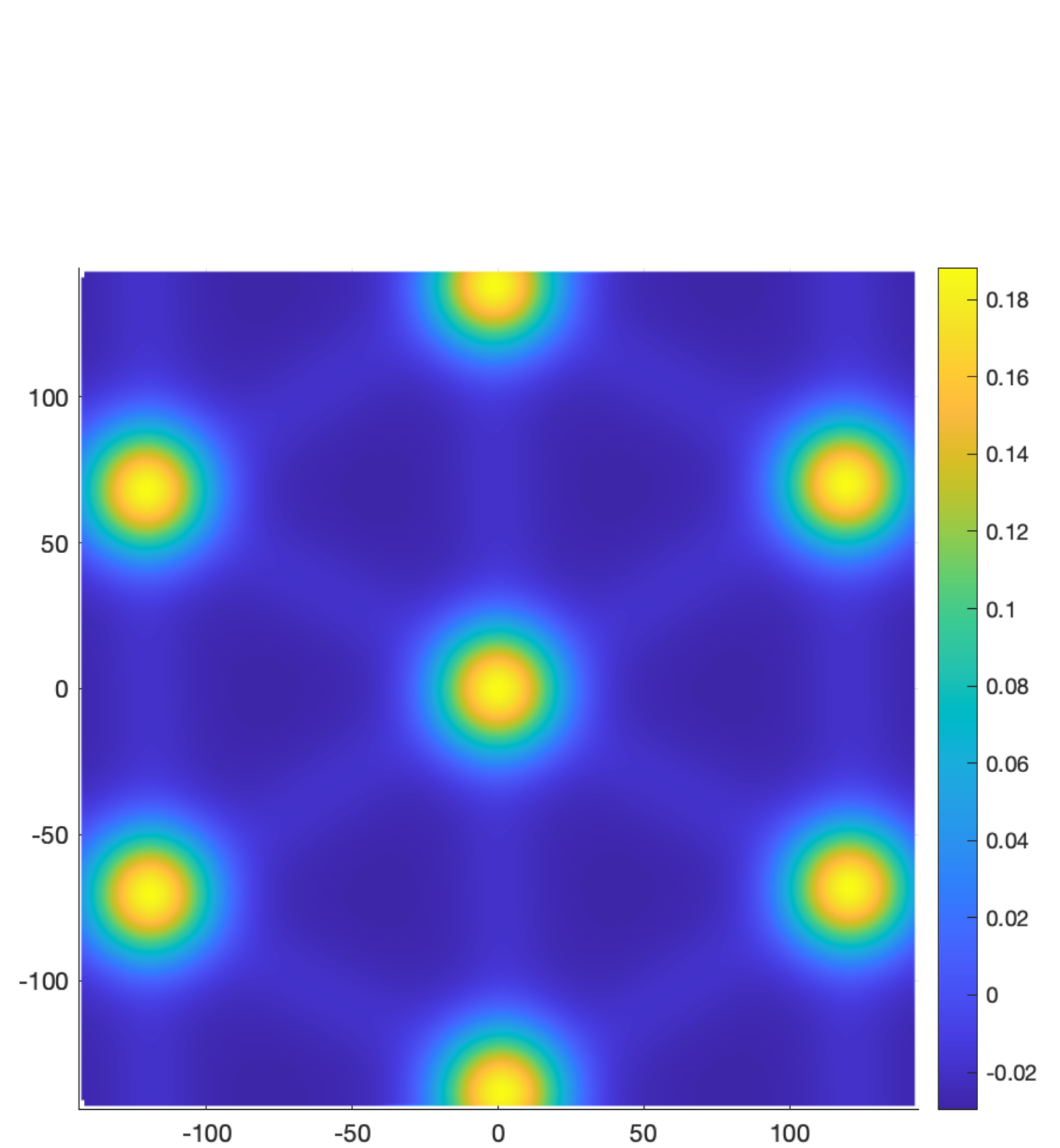}
  %\caption{eta\_LAMMPS\_full\_KCZ\_large\_comp.pdf.}
  \caption{$\eta$, Discrete Model.}  
\label{fr58}
\end{subfigure}
\hspace*{.1\linewidth}
\begin{subfigure}{0.3\textwidth}
  %trim=left botm right top ** THESE ARE MARGINS **
  \includegraphics[width=\textwidth, clip, trim=1.25in 3in 1.25in 1.5in]
                  {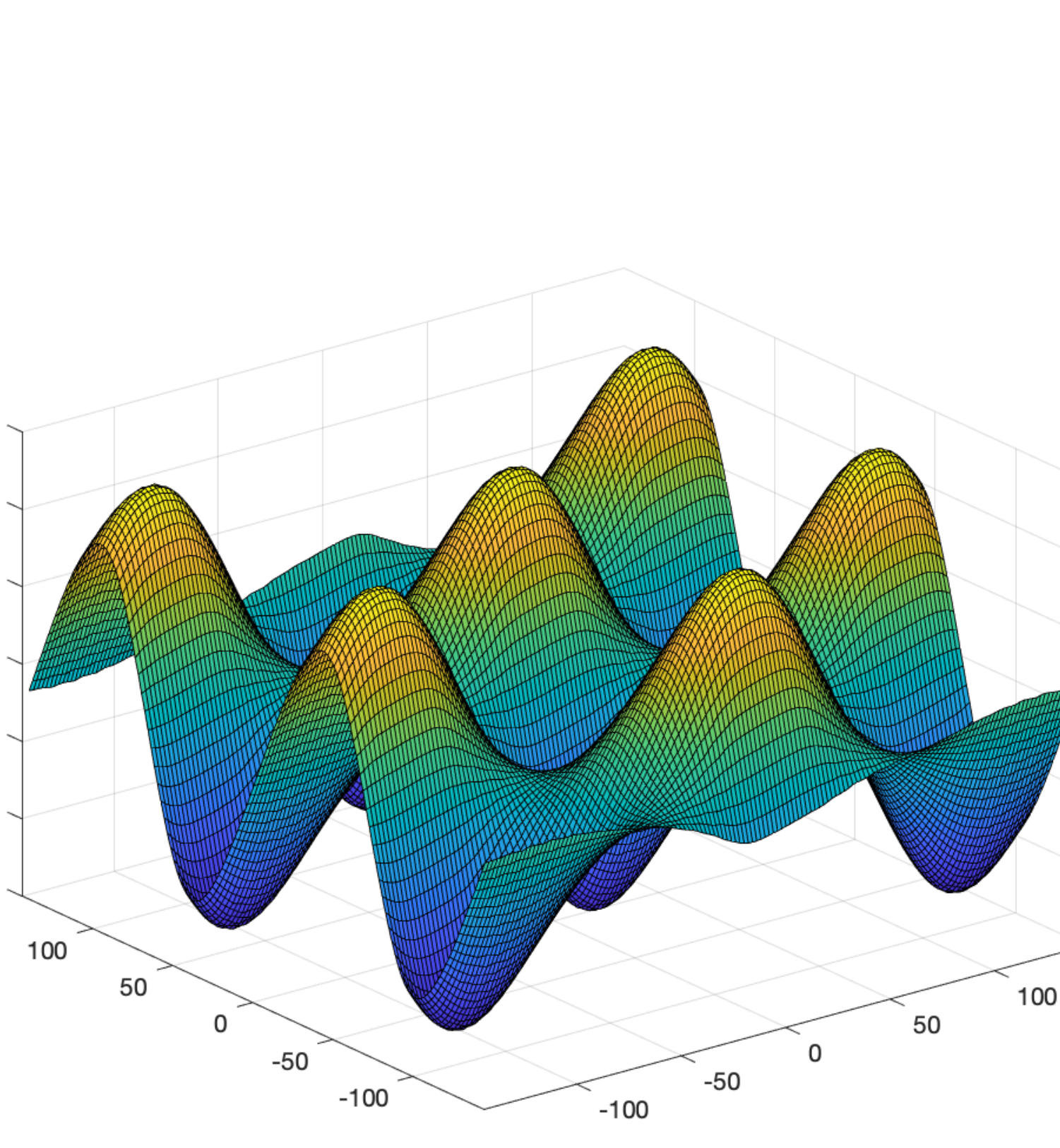}
  %\caption{xi1\_LAMMPS\_full\_KCZ\_large\_comp.pdf.}
  \caption{$\xi_{1}$, Discrete Model.}
  \label{fr61}
\end{subfigure}

\vspace*{-.25in}

\begin{subfigure}{0.3\textwidth}
  \includegraphics[width=\textwidth, clip, trim=1.25in 3in 1.25in 2in]
                  {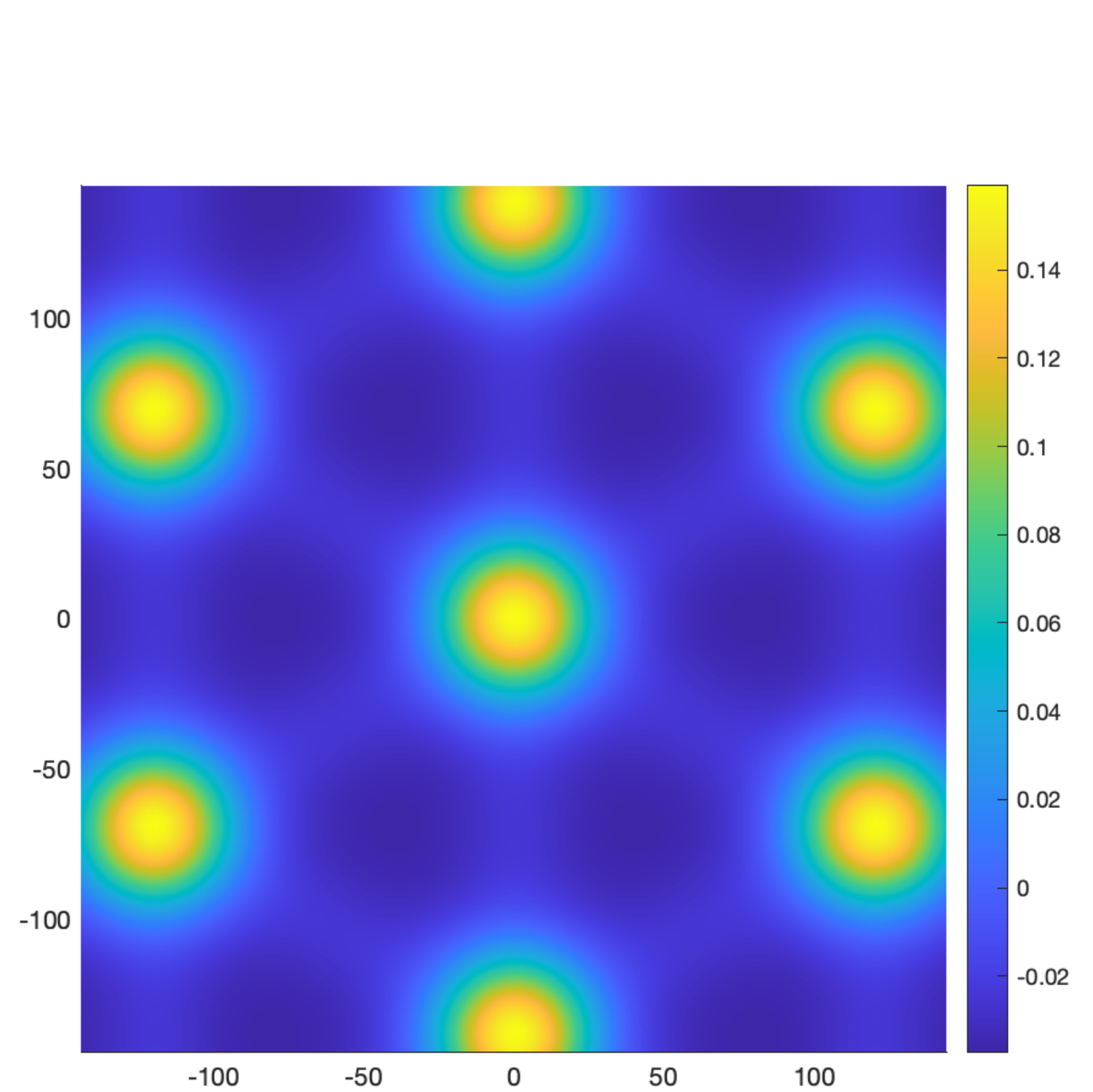}
  %\caption{eta\_COMSOL\_full\_KSZ\_large\_comp.pdf.}
  \caption{$\eta$, Continuum Model.}
  \label{fr59}
\end{subfigure}
\hspace*{.1\linewidth}
\begin{subfigure}{0.3\textwidth}
  \includegraphics[width=\textwidth, clip, trim=1.25in 3in 1.25in 2in]
                  {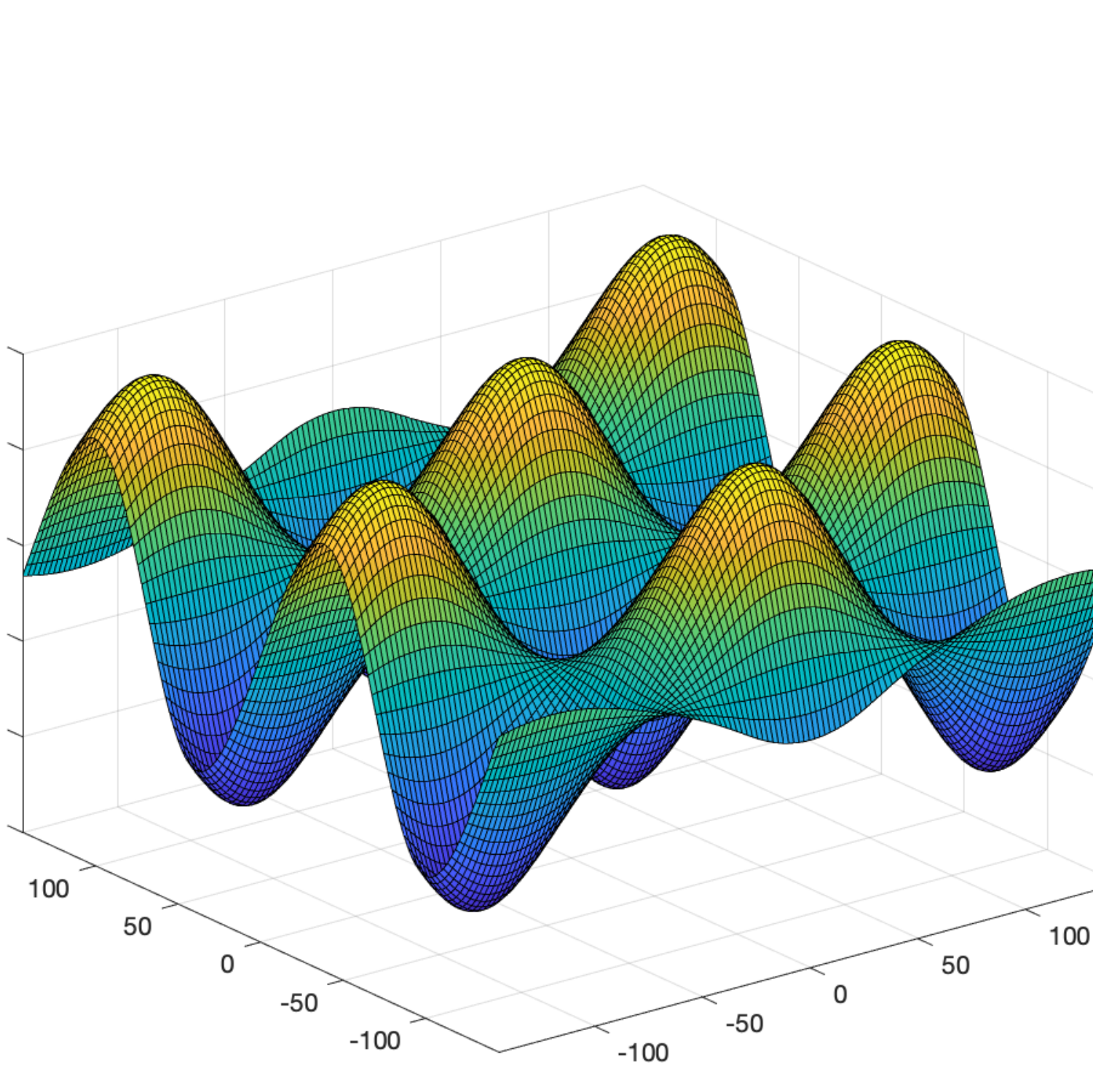}
  %\caption{xi1\_COMSOL\_full\_KSZ\_large\_comp.pdf.}
  \caption{$\xi_{1}$, Continuum Model.}
  \label{fr62}
\end{subfigure}

\caption{Simulation Results for the Kolmogorov-Crespi Potential,
  Large Elastic Constants.}
\label{fr60}
\end{figure}

In Figure~\ref{fr60}, we show results based on the
Kolmogorov-Crespi-z potential for the case when the elastic constants
are relatively large.  The key plot here is Figure~\ref{fr61}, which
shows that the discrete model predicts smooth deformations.  Comparing
the LAMMPS results to the COMSOL results in this set of figures, we
see that we get a good match between the predictions of the discrete
and continuum models.  Note in particular that in both
Figures~\ref{fr58} and \ref{fr59}, the secondary wrinkles deflect
away from the fixed lattice, i.e., the wrinkles are higher than the
surrounding commensurate regions.

\begin{figure}[!t]
\centering
\begin{subfigure}{0.3\textwidth}
  %trim=left botm right top ** THESE ARE MARGINS **
  \includegraphics[width=\textwidth, clip, trim=1.25in 3in 1.25in 3.in]
                  {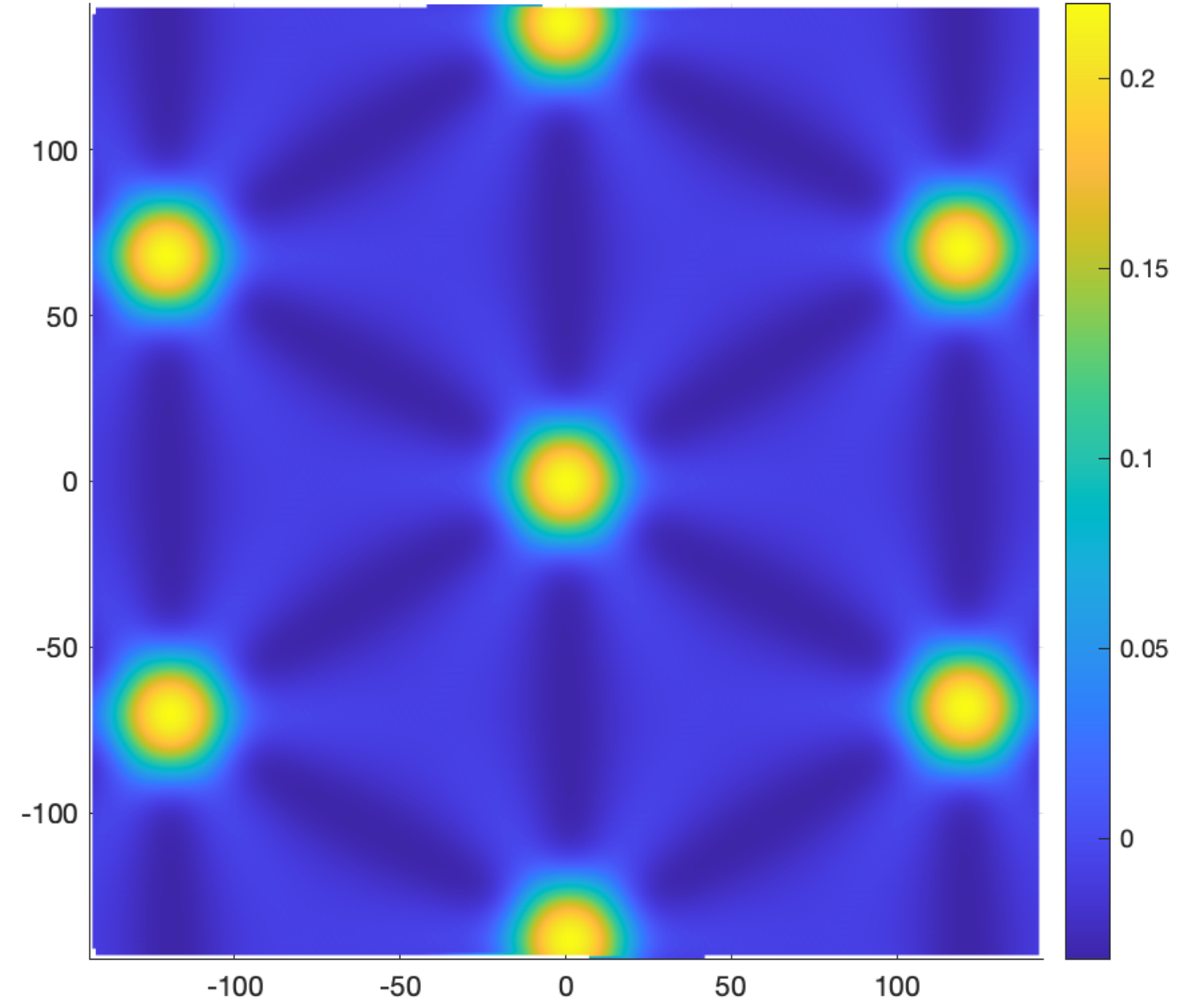}
  %\caption{eta\_LAMMPS\_full\_KCZ\_small\_comp.pdf.}
  \caption{$\eta$, Discrete Model.}  
  \label{fr64}
\end{subfigure}
\hspace*{.1\linewidth}
\begin{subfigure}{0.3\textwidth}
  %trim=left botm right top ** THESE ARE MARGINS **
  \includegraphics[width=\textwidth, clip, trim=1.25in 3in 1.25in 3.in]
                  {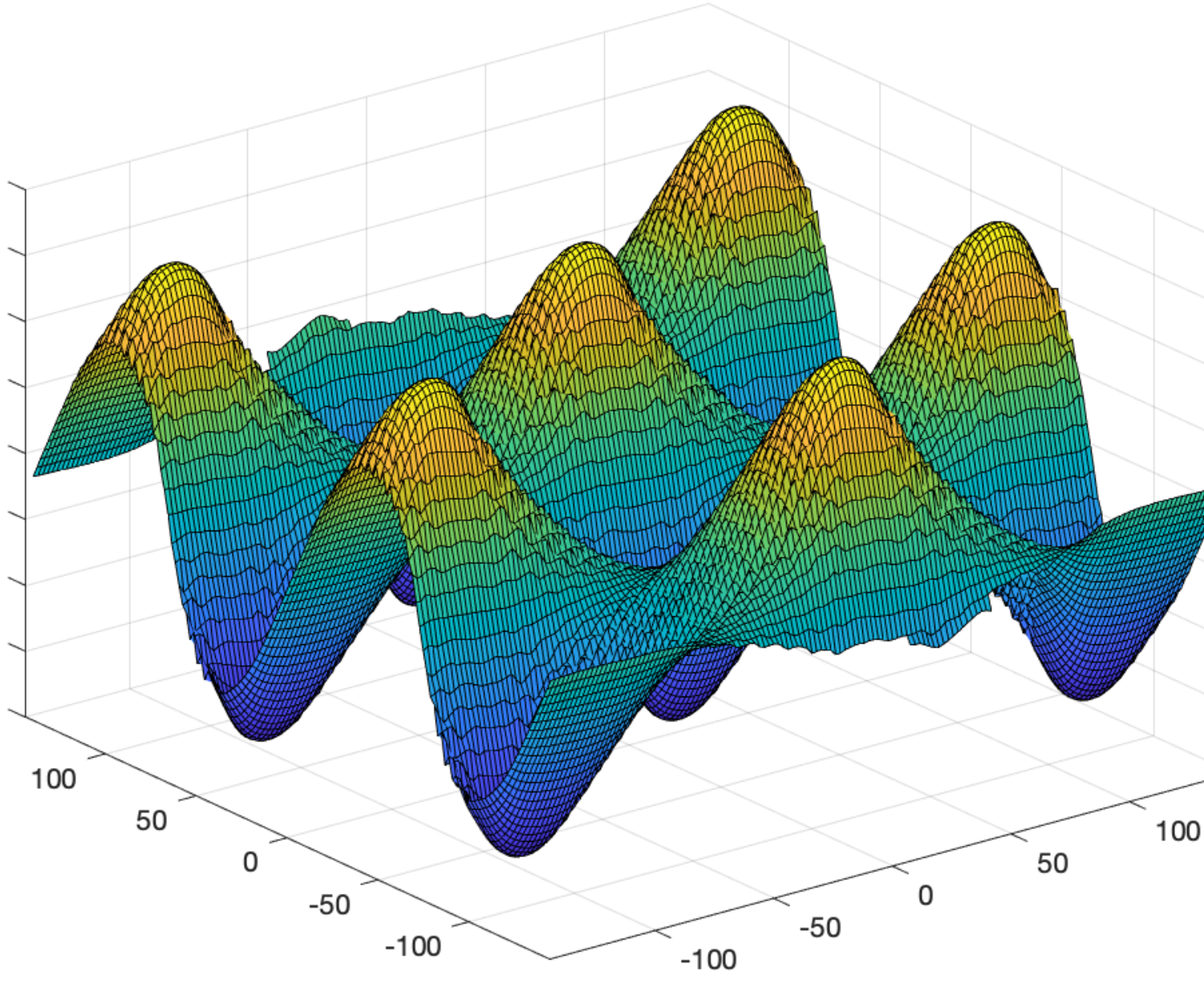}
  %\caption{xi1\_LAMMPS\_full\_KCZ\_small\_comp.pdf.}
  \caption{$\xi_{1}$, Discrete Model.}
  \label{fr67}
\end{subfigure}

\vspace*{-.35in}

\begin{subfigure}{0.3\textwidth}
  \includegraphics[width=\textwidth, clip, trim=1.25in 3in 1.25in 2in]
                  {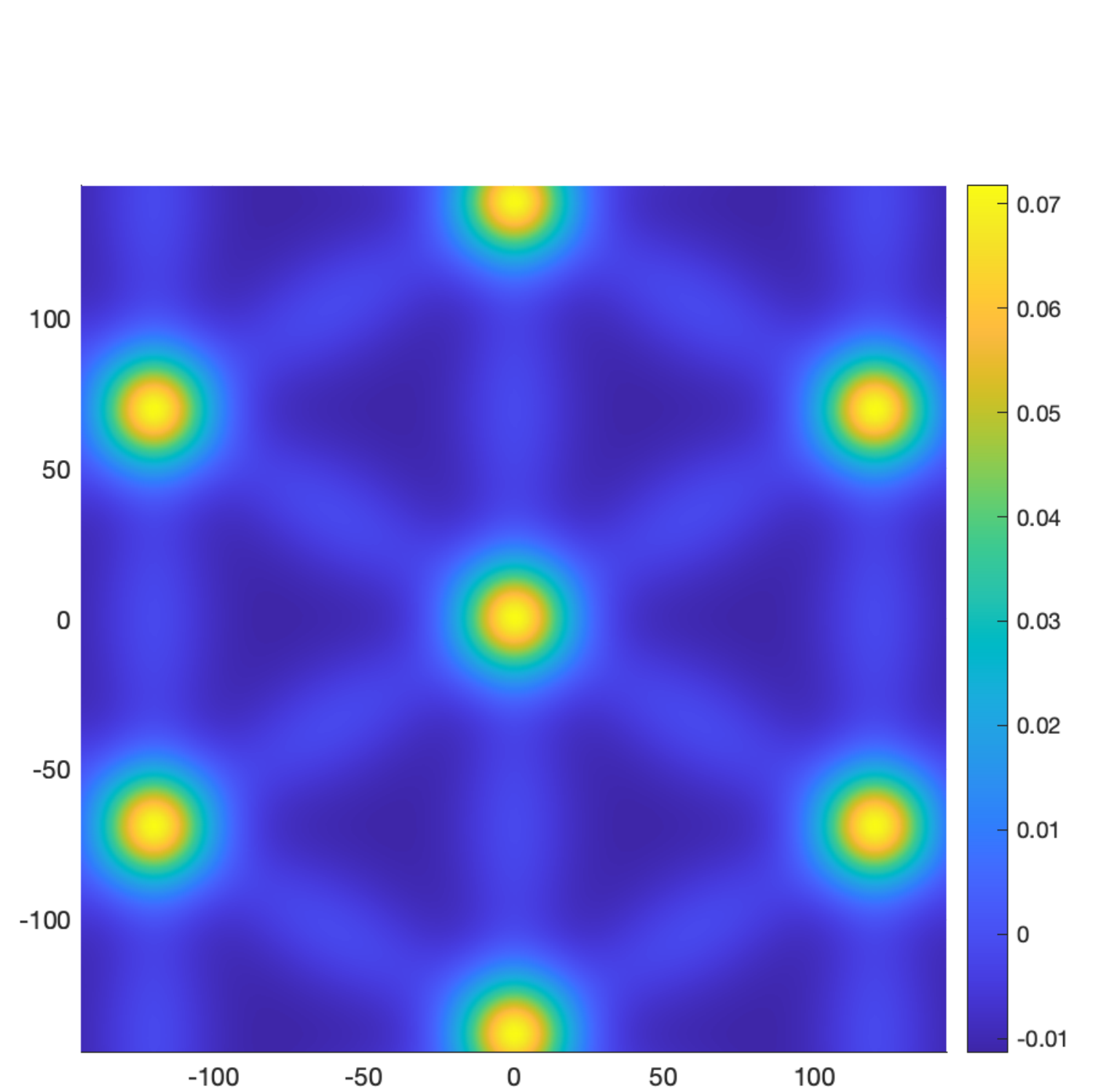}
  %\caption{eta\_COMSOL\_full\_KSZ\_small\_comp.pdf.}
  \caption{$\eta$, Continuum Model.}
  \label{fr65}
\end{subfigure}
\hspace*{.1\linewidth}
\begin{subfigure}{0.3\textwidth}
  \includegraphics[width=\textwidth, clip, trim=1.25in 3in 1.25in 2in]
                  {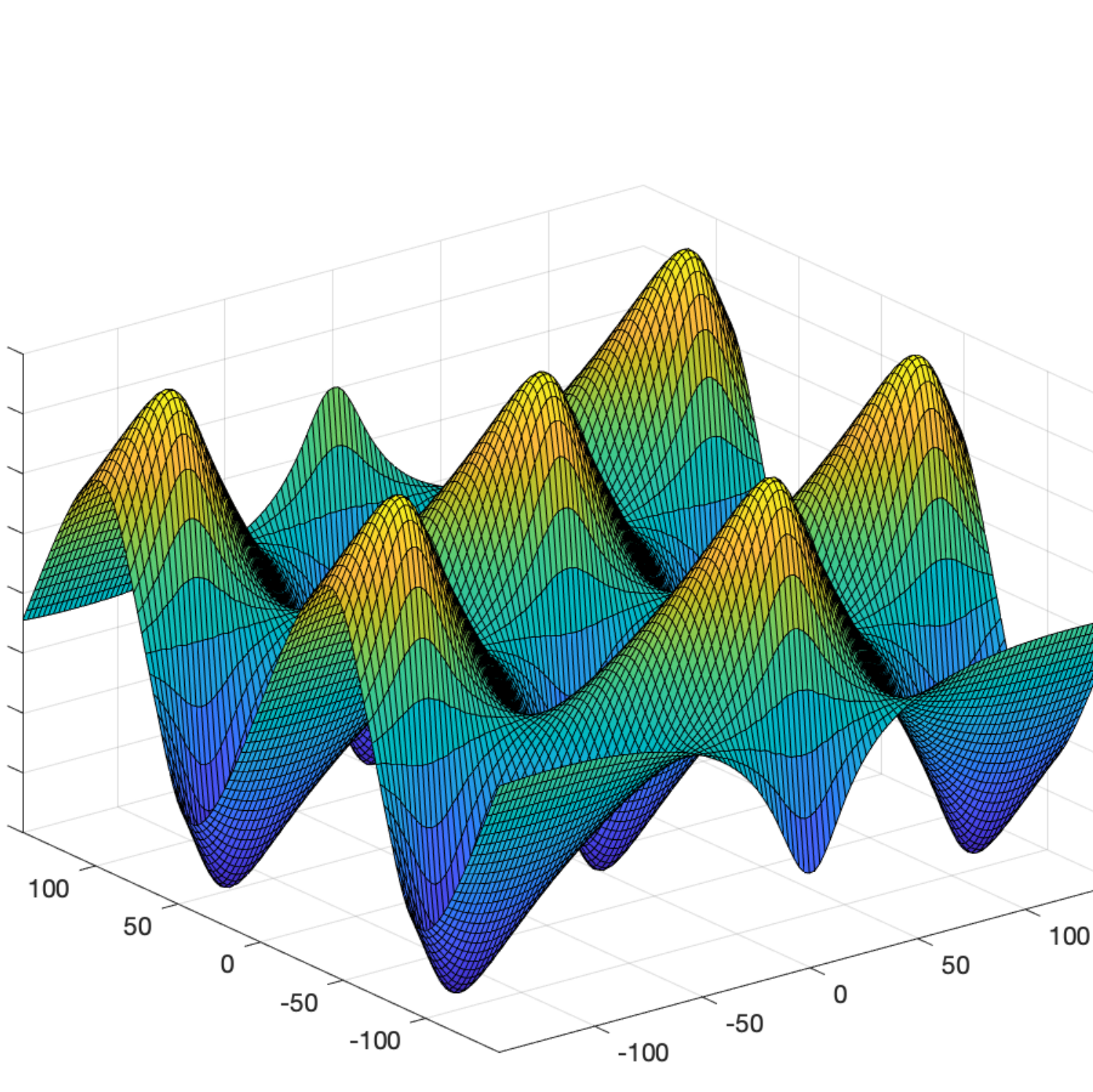}
  %\caption{xi1\_COMSOL\_full\_KSZ\_small\_comp.pdf.}
  \caption{$\xi_{1}$, Continuum Model.}
  \label{fr68}
\end{subfigure}

\vspace*{-.1in}

\caption{Simulation Results for the Kolmogorov-Crespi Potential,
  Small Elastic Constants.}
\label{fr66}
\end{figure}

In Figures~\ref{fr66}, we show results based on the
Kolmogorov-Crespi-z potential for the case when the elastic constants
are relatively small.  In this case, we get a poor match between the
predictions of the discrete and continuum models.
In particular,
Figure~\ref{fr64} indicates that the primary and secondary wrinkles
deflect toward the fixed lattice for the discrete model, while
Figure~\ref{fr65} indicates that the primary and secondary wrinkles
deflect away from the fixed lattice for the continuum model.
A key observation here is that when the elastic constants are small,
the deformed configuration predicted by the discrete simulation
exhibits small-scale spatial oscillation, as can be seen in
Figure~\ref{fr67}.

To explain these results, recall that a basic assumptions justifying
our discrete-to-continuum modeling procedure is that the discrete
lattice can be embedded in a smooth surface.  The small scale
oscillations we see in Figure~\ref{fr67} suggest that this assumption
is violated.  As a consequence, we do not expect to get a good match
between the predictions of the two models.  The results in Figure~\ref{fr60}
suggest that sufficiently large elastic constants can
suppress these small scale oscillations, in which case a good match
between the discrete and the continuum is attained.

%2.19 (a) \ref{fr67} compare to 2.21 (b) \ref{fr74}.
%2.18 (a) \ref{fr64} compare to 2.21 (a) \ref{fr73}.

\begin{figure}[!b]
\centering
\begin{subfigure}{0.3\textwidth}
  %trim=left botm right top ** THESE ARE MARGINS **
  \includegraphics[width=\textwidth, clip, trim=1.25in 3in 1.25in 3.in]
                  {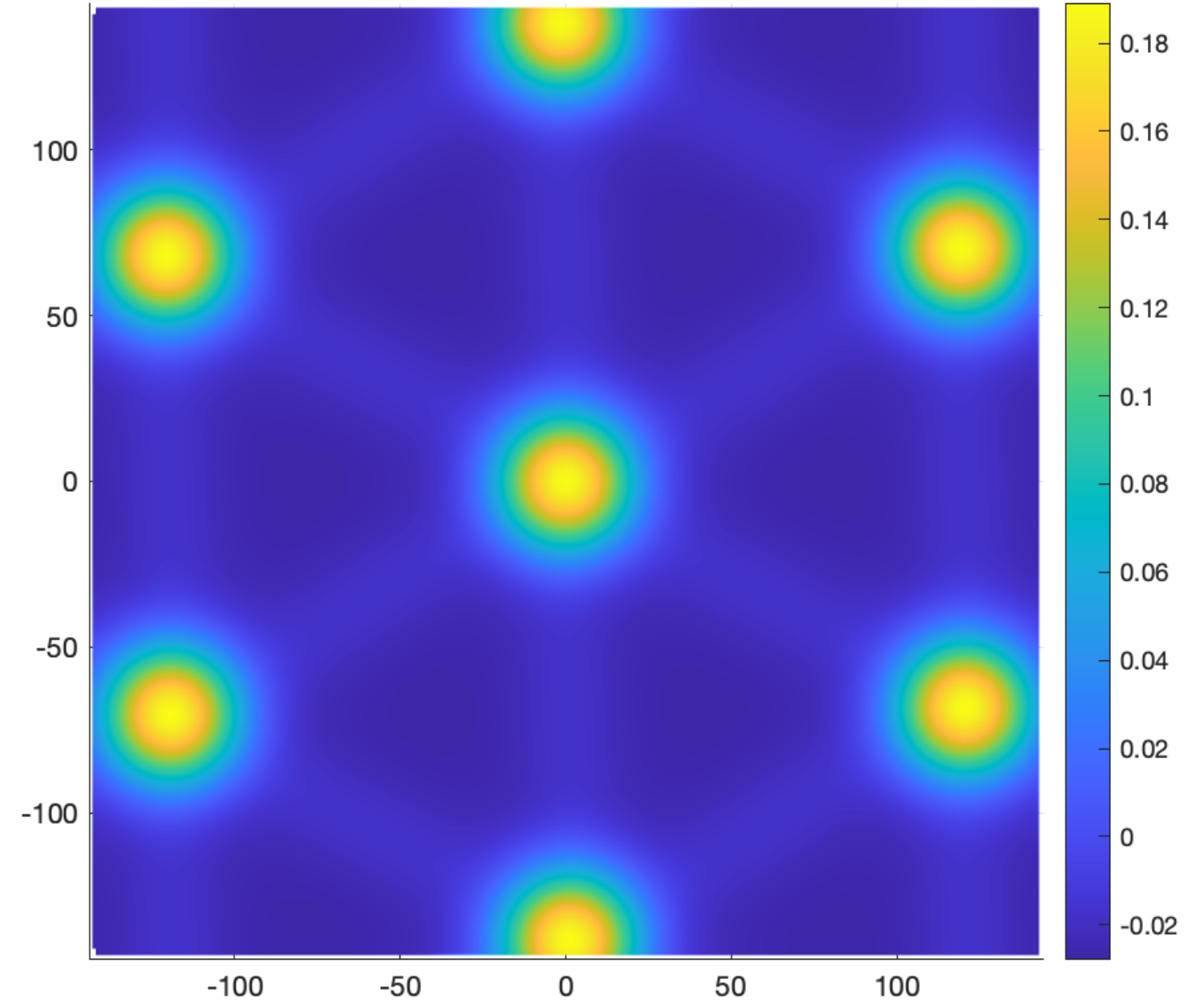}
  %\caption{eta\_LAMMPS\_full\_KCF\_large\_comp.pdf.}
  \caption{$\eta$, Large Elastic Constants.}  
  \label{fr70}
\end{subfigure}
\hspace*{.1\linewidth}
\begin{subfigure}{0.3\textwidth}
  \includegraphics[width=\textwidth, clip, trim=1.25in 3in 1.25in 3.in]
                  {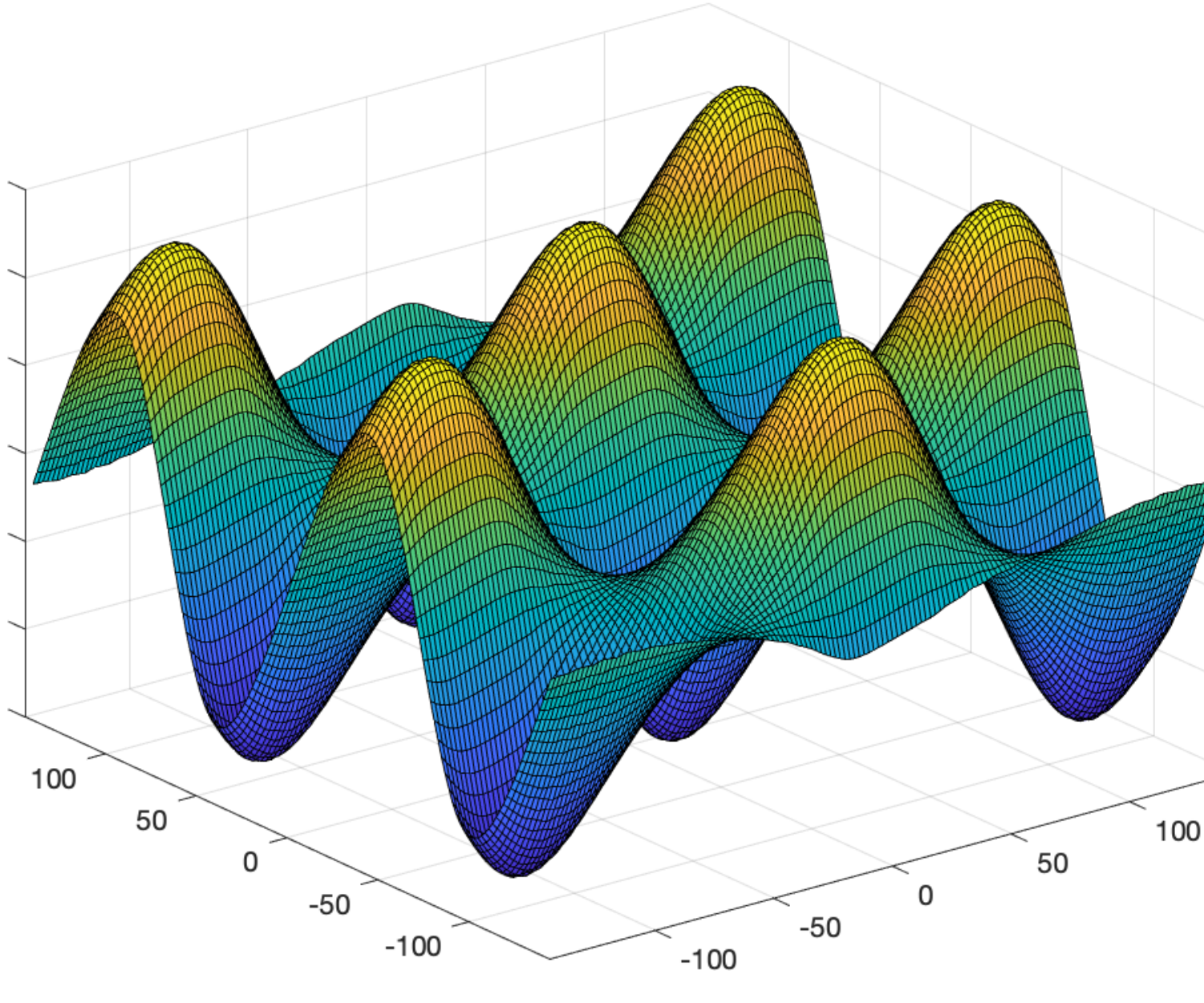}
  %\caption{xi1\_LAMMPS\_full\_KCF\_large\_comp.pdf.}
  \caption{$\xi_{1}$, Large Elastic Constants.}  
  \label{fr71}
\end{subfigure}

\vspace*{-.35in}

\begin{subfigure}{0.3\textwidth}
  %trim=left botm right top ** THESE ARE MARGINS **
  \includegraphics[width=\textwidth, clip, trim=1.25in 3in 1.25in 2in]
                  {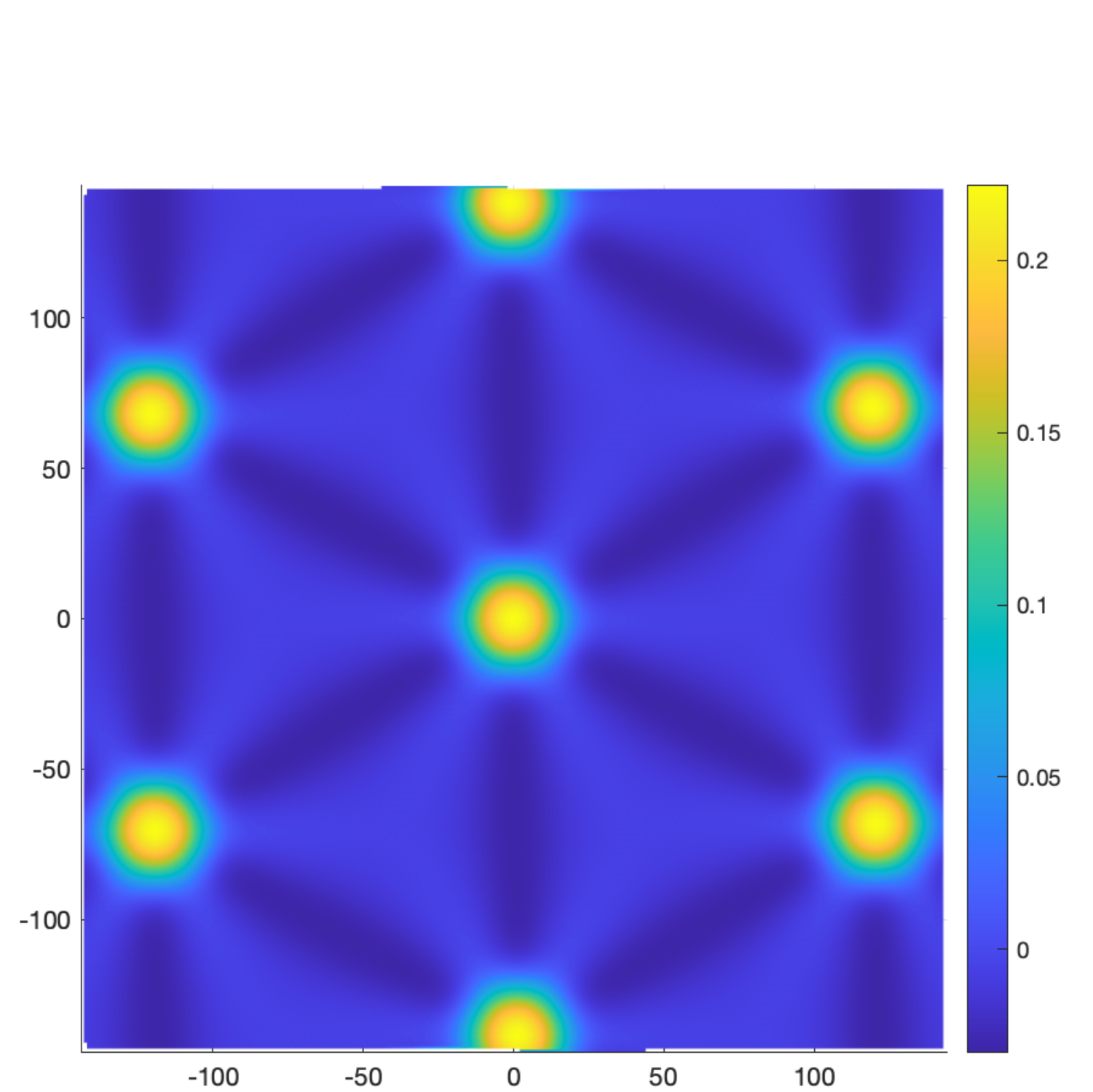}
  %\caption{eta\_LAMMPS\_full\_KCF\_small\_comp.pdf.}
  \caption{$\eta$, Small Elastic Constants.}  
  \label{fr73}
\end{subfigure}
\hspace*{.1\linewidth}
\begin{subfigure}{0.3\textwidth}
  \includegraphics[width=\textwidth, clip, trim=1.25in 3in 1.25in 2in]
                  {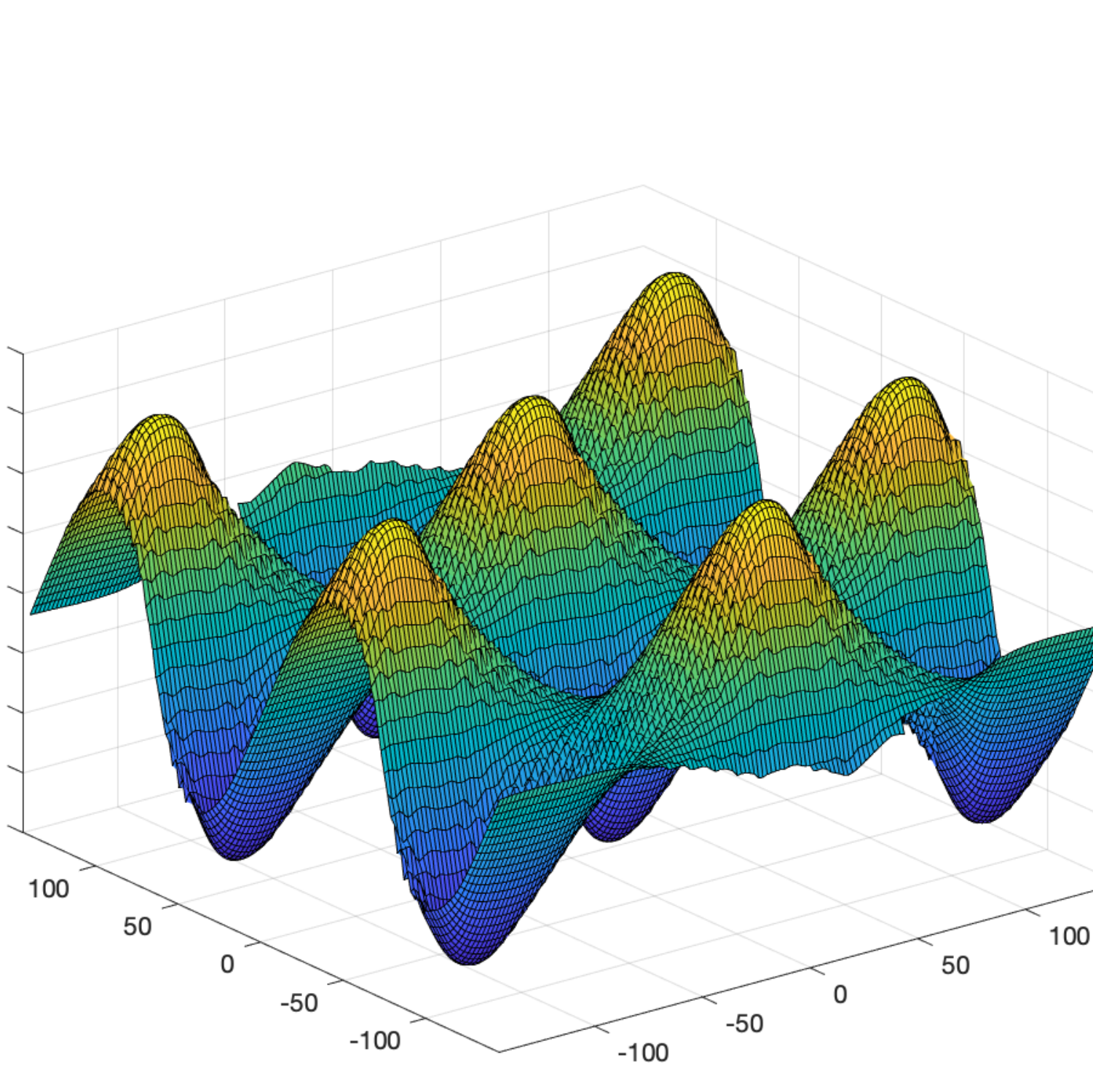}
  %\caption{xi1\_LAMMPS\_full\_KCF\_small\_comp.pdf.}
  \caption{$\xi_{1}$, Small Elastic Constants.}  
  \label{fr74}
\end{subfigure}

\vspace*{-.1in}

\caption{Simulation Results for the Discrete Model Using the Full Kolmogorov-Crespi Potential.}

\label{fr72}
\end{figure}

One could conjecture that the problem is with the use of the
Kolmogorov-Crespi-z potential.  To show that this is not the case, we
compare the predictions of the discrete model using the full
Kolmogorov-Crespi potential versus the predictions of the
Kolmogorov-Crespi-z potential.
See Figure~\ref{fr72}.
Figures~\ref{fr70} and \ref{fr71} show the predictions of the discrete
model using the full Kolmogorov-Crespi potential and relatively large
elastic constants.  These simulations use the same parameter values as
thoses used for Figure~\ref{fr60}.
Figures~\ref{fr73} and \ref{fr74} show the
predictions of the discrete model using the full Kolmogorov-Crespi
potential and relatively small elastic constants.  These simulations use the same parameter values as
thoses used for Figure~\ref{fr66}.
By comparing
Figures~\ref{fr70} and \ref{fr71} to Figures~\ref{fr58} and \ref{fr61}
and by comparing Figures~\ref{fr73} and \ref{fr74} to
Figures~\ref{fr64} and \ref{fr67}, we see the same predictions from
the discrete model if we use the full Kolmogorov-Crespi potential
versus the Kolmogorov-Crespi-z potential.

\begin{figure}
\centering
\begin{subfigure}{0.3\textwidth}
  %trim=left botm right top ** THESE ARE MARGINS **
  \includegraphics[width=\textwidth, clip, trim=1.25in 3in 1.25in 3.2in]
                  {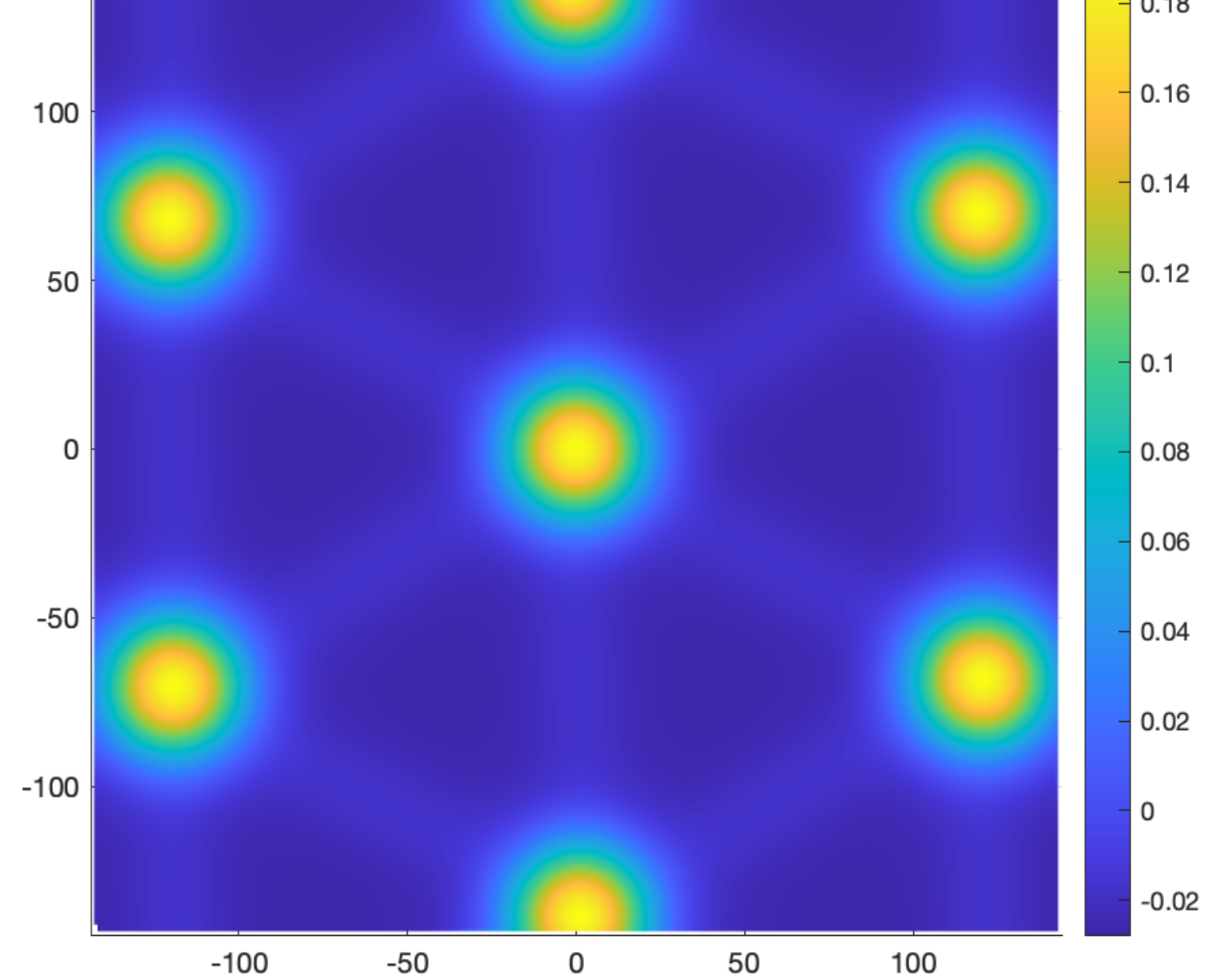}
  %\caption{eta\_LAMMPS\_full\_KCF\_large.pdf.}
  \caption{$\eta$, Discrete Model.}  
  \label{fr80}
\end{subfigure}
\hfill
\begin{subfigure}{0.3\textwidth}
  %trim=left botm right top ** THESE ARE MARGINS **
  \includegraphics[width=\textwidth, clip, trim=1.25in 3in 1.25in 3.2in]
                  {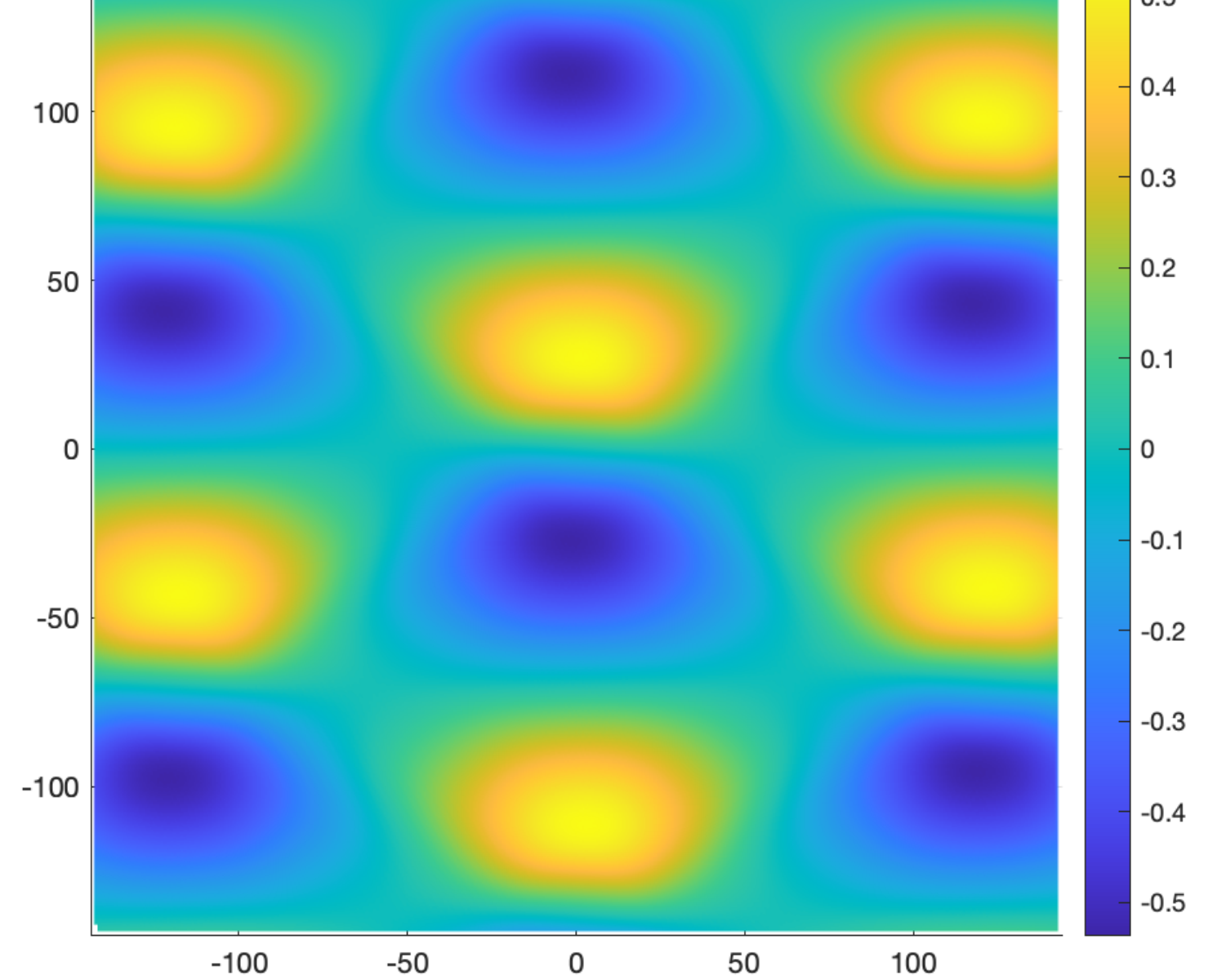}
  %\caption{xi1\_LAMMPS\_full\_KCF\_large.pdf.}
  \caption{$\xi_{1}$, Discrete Model.}
  \label{fr83}
\end{subfigure}
\hfill
\begin{subfigure}{0.3\textwidth}
  %trim=left botm right top ** THESE ARE MARGINS **
  \includegraphics[width=\textwidth, clip, trim=1.25in 3in 1.25in 3.2in]
                  {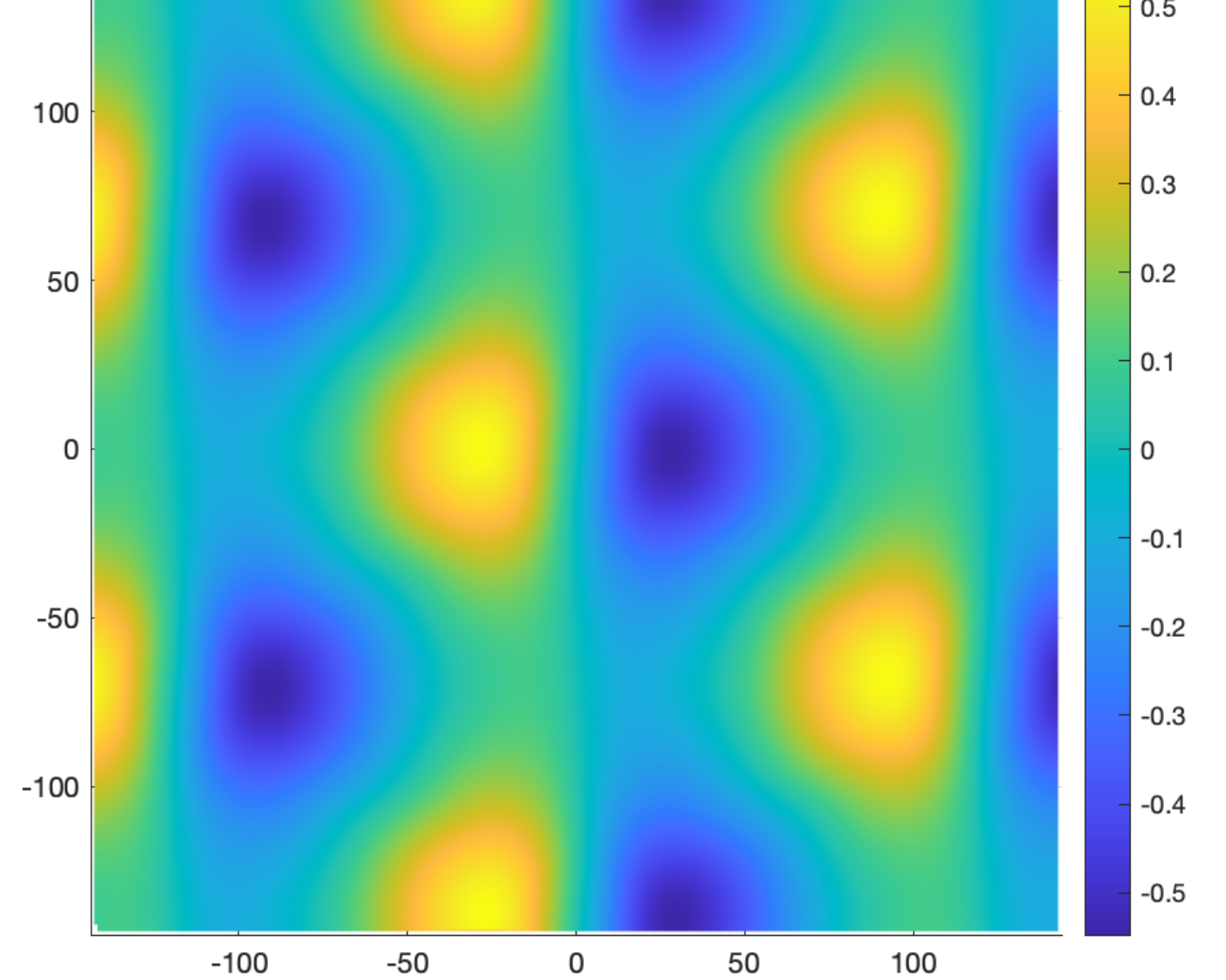}
  %\caption{xi2\_LAMMPS\_full\_KCF\_large.pdf.}
  \caption{$\xi_{2}$, Discrete Model.}
  \label{fr86}
\end{subfigure}

\vspace*{-.5in}

\begin{subfigure}{0.3\textwidth}
  \includegraphics[width=\textwidth, clip, trim=1.25in 3in 1.25in 1.5in]
                  {./Figures-results/KC-3D-results/eta_COMSOL_full.pdf}
  %\caption{eta\_COMSOLS\_full.pdf.}
  \caption{$\eta$, Continuum Model.}
  \label{fr81}
\end{subfigure}
\hfill
\begin{subfigure}{0.3\textwidth}
  \includegraphics[width=\textwidth, clip, trim=1.25in 3in 1.25in 1.5in]
                  {./Figures-results/KC-3D-results/xi1_COMSOL_full.pdf}
  %\caption{xi1\_COMSOLS\_full.pdf.}
  \caption{$\xi_{1}$, Continuum Model.}
  \label{fr84}
\end{subfigure}
\hfill
\begin{subfigure}{0.3\textwidth}
  \includegraphics[width=\textwidth, clip, trim=1.25in 3in 1.25in 1.5in]
                  {./Figures-results/KC-3D-results/xi2_COMSOL_full.pdf}
  %\caption{xi2\_COMSOLS\_full.pdf.}
  \caption{$\xi_{2}$, Continuum Model.}
  \label{fr87}
\end{subfigure}

\caption{Simulation Results for the Kolmogorov-Crespi Potential, Large
Elastic Constants.}

\label{fr82}
\end{figure}

In Figure~\ref{fr82} we show the predictions of the discrete
and continuum models for relatively large elastic constants.
These simulations use the same parameter values as
thoses used for Figure~\ref{fr60}.
Note
that here we are using the full Kolmogorov-Crespi potential.  As
expected from the above discussion, we see a good match between the
discrete and the continuum.

%%%%%%%%%%%%%%%%%%%%%%%%%%%%%%%%%%%%%%%%%%%%%%%%%%%%%%%%%%%%%%%%%%%%%%%%%%%%%%%
% Parameters

% horizontal
\begin{table}[h!]
\hspace*{.05\linewidth}
  \begin{subtable}[h!]{.9\linewidth}
    \label{tt11}
    \begin{tabular}{c|c|c|c|c|c|c|c|c|c}
               &
      $h_{1}$, $h_{2}$ & 
      $L$      & 
      $\theta$ &           
      $k_{s}$   &
      $k_{t}$   &
      $k_{d}$   &
      $\sigma$ & 
      $\omega$ &  
      $\delta$ \\ \hline
      Figures~\ref{fr60}, \ref{fr72}(a),(b), \ref{fr82} &   % large 
      $2.46$     &
      $120$   & 
      $-2.03$ & 
      $3.23$   &
      $0.75$   &
      $12$   &
      $3.34$  & 
      $0.01$  & 
      $0.58$  \\           
      Figures~\ref{fr66}, \ref{fr72}(c),(d)      &   % small
      $2.46$     &
      $120$   & 
      $-2.03$ & 
      $0.403$   &
      $0.015$   &
      $1.20$   &
      $3.34$  & 
      $0.01$  & 
      $0.58$ 
    \end{tabular}
  \caption{Parameters for the Discrete Model.  All lengths are in
    \r{A}.  $\theta$ is in degrees.  $k_{s}$, $k_{t}$, $k_{d}$, and $\omega$ are in eV.} %\hspace*{1.25in}
  \end{subtable}

\hspace*{.07\linewidth}
  \begin{subtable}[h!]{.86\linewidth}
    \label{tt12}
    \begin{tabular}{c|c|c|c|c|c|c|c|c|c}
      &
      $\varepsilon$ & 
      $\delta_{1}$, $\delta_{2}$   &
      $k$           & 
      $m$           & 
      $\alpha$      & 
      $\Theta$      & 
      $\gamma_{s}$   &
      $\gamma_{t}$   &
      $\gamma_{d}$   \\ \hline
      Figures~\ref{fr60}, \ref{fr72}(a),(b), \ref{fr82} &   % large 
      0.028    &  
      $0.74$    & 
      $2$      &  
      $-1$       &   
      $0$      &  
      $-1.28$  &  
      $123.66$   & 
      $306.64$    & 
      $10.39$     \\    
      Figures~\ref{fr66}, \ref{fr72}(c),(d)      &   % small
      $0.028$    &  
      $0.74$    & 
      $2$      &  
      $-1$       &   
      $0$      &  
      $-1.28$  &  
      $15.46$   & 
      $6.13$    & 
      $1.04$
    \end{tabular}
    \caption{Dimensionless Parameters for the Continuum Model.
      $\Theta$ is in radians.} %\hspace*{1.25in}
  \end{subtable}

  \caption{Parameter Values for Additional Simulations using
    KC Potential.  \label{tt13}}
\end{table}

% eta 

%% xi1
%\begin{figure}
%\centering
%
%
%\caption{xi1 full KCZ large comp}
%\label{fr63}
%\end{figure}

%%%%%%%%%%%%

% eta 

%\vspace*{-1.25in}

%% xi1
%\begin{figure}
%\centering
%
%\hfill
%
%\caption{xi1 full KCZ small comp}
%\label{fr69}
%\end{figure}

%%%%%%%%%%%%%

% 

%%%%%%%%%%%%%%
%
%\begin{figure}
%\centering
%
%\caption{LAMMPS\_full\_KCF\_small\_comp.pdf}
%\label{fr75}
%\end{figure}

%%%%%%%%%%%%%%
%
%\begin{figure}
%\centering
%\begin{subfigure}{0.4\textwidth}
%  %trim=left botm right top ** THESE ARE MARGINS **
%  \includegraphics[width=\textwidth, clip, trim=1.25in 1.75in 1.25in 1.5in]
%                  {./Figures-results/KCZ-vs-KCF-results/eta_LAMMPS_full_KCF_large.pdf}
%  \caption{eta\_LAMMPS\_full\_KCF\_large.pdf.}
%  \label{fr76}
%\end{subfigure}
%\hfill
%\begin{subfigure}{0.4\textwidth}
%  \includegraphics[width=\textwidth, clip, trim=1.25in 1.75in 1.25in 1.5in]
%                  {./Figures-results/KCZ-vs-KCF-results/xi1_LAMMPS_full_KCF_large.pdf}
%  \caption{xi1\_LAMMPS\_full\_KCF\_large.pdf.}
%  \label{fr77}
%\end{subfigure}
%
%\begin{subfigure}{0.4\textwidth}
%  \includegraphics[width=\textwidth, clip, trim=1.25in 1.75in 1.25in 1.5in]
%                  {./Figures-results/KCZ-vs-KCF-results/xi2_LAMMPS_full_KCF_large.pdf}
%  \caption{xi2\_LAMMPS\_full\_KCF\_large.pdf.}
%  \label{fr78}
%\end{subfigure}
%
%\caption{LAMMPS\_full\_KCF\_large}
%\label{fr79}
%\end{figure}

%%%%%%%%%%%%%%%%%%%%%%%%%%%%%%%%%%%%%%%%%%

% eta full

%% xi1 full
%\begin{figure}
%\centering
%
%\caption{xi1 full}
%\label{fr85}
%\end{figure}

%% xi2 full
%\begin{figure}
%\centering
%
%\hfill
%
%\caption{xi2 full}
%\label{fr88}
%\end{figure}

%\input{results-v9.tex}

%\section{Numerical Results}
%In this section, we compare the atomistic and continuum models. For that, we numerically %solve the discrete system and the
%system of Euler-Lagrange equations \eqref{e17} both subject to periodic boundary %conditions to explore the behavior of their minimizers. In order to have a fair %comparison, we will first find conditions and relationships on all the parameters %involved on both atomistic and continuum models. 

\section{Conclusions}\label{s:conclusion}

We derive a continuum model of a graphene bilayer in which one layer
is fixed.  We start with a discrete energy containing elastic terms
and a weak interaction term.  After expanding these terms in a small
geometric parameter, we truncate the expansions and approximate sums
with integrals to arrive at a continuum energy having a
Ginzburg-Landau structure.

To validate our modeling, we perform numerical simulations to compare
the predictions of the original discrete model and our continuum
model.  This comparison shows excellent agreement between the two models
if the dimensionless elastics constants that appear in the continuum
model are sufficiently large.
The continuum model predicts spatial patterns that have been observed
for twisted graphene bilayers in other papers
\cite{van2015relaxation,van2014moire,PhysRevB.96.075311,jain2016structure,enaldiev2020stacking}.
For the out-of-plane displacement, the model predicts hot spots and
wrinkles connecting neighboring hot spots.  These wrinkles form domain
walls between relatively large triangular commensurate regions.

For certain cases, we see discrepancies between what the discrete and
continuum models predict for the out-of-plane displacements of some
hot spots and wrinkles.  This occurs for both the Lennard-Jones
potential and the Kolmogorov-Crespi potential.  Surface plots of
solutions show small-scale spatial oscillations in the horizontal
displacement when the elastic constants are not sufficiently large
relative to the strength of the weak interaction.  The existence of
these oscillations violates a basic assumption of our
discrete-to-continuum modeling procedure, and we believe that this
explains why the discrepancies occur.  For the Lennard-Jones
potential, decreasing $\omega$, which controls the strength of the
weak interaction, increases the dimensionless elastic constants that
appear in the continuum model.  In this case, we see a good match for
the out-of-plane displacement of all hot spots and wrinkles.
Likewise, we see a good match for the Kolmogorov-Crespi potential when
we directly increase the size of the elastic constants in the discrete
model.

In this paper, we apply our discrete-to-continuum modeling procedure
to a discrete energy in which the weak-interaction term is based on
either the Lennard-Jones potential or a version of the
Kolmogorov-Crespi potential that assumes that the layers are locally
parallel.
Each of these potentials has the feature that the interaction between
two non-bonded atoms is computed from a simple function of the positions
of the two atoms.  In particular, neither of these potentials requires
additional information about the layers near the atoms, for example,
information about the local orientation of the normals.
Our approach is based on finding the local horizontal offset in the
reference configuration.  We decompose this offset into contributions
from the lattice mismatch of the layers and from the misalignment
between the layers.  
Because of the relatively simple form of the potentials we work with
here, we can in a straightforward way incorporate these contributions
as parameters in the final form of the potential in the continuum
model.
It would be of interest to explore whether this approach can be
adapted to more complicated potentials.  For example, the full
Kolmogorov-Crespi potential depends not just on the positions of the
interacting atoms but also on the local orientation of the lattices
near the atoms.

\section{Acknowledgment}
This work was supported by the National Science Foundation grant DMS-1615952.

\bibliographystyle{ieeetr}
\bibliography{gamma-converg-references}	

\section{Supplemental Material}

\subsection{Asymptotic Expressions}
\label{appe1}

In this section, we expand in $\varepsilon$ all contributions to the discrete energy that led to the asymptotic expressions \eqref{stretchs}--\eqref{dihes} in Section \ref{elencon}. Recall that ${\bf q}^k_{ij}$ and ${\bf b}_{ij}^p$ are nondimensional quantities in all the calculations that follow.

\subsubsection{Extensional Springs}

We expand the first term in \eqref{eq:DiscreteE_s} in $\varepsilon$.  We first expand 
$\bb^1_{ij}$.  Since all of the bonds relate to ${\boldsymbol\chi}^2_{ij}$, we will use it for the expansion. We have that
\begin{align}
\bb^1_{ij}
  &=
  \bq^1_{ij} - \bq^2_{ij} \nonumber \\
  &=
  \left({\boldsymbol\chi}^1_{ij}
  +
  \varepsilon\boldsymbol \xi({\boldsymbol\chi}^1_{ij}),
  \varepsilon
  +
  \varepsilon\eta({\boldsymbol\chi}^1_{ij})\right)  -
 \left(
 {\boldsymbol\chi}^2_{ij}
  +
  \varepsilon\boldsymbol\xi({\boldsymbol\chi}^2_{ij}),
  \varepsilon
  +
  \varepsilon\eta({\boldsymbol\chi}^2_{ij})  
  \right)  \nonumber\\
 & =
 \left({\boldsymbol\chi}^1_{ij} - {\boldsymbol\chi}^2_{ij}
 +   \varepsilon \left (\boldsymbol{\xi}\left({\boldsymbol\chi}^1_{ij}\right) - \boldsymbol{\xi}\left({\boldsymbol\chi}^2_{ij}\right) \right), \, 
  \varepsilon \left(\eta \left({\boldsymbol\chi}^1_{ij}\right) - \eta \left({\boldsymbol\chi}^2_{ij}\right) \right)
 \right)\nonumber\\
& = \left({\boldsymbol\chi}^1_{ij} - {\boldsymbol\chi}^2_{ij} 
 + \varepsilon \left( \nabla \boldsymbol{\xi}  \left({\boldsymbol\chi}^1_{ij} - {\boldsymbol\chi}^2_{ij}\right)
    + \frac{1}{2} \left({\boldsymbol\chi}^1_{ij} - {\boldsymbol\chi}^2_{ij}\right) \cdot H_{\boldsymbol \xi} \cdot \left({\boldsymbol\chi}^1_{ij} - {\boldsymbol\chi}^2_{ij}\right) 
 + \mbox{h.o.t.} \right), \right.  \nonumber \\
& \left. \qquad \qquad   \qquad \qquad   \qquad \qquad \varepsilon \left (\nabla \eta \cdot \left({\boldsymbol\chi}^1_{ij} - {\boldsymbol\chi}^2_{ij}\right) + \frac{1}{2}\left({\boldsymbol\chi}^1_{ij} - {\boldsymbol\chi}^2_{ij}\right) \cdot H_\eta \cdot \left({\boldsymbol\chi}^1_{ij} - {\boldsymbol\chi}^2_{ij}\right)  + \mbox{h.o.t.} \right) \right). \nonumber
\end{align}
Since 
\begin{align}
{\boldsymbol\chi}^1_{ij} - {\boldsymbol\chi}^2_{ij} & =  \varepsilon\delta_{2}\left(i+\frac{1}{3}\right)\ba_1 + \left(j+\frac{1}{3}\right)\ba_2 - \varepsilon\delta_{2}\left( \left(i+\frac{2}{3}\right)\ba_1 + \left(j+\frac{2}{3}\right)\ba_2\right) \nonumber \\
& = \varepsilon \delta_2 \left(-\frac{1}{3} (\ba_1 + \ba_2)\right) = \varepsilon \delta_2 \left(-\frac{1}{2},-\frac{\sqrt{3}}{6}\right) = \varepsilon \delta_2 (v_{11}, v_{12}) = \varepsilon \delta_2 \bold{v}_1, \nonumber
\end{align}
then,
\begin{align}
\bb^1_{ij} & = \varepsilon \delta_2 \left(v_{11} + \varepsilon \nabla \xi_1 \cdot \bold{v}_1
    + \frac{\varepsilon^2 \delta_2}{2}\bold{v}_1 \cdot H_{\xi_1} \cdot \bold{v}_1
 +  \mathcal{O}(\varepsilon^3),  v_{12} + \varepsilon \nabla \xi_2 \cdot \bold{v}_1
    + \frac{\varepsilon^2 \delta_2}{2}\bold{v}_1 \cdot H_{\xi_2} \cdot \bold{v}_1
 +  \mathcal{O}(\varepsilon^3) ,\, \right.\nonumber \\
& \left. \qquad \qquad \qquad \qquad \qquad \qquad  \qquad \qquad \qquad \qquad \qquad \qquad\varepsilon \nabla \eta \cdot \bold{v}_1+ \frac{\varepsilon^2 \delta_2}{2}\bold{v}_1 \cdot H_\eta \cdot \bold{v}_1  +  \mathcal{O}(\varepsilon^3)\right).
\label{ede1}
\end{align}
Similarly,
\begin{align}
\bb^2_{ij}
&= \varepsilon \delta_2 \left(v_{21} + \varepsilon \nabla \xi_1 \cdot \bold{v}_2
    + \frac{ \varepsilon^2 \delta_2}{2}\bold{v}_2 \cdot H_{\xi_1} \cdot \bold{v}_2
 +  \mathcal{O}(\varepsilon^3),  v_{22} + \varepsilon \nabla \xi_2 \cdot \bold{v}_2
    +  \frac{\varepsilon^2 \delta_2}{2}\bold{v}_2 \cdot H_{\xi_2} \cdot  \bold{v}_2
 +  \mathcal{O}(\varepsilon^3) ,\, \right.\nonumber \\
& \left. \qquad \qquad \qquad \qquad \qquad \qquad  \qquad \qquad \qquad \qquad \qquad \qquad\varepsilon \nabla \eta \cdot \bold{v}_2+ \frac{\varepsilon^2 \delta_2}{2}\bold{v}_2 \cdot H_\eta \cdot\bold{v}_2  +  \mathcal{O}(\varepsilon^3)\right),
\label{ede2}
\end{align}
where $\bold{v}_2  = (v_{21}, v_{22}) = \left(0, 1/ \sqrt{3}\right)$
and
\begin{align}
\bb^3_{ij}
&= \varepsilon \delta_2 \left(v_{31} + \varepsilon \nabla \xi_1 \cdot \bold{v}_3
    +  \frac{\varepsilon^2 \delta_2}{2}\bold{v}_3 \cdot H_{\xi_1} \cdot \bold{v}_3
 +  \mathcal{O}(\varepsilon^3),  v_{32} + \varepsilon \nabla \xi_2 \cdot \bold{v}_3
    + \frac{\varepsilon^2 \delta_2 }{2}\bold{v}_3 \cdot H_{\xi_2} \cdot \bold{v}_3
 +  \mathcal{O}(\varepsilon^3) ,\, \right.\nonumber \\
& \left. \qquad \qquad \qquad \qquad \qquad \qquad  \qquad \qquad \qquad \qquad \qquad \qquad\varepsilon \nabla \eta \cdot \bold{v}_3+\frac{\varepsilon^2 \delta_2 }{2}\bold{v}_3 \cdot H_\eta \cdot \bold{v}_3  + \mathcal{O}(\varepsilon^3)\right),
\label{ede3}
\end{align}
where $\bold{v}_3  = (v_{31}, v_{32}) = \left(\frac{1}{2},  -\frac{\sqrt{3}}{6}\right)$. Note that in \eqref{ede1} and in the expressions that follow, all
partial derivatives are evaluated at ${\boldsymbol\chi}^2_{ij}$.

We next expand
\begin{align}
\|\bb^1_{ij}\| 
&= \varepsilon \delta_{2}  \|\bold{v}_1\| + \frac{\varepsilon^2\delta_{2}}{\|\bold{v}_1\|} \left(v_{11} \nabla \xi_1 \cdot \bold{v}_1 + v_{12} \nabla \xi_2 \cdot \bold {v}_1\right)  + \frac{\varepsilon^3 \delta_{2} }{2\|\bold{v}_1\|} \left( \left(\nabla \xi_1 \cdot \bold{v}_1\right)^2 + \left(\nabla \xi_2 \cdot \bold {v}_1\right)^2 + \left(\nabla \eta \cdot \bold {v}_1\right)^2 \right. \nonumber \\
& \left. \qquad \qquad \qquad
 - \frac{1}{\|\bold{v}_1\|^2}\left(v_{11} \nabla \xi_1 \cdot \bold{v}_1 + v_{12} \nabla \xi_2 \cdot \bold {v}_1\right)^2 + \delta_2\left( v_{11}\bold{v}_1 \cdot H_{\xi_1} \cdot \bold{v}_1 
+v_{12}\bold{v}_1 \cdot H_{\xi_2} \cdot \bold{v}_1  \right) \right) +  \mathcal{O}(\varepsilon^{4}).
\label{ede4}
\end{align}

Using that $\|\bold{v}_1\|=\frac{1}{\sqrt{3}}$,  $v_{11} \nabla \xi_1 \cdot \bold{v}_1 + v_{12} \nabla \xi_2 \cdot \bold {v}_1 = \bold{v}_1 \cdot \nabla \boldsymbol \xi \cdot \bold{v}_1 $, and that $\left(\nabla \xi_1 \cdot \bold{v}_1\right)^2 + \left(\nabla \xi_2 \cdot \bold {v}_1\right)^2 = |\nabla \boldsymbol{\xi}\bold{v}_1|^2$, we have that

\begin{align}
 \frac{\|\bb^1_{ij}\|-\frac{\varepsilon\delta_{2}}{\sqrt{3}}}{\frac{\varepsilon\delta_{2}}{\sqrt{3}}} 
 &=   \frac{\varepsilon}{\|\bold{v}_1\|^2}  \bold{v}_1 \cdot \nabla \boldsymbol \xi \cdot \bold{v}_1  + \frac{\varepsilon^2 }{2\|\bold{v}_1\|^2} \left( |\nabla \boldsymbol{\xi} \cdot \bold{v}_1|^2 + \left(\nabla \eta \cdot \bold {v}_1\right)^2 \right. \nonumber \\
& \left. \qquad \qquad \qquad
 - \frac{1}{\|\bold{v}_1\|^2}\left( \bold{v}_1 \cdot \nabla \boldsymbol \xi  \cdot \bold{v}_1 \right)^2 + \delta_2\left( v_{11}\bold{v}_1 \cdot H_{\xi_1} \cdot \bold{v}_1 
+v_{12}\bold{v}_1 \cdot H_{\xi_2} \cdot \bold{v}_1  \right) \right) +  \mathcal{O}(\varepsilon^{3}).
  \label{ede5}
\end{align}
\begin{align}
  \left( \frac{\|\bb^1_{ij}\|-\frac{\varepsilon\delta_{2}}{\sqrt{3}}}{\frac{\varepsilon\delta_{2}}{\sqrt{3}}}\right)^2 
 &=  \frac{\varepsilon^2}{\|\bold{v}_1\|^4} \left( \bold{v}_1 \cdot \nabla \boldsymbol \xi \bold{v}_1 \right)^2  + \frac{\varepsilon^3 }{\|\bold{v}_1\|^4} \bold{v}_1 \cdot \nabla \boldsymbol \xi \bold{v}_1 \left(  |\nabla \boldsymbol{\xi}\bold{v}_1|^2 + \left(\nabla \eta \cdot \bold {v}_1\right)^2 \right. \nonumber \\
& \left. \qquad \qquad \qquad
 + \bold{v}_1 \cdot \left( - \frac{1}{\|\bold{v}_1\|^2}\nabla \boldsymbol{\xi} (\bold{v}_1 \otimes
\bold{v}_1 ) \nabla \boldsymbol{\xi}+ \delta_2 v_{11}H_{\xi_1} 
+\delta_2 v_{12} H_{\xi_2}\right)\bold{v}_1 \right) +  \mathcal{O}(\varepsilon^{4}).
  \label{ede6}
\end{align}

A similar computation for the second term in \eqref{eq:DiscreteE_s}
yields
\begin{align}
  \left( \frac{\|\bb^2_{ij}\|-\frac{\varepsilon\delta_{2}}{\sqrt{3}}}{\frac{\varepsilon\delta_{2}}{\sqrt{3}}}\right)^2
  &=  \frac{\varepsilon^2}{\|\bold{v}_2\|^4} \left( \bold{v}_2 \cdot \nabla \boldsymbol \xi \bold{v}_2 \right)^2  + \frac{\varepsilon^3 }{\|\bold{v}_2\|^4} \bold{v}_2 \cdot \nabla \boldsymbol \xi \bold{v}_2 \left(  |\nabla \boldsymbol{\xi}\bold{v}_2|^2 + \left(\nabla \eta \cdot \bold {v}_2\right)^2 \right. \nonumber \\
& \left. \qquad \qquad \qquad
 + \bold{v}_2 \cdot \left( - \frac{1}{\|\bold{v}_2\|^2}\nabla \boldsymbol{\xi} (\bold{v}_2 \otimes
\bold{v}_2) \nabla \boldsymbol{\xi}+ \delta_2 v_{21}H_{\xi_1} 
+\delta_2 v_{22} H_{\xi_2}\right)\bold{v}_2 \right) +  \mathcal{O}(\varepsilon^{4}).
  \label{ede7}
\end{align}
and third term
\begin{align}
 \left( \frac{\|\bb^3_{ij}\|-\frac{\varepsilon\delta_{2}}{\sqrt{3}}}{\frac{\varepsilon\delta_{2}}{\sqrt{3}}}\right)^2
  &=  \frac{\varepsilon^2}{\|\bold{v}_3\|^4} \left( \bold{v}_3 \cdot \nabla \boldsymbol \xi \bold{v}_3 \right)^2  + \frac{\varepsilon^3 }{\|\bold{v}_3\|^4} \bold{v}_3 \cdot  \nabla \boldsymbol \xi \bold{v}_3 \left(  |\nabla \boldsymbol{\xi}\bold{v}_3|^2 + \left(\nabla \eta \cdot \bold {v}_3\right)^2 \right. \nonumber \\
& \left. \qquad \qquad \qquad
 + \bold{v}_3 \cdot \left( - \frac{1}{\|\bold{v}_3\|^2}\nabla \boldsymbol{\xi} (\bold{v}_3 \otimes
\bold{v}_3 ) \nabla \boldsymbol{\xi}+ \delta_2 v_{31}H_{\xi_1} 
+\delta_2 v_{32} H_{\xi_2}\right)\bold{v}_3 \right) +  \mathcal{O}(\varepsilon^{4}).
  \label{ede8}
\end{align}

We now combine \eqref{ede6}, \eqref{ede7}, and \eqref{ede8}, which yields
\begin{align}
   \mathcal E_s[\boldsymbol\xi,\eta] &=
  \sum_{i,j=1}^{N_2} \frac{9 k_s}{2\omega}
 \left\{\varepsilon^3 \left[  \left( \bold{v}_1 \cdot \nabla \boldsymbol \xi \bold{v}_1 \right)^2   +  \left( \bold{v}_2 \cdot \nabla \boldsymbol \xi \bold{v}_2 \right)^2  + \left( \bold{v}_3 \cdot \nabla \boldsymbol \xi \bold{v}_3 \right)^2 \right] \right. \nonumber \\
&\qquad \qquad \quad + \varepsilon^4 \left[ \bold{v}_1 \cdot \nabla \boldsymbol \xi \bold{v}_1 \left( |\nabla \boldsymbol{\xi}\bold{v}_1|^2 + \left(\nabla \eta \cdot \bold {v}_1\right)^2 \right)+ \bold{v}_2 \cdot \nabla \boldsymbol \xi \bold{v}_2 \left(|\nabla \boldsymbol{\xi}\bold{v}_2|^2 + \left(\nabla \eta \cdot \bold {v}_2\right)^2 \right) \right. \nonumber \\
& \qquad \qquad \qquad \qquad +  \bold{v}_3 \cdot \nabla \boldsymbol \xi \bold{v}_3 \left(|\nabla \boldsymbol{\xi}\bold{v}_3|^2 + \left(\nabla \eta \cdot \bold {v}_3\right)^2 \right)  \nonumber \\
& \qquad \qquad \qquad \qquad 
 + \bold{v}_1 \cdot \left( - \frac{1}{\|\bold{v}_1\|^2}\nabla \boldsymbol{\xi} (\bold{v}_1 \otimes
\bold{v}_1 ) \nabla \boldsymbol{\xi}+ \delta_2 v_{11}H_{\xi_1} 
+\delta_2 v_{12} H_{\xi_2}\right)\bold{v}_1\nonumber\\
& \qquad \qquad \qquad \qquad
 + \bold{v}_2 \cdot \left( - \frac{1}{\|\bold{v}_2\|^2}\nabla \boldsymbol{\xi} (\bold{v}_2 \otimes
\bold{v}_2 ) \nabla \boldsymbol{\xi}+ \delta_2 v_{21}H_{\xi_1} 
+\delta_2 v_{22} H_{\xi_2}\right)\bold{v}_2 \nonumber\\
& \qquad \qquad \qquad \qquad \left. \left.
 + \bold{v}_3 \cdot \left( - \frac{1}{\|\bold{v}_3\|^2}\nabla \boldsymbol{\xi} (\bold{v}_3 \otimes
\bold{v}_3 ) \nabla \boldsymbol{\xi}+ \delta_2 v_{31}H_{\xi_1} 
+\delta_2 v_{32} H_{\xi_2}\right)\bold{v}_3\right] + \mathcal{O}(\varepsilon^{5}) \right\} .
 \label{ede9}
\end{align}
Notice that $\left(\nabla \eta \cdot \bold {v}_1\right)^2 = \bold {v}_1\cdot (\nabla \eta \otimes \nabla \eta) \cdot  \bold {v}_1 $ and the same for the other similar terms. To transform this sum to an integral, we use that the area of the small cells have dimension $\frac{\sqrt{3}}{2}\delta_2^2 \varepsilon^2$, and define $\gamma_s = \frac{6 \sqrt{3} k_s}{\omega \delta_2^2}$.

\subsubsection{Torsional Springs}

We expand the first term in \eqref{ede2.6} in $\varepsilon$ using the expansions \eqref{ede1} and \eqref{ede2}, and then \eqref{ede4} for the norm.

\begin{align}
\bb^1_{ij}\cdot\bb^2_{ij}
 &=
 \varepsilon^{2}\delta_{2}^{2} \left\{ \bold{v}_1\cdot  \bold{v}_2 +  \varepsilon \left( \bold{v}_1  \nabla \boldsymbol{\xi} \cdot \bold{v}_2 + \bold{v}_2 \nabla \boldsymbol{\xi} \cdot \bold{v}_1 \right) + \right. \nonumber \\
&\qquad \qquad
 \left. \varepsilon^2 \left ((\nabla \xi_1 \cdot \bold{v}_1)(\nabla \xi_1 \cdot \bold{v}_2) + (\nabla \xi_2 \cdot \bold{v}_1)(\nabla \xi_2 \cdot \bold{v}_2) +  (\nabla \eta \cdot \bold{v}_1)(\nabla \eta \cdot \bold{v}_2) \right) +\mathcal{O}(\varepsilon^{2})
  \right\}. \nonumber \\ 
 &=
 \varepsilon^{2}\delta_{2}^{2} \left\{ \bold{v}_1\cdot  \bold{v}_2 +  \varepsilon \left( \bold{v}_1  \nabla \boldsymbol{\xi} \cdot \bold{v}_2 + \bold{v}_2 \nabla \boldsymbol{\xi} \cdot \bold{v}_1 \right) + \right. \nonumber \\
&\qquad \qquad
 \left. \varepsilon^2 \bold{v}_1  \left ((\nabla \xi_1 \otimes \nabla \xi_1) + (\nabla \xi_2 \otimes \nabla \xi_2) + (\nabla \eta \otimes \nabla \eta )  \right) \cdot \bold{v}_2 +\mathcal{O}(\varepsilon^{2})
  \right\}.
\label{ede10}
\end{align}

Notice that here, there are other terms of order $\varepsilon^2$ that are left out by following  the work in \cite{espanol2018discrete} for the square case, where we learned that we do not need them to still be able to model what we want.

We also have that
$$
  \|\bb^1_{ij}\| \|\bb^2_{ij}\| =
  \varepsilon^{2}\delta_{2}^{2}
  \left(  \|\bold{v}_1\| \|\bold{v}_2\| + \varepsilon \left(\bold{v}_1 \nabla \boldsymbol{\xi} \cdot \bold{v}_1 + \bold{v}_2 \nabla \boldsymbol{\xi} \cdot \bold{v}_2 \right)
    +
     \mathcal{O}(\varepsilon^2) \right) = \varepsilon^{2}\delta_{2}^{2}\|\bold{v}_1\| \|\bold{v}_2\| \left(1 +  \mathcal{O}(\varepsilon) \right)
$$
and therefore, 
\begin{equation}
  \left(\|\bb^1_{ij}\| \|\bb^2_{ij}\|\right)^{-1}
  =
\frac{  \varepsilon^{-2}\delta_{2}^{-2}}
{\|\bold{v}_1\| \|\bold{v}_2\|}\left(1 +  \mathcal{O}(\varepsilon) \right).
\label{ede19}
\end{equation}

Then,

\begin{align}
\frac{ \bb^1_{ij}\cdot\bb^2_{ij}}{\|\bb^1_{ij}\| \|\bb^2_{ij}\|} & = \frac{\bold{v}_1\cdot \bold{v}_2}{\|\bold{v}_1\|\|\bold{v}_2\|} + \varepsilon \left( \frac{\bold{v}_2 \cdot \nabla \boldsymbol\xi \bold{v}_1}{\|\bold{v}_1\|\|\bold{v}_2\|} + \frac{\bold{v}_1\cdot \nabla \boldsymbol\xi \bold{v}_2}{\|\bold{v}_1\|\|\bold{v}_2\|}\right) + \nonumber \\ 
& \qquad  \frac{\varepsilon^2}{\|\bold{v}_1\|\|\bold{v}_2\|}  \bold{v}_1 \cdot \left( \nabla \xi_1 \otimes \nabla \xi_1 + \nabla \xi_2\otimes \nabla \xi_2  + \nabla \eta \otimes \nabla \eta \right)  \cdot \bold{v}_2 +  \mathcal{O}(\varepsilon^2) .
\end{align}

Using that $\frac{\bold{v}_1\cdot \bold{v}_2}{\|\bold{v}_1\|\|\bold{v}_2\|}=-1/2$ and that $\|\bold{v}_1\|=\|\bold{v}_2\|=1/\sqrt{3}$, we obtain that
\begin{align}\label{ede7.11}
\left(\frac{ \bb^1_{ij}\cdot\bb^2_{ij}}{\|\bb^1_{ij}\| \|\bb^2_{ij}\|} +\frac{1}{2} \right)^2& = 9\varepsilon^2 \left( \bold{v}_2 \cdot \nabla \boldsymbol\xi \bold{v}_1+ \bold{v}_1\cdot \nabla \boldsymbol\xi \bold{v}_2 \right)^2 + \nonumber \\ & 18\varepsilon^3 \left( \bold{v}_2 \cdot \nabla \boldsymbol\xi \bold{v}_1+ \bold{v}_1\cdot \nabla \boldsymbol\xi \bold{v}_2 \right) \left( \bold{v}_1 \cdot \left( \nabla \xi_1 \otimes \nabla \xi_1 + \nabla \xi_2\otimes \nabla \xi_2  + \nabla \eta \otimes \nabla \eta \right)  \cdot \bold{v}_2\right) + \mathcal{O}(\varepsilon^3)
\end{align}

Similarly, 
\begin{align}\label{ede7.12}
\left(\frac{ \bb^2_{ij}\cdot\bb^3_{ij}}{\|\bb^2_{ij}\| \|\bb^2_{ij}\|} +\frac{1}{2} \right)^2& = 9\varepsilon^2 \left( \bold{v}_2 \cdot \nabla \boldsymbol\xi \bold{v}_3+ \bold{v}_3\cdot \nabla \boldsymbol\xi \bold{v}_2 \right)^2 + \nonumber \\ & 18\varepsilon^3 \left( \bold{v}_2 \cdot \nabla \boldsymbol\xi \bold{v}_3+ \bold{v}_3\cdot \nabla \boldsymbol\xi \bold{v}_2 \right) \left( \bold{v}_2 \cdot \left( \nabla \xi_1 \otimes \nabla \xi_1 + \nabla \xi_2\otimes \nabla \xi_2  + \nabla \eta \otimes \nabla \eta \right)  \cdot \bold{v}_3\right) + \mathcal{O}(\varepsilon^3)
\end{align}
and

\begin{align}\label{ede7.13}
\left(\frac{ \bb^1_{ij}\cdot\bb^3_{ij}}{\|\bb^1_{ij}\| \|\bb^3_{ij}\|} +\frac{1}{2} \right)^2& =  9\varepsilon^2 \left( \bold{v}_3 \cdot \nabla \boldsymbol\xi \bold{v}_1+ \bold{v}_1\cdot \nabla \boldsymbol\xi \bold{v}_3 \right)^2 + \nonumber \\ & 18\varepsilon^3 \left( \bold{v}_3 \cdot \nabla \boldsymbol\xi \bold{v}_1+ \bold{v}_1\cdot \nabla \boldsymbol\xi \bold{v}_3 \right) \left( \bold{v}_1 \cdot \left( \nabla \xi_1 \otimes \nabla \xi_1 + \nabla \xi_2\otimes \nabla \xi_2  + \nabla \eta \otimes \nabla \eta \right)  \cdot \bold{v}_3\right) + \mathcal{O}(\varepsilon^3)
\end{align}

For the following term, we need the expansion of $\bb^3_{i-1j}$. For it, we can write
\begin{equation}
 \bb^3_{i-1j} = 
  \bq^1_{ij}-\bq^2_{i-1j} =
  (\bq^1_{ij}-\bq^2_{ij})-(\bq^2_{i-1j}-\bq^2_{ij})=\bb^1_{ij}-(\bq^2_{i-1j}-\bq^2_{ij}). 
  \label{ede3.1}
\end{equation}

\begin{align}
\bq^2_{i-1j}-\bq^2_{ij} 
&=
  \left({\boldsymbol\chi}^2_{i-1j}
  +
  \varepsilon\boldsymbol \xi({\boldsymbol\chi}^2_{i-1j}),
  \varepsilon
  +
  \varepsilon\eta({\boldsymbol\chi}^2_{i-1j})\right)  -
 \left(
 {\boldsymbol\chi}^2_{ij}
  +
  \varepsilon\boldsymbol\xi({\boldsymbol\chi}^2_{ij}),
  \varepsilon
  +
  \varepsilon\eta({\boldsymbol\chi}^2_{ij})  
  \right)  \nonumber\\ 
& = \left({\boldsymbol\chi}^2_{i-1j} - {\boldsymbol\chi}^2_{ij} 
 + \varepsilon \left( \nabla \boldsymbol{\xi} \cdot \left({\boldsymbol\chi}^2_{i-1j} - {\boldsymbol\chi}^2_{ij}\right)
    + \frac{1}{2} \left({\boldsymbol\chi}^2_{i-1j} - {\boldsymbol\chi}^2_{ij}\right)\cdot H_{\boldsymbol \xi}  \left({\boldsymbol\chi}^2_{i-1j} - {\boldsymbol\chi}^2_{ij}\right) 
 + \mbox{h.o.t.} \right), \right.  \nonumber \\
& \left. \qquad \qquad   \qquad \qquad   \qquad \qquad \varepsilon \left (\nabla \eta \cdot \left({\boldsymbol\chi}^2_{i-1j} - {\boldsymbol\chi}^2_{ij}\right) + \frac{1}{2}\left({\boldsymbol\chi}^2_{i-1j} - {\boldsymbol\chi}^2_{ij}\right) \cdot H_\eta \left({\boldsymbol\chi}^2_{i-1j} - {\boldsymbol\chi}^2_{ij}\right)  + \mbox{h.o.t.} \right) \right) \nonumber\\ 
& = \left(-\varepsilon \delta_2 \ba_1
 + \varepsilon^2 \delta_2 \left( -\nabla \boldsymbol{\xi} \cdot \ba_1
    + \frac{1}{2} \varepsilon \delta_2 \ba_1 \cdot H_{\boldsymbol \xi}  \ba_1
 + \mbox{h.o.t.} \right), \right.  \nonumber \\
& \left. \qquad \qquad   \qquad \qquad   \qquad \qquad \varepsilon^2 \delta_2 \left (-\nabla \eta \cdot \ba_1+ \varepsilon \delta_2\frac{1}{2} \ba_1\cdot H_\eta \ba_1  + \mbox{h.o.t.} \right) \right)
 \label{ede12}
\end{align}

\begin{align}
 \bb^3_{i-1j} & = \varepsilon \delta_2 \left(v_{11}+1 + \varepsilon \nabla \xi_1 \cdot (\bold{v}_1+\ba_1)
    + \varepsilon^2 \delta_2 \frac{1}{2}\left( \bold{v}_1H_{\xi_1}  \bold{v}_1 - \ba_1H_{\xi_1}  \ba_1\right)
 +  \mathcal{O}(\varepsilon^3),\, \right.\nonumber \\
& \qquad \qquad v_{12} + \varepsilon \nabla \xi_2 \cdot (\bold{v}_1 + \ba_1)
    + \varepsilon^2 \delta_2 \frac{1}{2}\left(\bold{v}_1H_{\xi_2}  \bold{v}_1- \ba_1H_{\xi_2}\ba_1\right) +  \mathcal{O}(\varepsilon^3),\, \nonumber \\
& \left. \qquad \qquad
  \varepsilon \nabla \eta \cdot (\bold{v}_1+\ba_1)+\varepsilon^2 \delta_2 \frac{1}{2} \left(\bold{v}_1 H_\eta \bold{v}_1 - \ba_1H_{\eta}  \ba_1\right)  +  \mathcal{O}(\varepsilon^3)\right).
\end{align}
But because $\bold{v}_1 + \ba_1 = \bold{v}_3$, we have that

\begin{align}
 \bb^3_{i-1j} & = \varepsilon \delta_2 \left(v_{31} + \varepsilon \nabla \xi_1 \cdot \bold{v}_3    + \varepsilon^2 \delta_2 \frac{1}{2}\left( \bold{v}_1H_{\xi_1}  \bold{v}_1 - \ba_1H_{\xi_1}  \ba_1\right)
 +  \mathcal{O}(\varepsilon^3),\, \right.\nonumber \\
& \qquad \qquad v_{32} + \varepsilon \nabla \xi_2 \cdot \bold{v}_3 
    + \varepsilon^2 \delta_2 \frac{1}{2}\left(\bold{v}_1H_{\xi_2}  \bold{v}_1- \ba_1H_{\xi_2}\ba_1\right) +  \mathcal{O}(\varepsilon^3),\, \nonumber \\
& \left. \qquad \qquad
  \varepsilon \nabla \eta \cdot \bold{v}_3+\varepsilon^2 \delta_2 \frac{1}{2} \left(\bold{v}_1 H_\eta \bold{v}_1 - \ba_1H_{\eta}  \ba_1\right)  +  \mathcal{O}(\varepsilon^3)\right).
\end{align}

Then, 
\begin{align}
\left( \frac{ \bb^3_{i-1j}\cdot\bb^1_{ij}}{\|\bb^3_{i-1j}\| \|\bb^1_{ij}\|} +\frac{1}{2} \right)^2 & =  9\varepsilon^2 \left( \bold{v}_3 \cdot \nabla \boldsymbol\xi \bold{v}_1+ \bold{v}_1\cdot \nabla \boldsymbol\xi \bold{v}_3 \right)^2 + \nonumber \\ & 18\varepsilon^3 \left( \bold{v}_3 \cdot \nabla \boldsymbol\xi \bold{v}_1+ \bold{v}_1\cdot \nabla \boldsymbol\xi \bold{v}_3 \right) \left( \bold{v}_1 \cdot \left( \nabla \xi_1 \otimes \nabla \xi_1 + \nabla \xi_2\otimes \nabla \xi_2  + \nabla \eta \otimes \nabla \eta \right)  \cdot \bold{v}_3\right) + \mathcal{O}(\varepsilon^3)
\label{ede18}
\end{align}

We also need 
\begin{align}
\bb^2_{ij-1} & =  \bq^1_{ij}-\bq^2_{ij-1} =
  (\bq^1_{ij}-\bq^2_{ij})-(\bq^2_{ij-1}-\bq^2_{ij})=\bb^1_{ij}-(\bq^2_{ij-1}-\bq^2_{ij}) \nonumber \\
& = \varepsilon \delta_2 \left(v_{11}+\frac{1}{2} + \varepsilon \nabla \xi_1 \cdot (\bold{v}_1+\ba_2)
    + \varepsilon^2 \delta_2 \frac{1}{2}\left( \bold{v}_1H_{\xi_1}  \bold{v}_1  - \ba_2H_{\xi_1}  \ba_2\right)
 +  \mathcal{O}(\varepsilon^3),\, \right.\nonumber \\
& \qquad \qquad v_{12} + \frac{\sqrt{3}}{2} + \varepsilon \nabla \xi_2 \cdot (\bold{v}_1 + \ba_2)
    + \varepsilon^2 \delta_2 \frac{1}{2}\left(\bold{v}_1H_{\xi_2}  \bold{v}_1 - \ba_2H_{\xi_2}\ba_2\right) +  \mathcal{O}(\varepsilon^3),\, \nonumber \\
& \left. \qquad \qquad
  \varepsilon \nabla \eta \cdot  (\bold{v}_1 + \ba_2)+\varepsilon^2 \delta_2 \frac{1}{2} \left(\bold{v}_1 H_\eta \bold{v}_1 - \ba_2H_{\eta}  \ba_2\right)  +  \mathcal{O}(\varepsilon^3)\right). 
\end{align}

But because $\bold{v}_1 + \ba_2 = \bold{v}_2$, we have that

\begin{align}
 \bb^2_{ij-1} & = \varepsilon \delta_2 \left(v_{21} + \varepsilon \nabla \xi_1 \cdot \bold{v}_2    + \varepsilon^2 \delta_2 \frac{1}{2}\left( \bold{v}_1H_{\xi_1}  \bold{v}_1 - \ba_2H_{\xi_1}  \ba_2\right)
 +  \mathcal{O}(\varepsilon^3),\, \right.\nonumber \\
& \qquad \qquad v_{22} + \varepsilon \nabla \xi_2 \cdot \bold{v}_2 
    + \varepsilon^2 \delta_2 \frac{1}{2}\left(\bold{v}_1H_{\xi_2}  \bold{v}_1- \ba_2H_{\xi_2}\ba_2\right) +  \mathcal{O}(\varepsilon^3),\, \nonumber \\
& \left. \qquad \qquad
  \varepsilon \nabla \eta \cdot \bold{v}_2+\varepsilon^2 \delta_2 \frac{1}{2} \left(\bold{v}_1 H_\eta \bold{v}_1 - \ba_2H_{\eta}  \ba_2\right)  +  \mathcal{O}(\varepsilon^3)\right).
\end{align}

\begin{align}
\left(\frac{ \bb^2_{ij-1}\cdot\bb^1_{ij}}{\|\bb^2_{ij-1}\| \|\bb^1_{ij}\|} +\frac{1}{2} \right)^2&= 9\varepsilon^2 \left( \bold{v}_2 \cdot \nabla \boldsymbol\xi \bold{v}_1+ \bold{v}_1\cdot \nabla \boldsymbol\xi \bold{v}_2 \right)^2 + \nonumber \\ & 18\varepsilon^3 \left( \bold{v}_2 \cdot \nabla \boldsymbol\xi \bold{v}_1+ \bold{v}_1\cdot \nabla \boldsymbol\xi \bold{v}_2 \right) \left( \bold{v}_1 \cdot \left( \nabla \xi_1 \otimes \nabla \xi_1 + \nabla \xi_2\otimes \nabla \xi_2  + \nabla \eta \otimes \nabla \eta \right)  \cdot \bold{v}_2\right) + \mathcal{O}(\varepsilon^3)
\label{ede21}
\end{align}

\begin{align}
\left(\frac{ \bb^3_{i-1j}\cdot\bb^2_{ij-1}}{\|\bb^3_{i-1j}\| \|\bb^2_{ij-1}\|} +\frac{1}{2} \right)^2&=  9\varepsilon^2 \left( \bold{v}_2 \cdot \nabla \boldsymbol\xi \bold{v}_3+ \bold{v}_3\cdot \nabla \boldsymbol\xi \bold{v}_2 \right)^2 + \nonumber \\ & 18\varepsilon^3 \left( \bold{v}_2 \cdot \nabla \boldsymbol\xi \bold{v}_3+ \bold{v}_3\cdot \nabla \boldsymbol\xi \bold{v}_2 \right) \left( \bold{v}_3 \cdot \left( \nabla \xi_1 \otimes \nabla \xi_1 + \nabla \xi_2\otimes \nabla \xi_2  + \nabla \eta \otimes \nabla \eta \right)  \cdot \bold{v}_2\right) + \mathcal{O}(\varepsilon^3)
\label{ede22}
\end{align}

Combining \eqref{ede2.6} with \eqref{ede7.11}, \eqref{ede7.12}, \eqref{ede7.13}, \eqref{ede18}, \eqref{ede21}, and \eqref{ede22}, we arrive at

\small
\begin{align}
  \mathcal E_t[\boldsymbol\xi,\eta] &\sim 
 \sum_{i,j=1}^{N_2} \frac{48 k_t}{\omega}\left\{
 \varepsilon^3 \left( \left(\frac{\bold{v}_1 \cdot \nabla \boldsymbol\xi \bold{v}_2 + \bold{v}_2\cdot \nabla \boldsymbol\xi \bold{v}_1}{2}\right)^2 +  \left(\frac{\bold{v}_2 \cdot \nabla \boldsymbol\xi \bold{v}_3+ \bold{v}_3\cdot \nabla \boldsymbol\xi \bold{v}_2}{2}\right)^2+ \left(\frac{\bold{v}_1 \cdot \nabla \boldsymbol\xi \bold{v}_3 + \bold{v}_3\cdot \nabla \boldsymbol\xi \bold{v}_1}{2}\right)^2\right) +\right. \nonumber \\
 & \qquad \qquad  \qquad \varepsilon^4 \left[\left(\frac{\bold{v}_1 \cdot \nabla \boldsymbol\xi \bold{v}_2+ \bold{v}_2\cdot \nabla \boldsymbol\xi \bold{v}_1}{2} \right) \left( \bold{v}_1 \cdot \left( \nabla \xi_1 \otimes \nabla \xi_1 + \nabla \xi_2\otimes \nabla \xi_2  + \nabla \eta \otimes \nabla \eta \right)  \cdot \bold{v}_2\right) \right. \nonumber \\
& \qquad \qquad  \qquad + \left(\frac{\bold{v}_2 \cdot \nabla \boldsymbol\xi \bold{v}_3+ \bold{v}_3\cdot \nabla \boldsymbol\xi \bold{v}_2}{2} \right) \left( \bold{v}_3 \cdot \left( \nabla \xi_1 \otimes \nabla \xi_1 + \nabla \xi_2\otimes \nabla \xi_2  + \nabla \eta \otimes \nabla \eta \right)  \cdot \bold{v}_2\right) \nonumber \\
 & \left. \left. \qquad \qquad  \qquad  + \left( \frac{\bold{v}_1 \cdot \nabla \boldsymbol\xi \bold{v}_3+ \bold{v}_3\cdot \nabla \boldsymbol\xi \bold{v}_1 }{2}\right) \left( \bold{v}_3 \cdot \left( \nabla \xi_1 \otimes \nabla \xi_1 + \nabla \xi_2\otimes \nabla \xi_2  + \nabla \eta \otimes \nabla \eta \right)  \cdot \bold{v}_1\right) \right] \right\}
  \label{ede26}
\end{align}
\normalsize
Again, to transform this sum to an integral, we use that the area of the small cells have dimension $\frac{\sqrt{3}}{2}\delta_2^2 \varepsilon^2$, and define $\gamma_t = \frac{4 \sqrt{3} k_t}{\omega \delta_2^2}$.

\subsubsection{Dihedral Springs}

We expand the third term in \eqref{e_dih} in $\varepsilon$.  To do
this, we need the expansions of $\bb^3_{i-1j+1}, \bb^1_{ij+1}, \bb^1_{i+1j}$ and $\bb^1_{i+1j+1}$.  First, we write
\begin{align}
  \bb^3_{i-1j+1}
  &=
  \bq^1_{ij+1}-\bq^2_{i-1j+1} =
  (\bq^1_{ij+1}-\bq^2_{ij}) - (\bq^2_{i-1j+1}-\bq^2_{ij}) = \bb^2_{ij} - (\bq^2_{i-1j+1}-\bq^2_{ij}) . 
  \label{ede30}
\end{align}
The first term in \eqref{ede30} is $\bb^2_{ij}$.  For the second, we have 
\begin{align}
\bq^2_{i-1j+1}-\bq^2_{ij}
&=
  \left({\boldsymbol\chi}^2_{i-1j+1}
  +
  \varepsilon\boldsymbol \xi({\boldsymbol\chi}^2_{i-1j+1}),
  \varepsilon
  +
  \varepsilon\eta({\boldsymbol\chi}^2_{i-1j+1})\right)  -
 \left(
 {\boldsymbol\chi}^2_{ij}
  +
  \varepsilon\boldsymbol\xi({\boldsymbol\chi}^2_{ij}),
  \varepsilon
  +
  \varepsilon\eta({\boldsymbol\chi}^2_{ij})  
  \right)  \nonumber\\ 
& = \left({\boldsymbol\chi}^2_{i-1j+1} - {\boldsymbol\chi}^2_{ij} 
 + \varepsilon \left( \nabla \boldsymbol{\xi} \cdot \left({\boldsymbol\chi}^2_{i-1j+1} - {\boldsymbol\chi}^2_{ij}\right)
    + \frac{1}{2} \left({\boldsymbol\chi}^2_{i-1j+1} - {\boldsymbol\chi}^2_{ij}\right) H_{\boldsymbol \xi}  \left({\boldsymbol\chi}^2_{i-1j+1} - {\boldsymbol\chi}^2_{ij}\right) 
 + \mbox{h.o.t.} \right), \right.  \nonumber \\
& \left. \qquad \qquad    \qquad \qquad \varepsilon \left (\nabla \eta \cdot \left({\boldsymbol\chi}^2_{i-1j+1} - {\boldsymbol\chi}^2_{ij}\right) + \frac{1}{2}\left({\boldsymbol\chi}^2_{i-1j+1} - {\boldsymbol\chi}^2_{ij}\right) H_\eta \left({\boldsymbol\chi}^2_{i-1j+1} - {\boldsymbol\chi}^2_{ij}\right)  + \mbox{h.o.t.} \right) \right) \nonumber\\ 
& = \left(\varepsilon \delta_2 (\ba_2 - \ba_1)
 + \varepsilon^2 \delta_2 \left( \nabla \boldsymbol{\xi} \cdot (\ba_2 - \ba_1)
    + \frac{1}{2} \varepsilon \delta_2 \ba_1 H_{\boldsymbol \xi} (\ba_2 - \ba_1)
 + \mbox{h.o.t.} \right), \right.  \nonumber \\
& \left. \qquad \qquad   \qquad \qquad   \qquad \qquad \varepsilon^2 \delta_2 \left (\nabla \eta \cdot (\ba_2 - \ba_1)+ \varepsilon \delta_2\frac{1}{2} (\ba_2 - \ba_1) H_\eta (\ba_2 - \ba_1)  + \mbox{h.o.t.} \right) \right)
  \label{ede31}
\end{align}

Because $\bold{v}_2 + \ba_1 - \ba_2 = \bold{v}_3$, we have that

\begin{align}
 \bb^3_{i-1j+1} & = \varepsilon \delta_2 \left(v_{31} + \varepsilon \nabla \xi_1 \cdot \bold{v}_3    + \varepsilon^2 \delta_2 \frac{1}{2}\left( \bold{v}_2H_{\xi_1}  \bold{v}_2 - (\ba_2-\ba_1)H_{\xi_1} (\ba_2-\ba_1)\right)
 +  \mathcal{O}(\varepsilon^3),\, \right.\nonumber \\
& \qquad \qquad v_{32} + \varepsilon \nabla \xi_2 \cdot \bold{v}_3 
    + \varepsilon^2 \delta_2 \frac{1}{2}\left(\bold{v}_1H_{\xi_2}  \bold{v}_2- (\ba_2-\ba_1)H_{\xi_2}(\ba_2-\ba_1)\right) +  \mathcal{O}(\varepsilon^3),\, \nonumber \\
& \left. \qquad \qquad
  \varepsilon \nabla \eta \cdot \bold{v}_3+\varepsilon^2 \delta_2 \frac{1}{2} \left(\bold{v}_2 H_\eta \bold{v}_2 - (\ba_2-\ba_1)H_{\eta}  (\ba_2-\ba_1)\right)  +  \mathcal{O}(\varepsilon^3)\right).
\end{align}

Now, we expand $\bb^1_{ij+1}$
\begin{align}
  \bb^1_{ij+1}
  &=
  \bq^1_{ij+1}-\bq^2_{ij+1} =
  (\bq^1_{ij+1}-\bq^2_{ij}) - (\bq^2_{ij+1}-\bq^2_{ij}) = \bb^2_{ij} - (\bq^2_{ij+1}-\bq^2_{ij}) . 
  \label{ede32}
\end{align}
The first term in \eqref{ede32} is $\bb^2_{ij}$.  For the second, we have 
\begin{align}
\bq^2_{ij+1}-\bq^2_{ij}
&=
  \left({\boldsymbol\chi}^2_{ij+1}
  +
  \varepsilon\boldsymbol \xi({\boldsymbol\chi}^2_{ij+1}),
  \varepsilon
  +
  \varepsilon\eta({\boldsymbol\chi}^2_{ij+1})\right)  -
 \left(
 {\boldsymbol\chi}^2_{ij}
  +
  \varepsilon\boldsymbol\xi({\boldsymbol\chi}^2_{ij}),
  \varepsilon
  +
  \varepsilon\eta({\boldsymbol\chi}^2_{ij})  
  \right)  \nonumber\\ 
& = \left({\boldsymbol\chi}^2_{ij+1} - {\boldsymbol\chi}^2_{ij} 
 + \varepsilon \left( \nabla \boldsymbol{\xi} \cdot \left({\boldsymbol\chi}^2_{ij+1} - {\boldsymbol\chi}^2_{ij}\right)
    + \frac{1}{2} \left({\boldsymbol\chi}^2_{ij+1} - {\boldsymbol\chi}^2_{ij}\right) H_{\boldsymbol \xi}  \left({\boldsymbol\chi}^2_{ij+1} - {\boldsymbol\chi}^2_{ij}\right) 
 + \mbox{h.o.t.} \right), \right.  \nonumber \\
& \left. \qquad \qquad    \qquad \qquad \varepsilon \left (\nabla \eta \cdot \left({\boldsymbol\chi}^2_{ij+1} - {\boldsymbol\chi}^2_{ij}\right) + \frac{1}{2}\left({\boldsymbol\chi}^2_{ij+1} - {\boldsymbol\chi}^2_{ij}\right) H_\eta \left({\boldsymbol\chi}^2_{ij+1} - {\boldsymbol\chi}^2_{ij}\right)  + \mbox{h.o.t.} \right) \right) \nonumber\\ 
& = \left(\varepsilon \delta_2 \ba_2 
 + \varepsilon^2 \delta_2 \left( \nabla \boldsymbol{\xi} \cdot \ba_2
    + \frac{1}{2} \varepsilon \delta_2 \ba_1 H_{\boldsymbol \xi} \ba_2 
 + \mbox{h.o.t.} \right), \right.  \nonumber \\
& \left. \qquad \qquad   \qquad \qquad   \qquad \qquad \varepsilon^2 \delta_2 \left (\nabla \eta \cdot \ba_2 + \varepsilon \delta_2\frac{1}{2} \ba_2 H_\eta \ba_2  + \mbox{h.o.t.} \right) \right)
  \label{ede33}
\end{align}

Because $\bold{v}_2 - \ba_2 = \bold{v}_1$, we have that

\begin{align}
 \bb^1_{ij+1} & = \varepsilon \delta_2 \left(v_{11} + \varepsilon \nabla \xi_1 \cdot \bold{v}_1    + \varepsilon^2 \delta_2 \frac{1}{2}\left( \bold{v}_2H_{\xi_1}  \bold{v}_2 - \ba_2H_{\xi_1} \ba_2\right)
 +  \mathcal{O}(\varepsilon^3),\, \right.\nonumber \\
& \qquad \qquad v_{12} + \varepsilon \nabla \xi_2 \cdot \bold{v}_1
    + \varepsilon^2 \delta_2 \frac{1}{2}\left(\bold{v}_2H_{\xi_2}  \bold{v}_2- \ba_2H_{\xi_2}\ba_2\right) +  \mathcal{O}(\varepsilon^3),\, \nonumber \\
& \left. \qquad \qquad
  \varepsilon \nabla \eta \cdot \bold{v}_1+\varepsilon^2 \delta_2 \frac{1}{2} \left(\bold{v}_2H_\eta \bold{v}_2 - \ba_2H_{\eta}  \ba_2\right)  +  \mathcal{O}(\varepsilon^3)\right).
\end{align}

Next, we expand $\bb^1_{i+1j}$
\begin{align}
  \bb^1_{i+1j}
  &=
  \bq^1_{i+1j}-\bq^2_{i+1j} =
  (\bq^1_{i+1j}-\bq^2_{ij}) - (\bq^2_{i+1j}-\bq^2_{ij}) = \bb^3_{ij} - (\bq^2_{i+1j}-\bq^2_{ij}). 
  \label{ede34}
\end{align}
For the second, we have 
\begin{align}
\bq^2_{i+1j}-\bq^2_{ij}
&=
  \left({\boldsymbol\chi}^2_{i+1j}
  +
  \varepsilon\boldsymbol \xi({\boldsymbol\chi}^2_{i+1j}),
  \varepsilon
  +
  \varepsilon\eta({\boldsymbol\chi}^2_{i+1j})\right)  -
 \left(
 {\boldsymbol\chi}^2_{ij}
  +
  \varepsilon\boldsymbol\xi({\boldsymbol\chi}^2_{ij}),
  \varepsilon
  +
  \varepsilon\eta({\boldsymbol\chi}^2_{ij})  
  \right)  \nonumber\\ 
& = \left({\boldsymbol\chi}^2_{i+1j} - {\boldsymbol\chi}^2_{ij} 
 + \varepsilon \left( \nabla \boldsymbol{\xi} \cdot \left({\boldsymbol\chi}^2_{i+1j} - {\boldsymbol\chi}^2_{ij}\right)
    + \frac{1}{2} \left({\boldsymbol\chi}^2_{i+1j} - {\boldsymbol\chi}^2_{ij}\right) H_{\boldsymbol \xi}  \left({\boldsymbol\chi}^2_{i+1j} - {\boldsymbol\chi}^2_{ij}\right) 
 + \mbox{h.o.t.} \right), \right.  \nonumber \\
& \left. \qquad \qquad    \qquad \qquad \varepsilon \left (\nabla \eta \cdot \left({\boldsymbol\chi}^2_{i+1j} - {\boldsymbol\chi}^2_{ij}\right) + \frac{1}{2}\left({\boldsymbol\chi}^2_{i+1j} - {\boldsymbol\chi}^2_{ij}\right) H_\eta \left({\boldsymbol\chi}^2_{i+1j} - {\boldsymbol\chi}^2_{ij}\right)  + \mbox{h.o.t.} \right) \right) \nonumber\\ 
& = \left(\varepsilon \delta_2 \ba_1 
 + \varepsilon^2 \delta_2 \left( \nabla \boldsymbol{\xi} \cdot \ba_1
    + \frac{1}{2} \varepsilon \delta_2 \ba_1 H_{\boldsymbol \xi} \ba_1
 + \mbox{h.o.t.} \right), \right.  \nonumber \\
& \left. \qquad \qquad   \qquad \qquad   \qquad \qquad \varepsilon^2 \delta_2 \left (\nabla \eta \cdot \ba_1 + \varepsilon \delta_2\frac{1}{2} \ba_1 H_\eta \ba_1  + \mbox{h.o.t.} \right) \right)
  \label{ede35}
\end{align}

Because $\bold{v}_3 - \ba_1 = \bold{v}_1$, we have that

\begin{align}
 \bb^1_{ij+1} & = \varepsilon \delta_2 \left(v_{11} + \varepsilon \nabla \xi_1 \cdot \bold{v}_1    + \varepsilon^2 \delta_2 \frac{1}{2}\left( \bold{v}_3H_{\xi_1}  \bold{v}_3 - \ba_1H_{\xi_1} \ba_1\right)
 +  \mathcal{O}(\varepsilon^3),\, \right.\nonumber \\
& \qquad \qquad v_{12} + \varepsilon \nabla \xi_2 \cdot \bold{v}_1
    + \varepsilon^2 \delta_2 \frac{1}{2}\left(\bold{v}_3H_{\xi_2}  \bold{v}_3- \ba_1H_{\xi_2}\ba_1\right) +  \mathcal{O}(\varepsilon^3),\, \nonumber \\
& \left. \qquad \qquad
  \varepsilon \nabla \eta \cdot \bold{v}_1+\varepsilon^2 \delta_2 \frac{1}{2} \left(\bold{v}_3 H_\eta \bold{v}_3 - \ba_1H_{\eta}  \ba_1\right)  +  \mathcal{O}(\varepsilon^3)\right).
\end{align}

Lastly, we expand $\bb^2_{i+1j-1}$
\begin{align}
  \bb^2_{i+1j-1}
  &=
  \bq^1_{i+1j}-\bq^2_{i+1j-1} =
  (\bq^1_{i+1j}-\bq^2_{ij}) - (\bq^2_{i+1j-1}-\bq^2_{ij}) = \bb^3_{ij}- (\bq^2_{i+1j-1}-\bq^2_{ij}). 
  \label{ede36}
\end{align}
For the first term, we have 
\begin{align}
 \bq^2_{i+1j-1}-\bq^2_{ij}
&=
  \left({\boldsymbol\chi}^1_{i+1j+1}
  +
  \varepsilon\boldsymbol \xi({\boldsymbol\chi}^2_{i+1j-1}),
  \varepsilon
  +
  \varepsilon\eta({\boldsymbol\chi}^2_{i+1j-1})\right)  -
 \left(
 {\boldsymbol\chi}^2_{ij}
  +
  \varepsilon\boldsymbol\xi({\boldsymbol\chi}^2_{ij}),
  \varepsilon
  +
  \varepsilon\eta({\boldsymbol\chi}^2_{ij})  
  \right)  \nonumber\\ 
& = \left({\boldsymbol\chi}^2_{i+1j-1} - {\boldsymbol\chi}^2_{ij} 
 + \varepsilon \left( \nabla \boldsymbol{\xi} \cdot \left({\boldsymbol\chi}^2_{i+1j-1} - {\boldsymbol\chi}^2_{ij}\right)
    + \frac{1}{2} \left({\boldsymbol\chi}^2_{i+1j-1} - {\boldsymbol\chi}^2_{ij}\right) H_{\boldsymbol \xi}  \left({\boldsymbol\chi}^2_{i+1j-1} - {\boldsymbol\chi}^2_{ij}\right) 
 + \mbox{h.o.t.} \right), \right.  \nonumber \\
& \left. \qquad \qquad    \qquad \qquad \varepsilon \left (\nabla \eta \cdot \left({\boldsymbol\chi}^2_{i+1j-1} - {\boldsymbol\chi}^2_{ij}\right) + \frac{1}{2}\left({\boldsymbol\chi}^2_{i+1j-1} - {\boldsymbol\chi}^2_{ij}\right) H_\eta \left({\boldsymbol\chi}^2_{i+1j-1} - {\boldsymbol\chi}^2_{ij}\right)  + \mbox{h.o.t.} \right) \right).
  \label{ede37}
\end{align}
Since
\begin{align}
{\boldsymbol\chi}^2_{i+1j-1} - {\boldsymbol\chi}^2_{ij} & =  \varepsilon\delta_{2}\left(i+1+\frac{2}{3}\right)\ba_1 + \left(j-1+\frac{2}{3}\right)\ba_2 - \varepsilon\delta_{2}\left( \left(i+\frac{2}{3}\right)\ba_1 + \left(j+\frac{2}{3}\right)\ba_2\right) \nonumber \\
& = \varepsilon \delta_2 (\ba_1 - \ba_2), 
\end{align}
then,
\begin{align}
\bq^2_{i+1j-1}-\bq^2_{ij}
& = \left(\varepsilon \delta_2(\ba_1 - \ba_2)
 + \varepsilon^2 \delta_2 \left( \nabla \boldsymbol{\xi} \cdot (\ba_1 - \ba_2)
    + \varepsilon \delta_2 \frac{4}{2} (\ba_1 - \ba_2) H_{\boldsymbol \xi} (\ba_1 - \ba_2)
 + \mbox{h.o.t.} \right), \right.  \nonumber \\
& \left. \qquad \qquad   \qquad \qquad   \qquad \qquad \varepsilon^2 \delta_2 \left (\nabla \eta \cdot (\ba_1 - \ba_2) + \varepsilon \delta_2 \frac{4}{2}(\ba_1 - \ba_2) H_\eta(\ba_1 - \ba_2) + \mbox{h.o.t.} \right) \right).
  \label{ede37bis}
\end{align}
Then, using that $ \bold{v}_3-(\ba_1 - \ba_2) =  \bold{v}_2$ we have that
\begin{align}
 \bb^2_{i+1j-1} & = \varepsilon \delta_2 \left(v_{21} + \varepsilon \nabla \xi_1 \cdot \bold{v}_2  + \varepsilon^2 \delta_2 \frac{1}{2}\left( \bold{v}_3 H_{\xi_1}  \bold{v}_3 - (\ba_1-\ba_2)H_{\xi_1} (\ba_1-\ba_2)\right)
 +  \mathcal{O}(\varepsilon^3), \right.\nonumber \\
& \qquad \qquad v_{22} + \varepsilon \nabla \xi_2 \cdot \bold{v}_2
    + \varepsilon^2 \delta_2 \frac{1}{2} \left(\bold{v}_3 H_{\xi_2}  \bold{v}_3 - (\ba_1-\ba_2) H_{\xi_2}(\ba_1-\ba_2) \right) +  \mathcal{O}(\varepsilon^3),\, \nonumber \\
& \left. \qquad \qquad
  \varepsilon \nabla \eta \cdot \bold{v}_2 + \varepsilon^2 \delta_2 \frac{1}{2} \left(\bold{v}_3 H_\eta \bold{v}_3 - (\ba_1-\ba_2)H_{\eta}  (\ba_1-\ba_2)\right)  +  \mathcal{O}(\varepsilon^3)\right).
\end{align}
Returning to \eqref{e_dih}, we now must expand
\begin{align}
\bb^1_{ij}\times\bb^2_{ij}
&=\varepsilon^2 \delta_2^2
\biggl( \varepsilon  \left( v_{12} \nabla \eta \cdot \bold{v}_2 - v_{22} \nabla \eta \cdot \bold{v}_1 \right) +
\biggr. \biggr. \nonumber \\
& \qquad \qquad \quad
 \varepsilon^2 \biggl[ (\nabla \xi_2 \otimes \nabla \eta) \bold{v}_2 \cdot \bold{v}_1 - (\nabla \xi_2 \otimes \nabla \eta) \bold{v}_1 \cdot \bold{v}_2 +\delta_2 \frac{1}{2} \left(v_{12} \bold{v}_2 H_{\eta} \bold{v}_2 - v_{22} \bold{v}_1 H_{\eta} \bold{v}_1\right)  \biggr] + \mathcal{O}(\varepsilon^3), \nonumber \\
& \qquad \qquad \varepsilon \left(v_{21} \nabla \eta \cdot \bold{v}_1-v_{11} \nabla \eta \cdot \bold{v}_2 \right) + 
\biggr. \biggr. \nonumber \\
&  \qquad \qquad \quad
\varepsilon^2 \biggl[ -(\nabla \xi_1 \otimes \nabla \eta) \bold{v}_2 \cdot \bold{v}_1 + (\nabla \xi_1 \otimes \nabla \eta) \bold{v}_1 \cdot \bold{v}_2 +\delta_2 \frac{1}{2} \left(-v_{11} \bold{v}_2 H_{\eta} \bold{v}_2 + v_{21} \bold{v}_1 H_{\eta} \bold{v}_1\right)  \biggr] + \mathcal{O}(\varepsilon^3), \nonumber \\
&  \qquad \qquad  v_{11}v_{22}- v_{12}v_{21} + \varepsilon \left( v_{11} \nabla \xi_2 \cdot \bold{v}_2  +  v_{22} \nabla \xi_1 \cdot \bold{v}_1  - v_{21} \nabla \xi_2 \cdot \bold{v}_1 - v_{12} \nabla \xi_1 \cdot \bold{v}_2 \right) + 
\biggr. \biggr. \nonumber \\
& \qquad \qquad \quad
\varepsilon^2 \biggl[ (\nabla \xi_1 \otimes \nabla \xi_2) \bold{v}_2 \cdot \bold{v}_1 - (\nabla \xi_1 \otimes \nabla \xi_2) \bold{v}_1 \cdot \bold{v}_2   \biggr. \nonumber \\
& \biggl. \biggl. \qquad \qquad \quad \quad
+\delta_2 \frac{1}{2} \left(v_{11} \bold{v}_2 H_{\xi_2} \bold{v}_2 + v_{22} \bold{v}_1 H_{\xi_1} \bold{v}_1 - v_{21} \bold{v}_1 H_{\xi_2} \bold{v}_1- v_{12} \bold{v}_2 H_{\xi_2} \bold{v}_2\right)  \biggr] + \mathcal{O}(\varepsilon^3) \biggr). 
  \label{ede39}
\end{align}

Next we compute
\begin{align}
  (\bb^1_{ij}&\times\bb^2_{ij})\cdot \bb^2_{ij-1}
  = \nonumber \\
 &
  \varepsilon^{3}\delta_{2}^{3}
\biggl[  \biggl( \varepsilon  \left( v_{12} \nabla \eta \cdot \bold{v}_2 - v_{22} \nabla \eta \cdot \bold{v}_1 \right) +
\biggr. \biggr. \nonumber \\
& \qquad \qquad \quad
 \varepsilon^2 \biggl[ (\nabla \xi_2 \otimes \nabla \eta) \bold{v}_2 \cdot \bold{v}_1 - (\nabla \xi_2 \otimes \nabla \eta) \bold{v}_1 \cdot \bold{v}_2 +\delta_2 \frac{1}{2} \left(v_{12} \bold{v}_2 H_{\eta} \bold{v}_2 - v_{22} \bold{v}_1 H_{\eta} \bold{v}_1\right)  \biggr] + \mathcal{O}(\varepsilon^3) \biggr) \times \nonumber \\
&
\left(v_{21} + \varepsilon \nabla \xi_1 \cdot \bold{v}_2    + \varepsilon^2 \delta_2 \frac{1}{2}\left( \bold{v}_1H_{\xi_1}  \bold{v}_1 - \ba_2H_{\xi_1}  \ba_2\right)  +  \mathcal{O}(\varepsilon^3)\right)+ \nonumber \\
& \qquad \qquad  \left(\varepsilon \left(v_{21} \nabla \eta \cdot \bold{v}_1-v_{11} \nabla \eta \cdot \bold{v}_2 \right) + 
\biggr. \biggr. \right. \nonumber \\
&  \qquad \qquad \quad \left.
\varepsilon^2 \biggl[- (\nabla \xi_1 \otimes \nabla \eta) \bold{v}_2 \cdot \bold{v}_1 + (\nabla \xi_1 \otimes \nabla \eta) \bold{v}_1 \cdot \bold{v}_2 +\delta_2 \frac{1}{2} \left(-v_{11} \bold{v}_2 H_{\eta} \bold{v}_2 + v_{21} \bold{v}_1 H_{\eta} \bold{v}_1\right)  \biggr] + \mathcal{O}(\varepsilon^3)\right) \times \nonumber \\
& \biggl( v_{22} + \varepsilon \nabla \xi_2 \cdot \bold{v}_2 
  + \varepsilon^2 \delta_2 \frac{1}{2}\left(\bold{v}_1H_{\xi_2}  \bold{v}_1- \ba_2H_{\xi_2}\ba_2\right) +  \mathcal{O}(\varepsilon^3) \biggr)+ \nonumber \\
&  \qquad \qquad \left(  v_{11}v_{22}- v_{12}v_{21} + \varepsilon \left( v_{11} \nabla \xi_2 \cdot \bold{v}_2  +  v_{22} \nabla \xi_1 \cdot \bold{v}_1  - v_{21} \nabla \xi_2 \cdot \bold{v}_1 - v_{12} \nabla \xi_1 \cdot \bold{v}_2 \right) + 
\biggr. \biggr. \right. \nonumber \\
& \qquad \qquad \quad
\varepsilon^2 \biggl[ (\nabla \xi_1 \otimes \nabla \xi_2) \bold{v}_2 \cdot \bold{v}_1 - (\nabla \xi_1 \otimes \nabla \xi_2) \bold{v}_1 \cdot \bold{v}_2   \biggr. \nonumber \\
& \biggl. \biggl. \qquad \qquad \quad \quad
+\delta_2 \frac{1}{2} \left(v_{11} \bold{v}_2 H_{\xi_2} \bold{v}_2 + v_{22} \bold{v}_1 H_{\xi_1} \bold{v}_1 - v_{21} \bold{v}_1 H_{\xi_2} \bold{v}_1- v_{12} \bold{v}_2 H_{\xi_2} \bold{v}_2\right)  \biggr] + \mathcal{O}(\varepsilon^3) \biggr) \times \nonumber \\ 
& \biggl. \left( \varepsilon \nabla \eta \cdot \bold{v}_2+\varepsilon^2 \delta_2 \frac{1}{2} \left(\bold{v}_1 H_\eta \bold{v}_1 - \ba_2H_{\eta}  \ba_2\right)  +  \mathcal{O}(\varepsilon^3) \right)
\biggr] \nonumber \\
& = \varepsilon^3\delta_2^3\left[( v_{11}v_{22}- v_{12}v_{21}) \varepsilon^2 \delta_2 \frac{1}{2}\left(\bold{v}_1 H_{\eta} \bold{v}_1 - \bold{v}_2 H_{\eta} \bold{v}_2 - \ba_2 H_{\eta} \ba_2 \right) + \mathcal{O}(\varepsilon^3) \right].
  \label{ede40}
\end{align}
Hence
\begin{equation}
  \left[
    (\bb^1_{ij}\times\bb^2_{ij})\cdot \bb^2_{ij-1}
  \right]^{2}
  = 
  \varepsilon^{6}\delta_{2}^{6}
  \left[  \varepsilon^{4}\delta_{2}^{2} ( v_{11}v_{22}- v_{12}v_{21})^2 \frac{1}{4} \left(\bold{v}_1 H_{\eta} \bold{v}_1 - \bold{v}_2 H_{\eta} \bold{v}_2 - \ba_2 H_{\eta} \ba_2 \right)^2 
   + \mathcal{O}(\varepsilon^{5})
  \right].          
  \label{ede41}
\end{equation}

Using \eqref{ede7}, we have
\begin{align}
  \|\bb^1_{ij}\times\bb^2_{ij}\|^{2}
  &=
  \varepsilon^{4}\delta_{2}^{4}
    (v_{11}v_{22}-v_{12}v_{21})^2 +\mathcal{O}(\varepsilon)=  \varepsilon^{4}\delta_{2}^{4}(det(\bold{v}_1,\bold{v}_2))^2  +\mathcal{O}(\varepsilon). \label{ede42}
\end{align}
and
\begin{equation}
  \|\bb^2_{ij-1}\|^{2}
 =
  \varepsilon^{2}\delta_{2}^{2}
  \left[
    \|\bold{v}_2\|^2 + 2\varepsilon (\nabla \boldsymbol{\xi}\bold{v}_2 \cdot \bold{v}_2 + \mathcal{O}(\varepsilon^{2}))
  \right].  \label{ede43}
\end{equation}

From \eqref{ede42} and \eqref{ede43} one checks that
\begin{align}
  \left[\|\bb^1_{ij}\times\bb^2_{ij}\|^{2}\|\bb^2_{ij-1}\|^{2}\right]^{-1}
  &=
  \left[ \varepsilon^{6}\delta_{2}^{6} (det(\bold{v}_1,\bold{v}_2)^2\|\bold{v}_2\|^2+\mathcal{O}(\varepsilon)) \right]^{-1}
  =
  \varepsilon^{-6}\delta_{2}^{-6} (det(\bold{v}_1,\bold{v}_2)^2\|\bold{v}_2\|^2)^{-1} (1+\mathcal{O}(\varepsilon)).
  \label{ede44}
\end{align}
Finally, combining \eqref{ede9} and \eqref{ede12}, we get the expansion
\begin{equation}
  \frac{{\left((\bb^1_{ij}\times\bb^2_{ij})\cdot\bb^2_{ij-1}\right)}^2}{{\|\bb^1_{ij}\times\bb^2_{ij}\|}^2{\|\bb^2_{ij-1}\|}^2}
  =
\frac{ \delta_{2}^{2}\varepsilon^{4}}{4\|\bold{v}_2\|^2} \left(\bold{v}_1 H_{\eta} \bold{v}_1 - \bold{v}_2 H_{\eta} \bold{v}_2 - \ba_2 H_{\eta} \ba_2 \right)^2
  +
  \mathcal{O}(\varepsilon^{5}).
  \label{ede45}
\end{equation}
So similar computations gives us the following equalities.
\begin{equation}
 \frac{{\left((\bb^1_{ij}\times\bb^2_{ij})\cdot\bb^3_{i-1j}\right)}^2}{{\|\bb^1_{ij}\times\bb^2_{ij}\|}^2{\|\bb^3_{i-1j}\|}^2} 
  =
\frac{ \delta_{2}^{2}\varepsilon^{4}}{4\|\bold{v}_3\|^2} \left(\bold{v}_2 H_{\eta} \bold{v}_2 +2 \bold{v}_1 H_{\eta} \bold{v}_1 - \ba_1 H_{\eta} \ba_1 \right)^2
  +
  \mathcal{O}(\varepsilon^{5}).
  \label{ede46}
\end{equation}
\begin{equation}
  \frac{{\left((\bb^1_{ij}\times\bb^3_{ij})\cdot\bb^2_{ij-1}\right)}^2}{{\|\bb^1_{ij}\times\bb^3_{ij}\|}^2{\|\bb^2_{ij-1}\|}^2}
  =
\frac{ \delta_{2}^{2}\varepsilon^{4}}{4\|\bold{v}_2\|^2} \left(\bold{v}_3 H_{\eta} \bold{v}_3 + 2\bold{v}_1 H_{\eta} \bold{v}_1 - \ba_2 H_{\eta} \ba_2 \right)^2
  +
  \mathcal{O}(\varepsilon^{5}).
  \label{ede47}
\end{equation}

\begin{equation}
  \frac{{\left((\bb^1_{ij}\times\bb^3_{ij})\cdot\bb^3_{i-1j}\right)}^2}{{\|\bb^1_{ij}\times\bb^3_{ij}\|}^2{\|\bb^3_{i-1j}\|}^2}
  =
\frac{ \delta_{2}^{2}\varepsilon^{4}}{4\|\bold{v}_3\|^2} \left(\bold{v}_1 H_{\eta} \bold{v}_1 - \bold{v}_3 H_{\eta} \bold{v}_3 - \ba_1 H_{\eta} \ba_1 \right)^2
  +
  \mathcal{O}(\varepsilon^{5}).
  \label{ede48.3}
\end{equation}

\begin{equation}
 \frac{{\left((\bb^2_{ij}\times\bb^3_{i-1j+1})\cdot\bb^1_{ij}\right)}^2}{{\|\bb^2_{ij}\times\bb^3_{i-1j+1}\|}^2{\|\bb^1_{ij}\|}^2}
  =
\frac{ \delta_{2}^{2}\varepsilon^{4}}{4\|\bold{v}_1\|^2} \left(\bold{v}_1 H_{\eta} \bold{v}_1 +2 \bold{v}_2 H_{\eta} \bold{v}_2 - (\ba_2-\ba_1) H_{\eta} (\ba_2-\ba_1) \right)^2
  +
  \mathcal{O}(\varepsilon^{5}).
  \label{ede49}
\end{equation}

\begin{equation}
  \frac{{\left((\bb^2_{ij}\times\bb^3_{i-1j+1})\cdot\bb^3_{ij}\right)}^2}{{\|\bb^2_{ij}\times\bb^3_{i-1j+1}\|}^2{\|\bb^3_{ij}\|}^2} 
  =
\frac{ \delta_{2}^{2}\varepsilon^{4}}{4\|\bold{v}_3\|^2} \left(\bold{v}_3 H_{\eta} \bold{v}_3 - \bold{v}_2 H_{\eta} \bold{v}_2 + (\ba_2-\ba_1) H_{\eta} (\ba_2-\ba_1) \right)^2
  +
  \mathcal{O}(\varepsilon^{5}).
  \label{ede50}
\end{equation}
 
\begin{equation}
  \frac{{\left((\bb^2_{ij}\times\bb^1_{ij+1})\cdot\bb^1_{ij}\right)}^2}{{\|\bb^2_{ij}\times\bb^1_{ij+1}\|}^2{\|\bb^1_{ij}\|}^2}
  =
\frac{ \delta_{2}^{2}\varepsilon^{4}}{4\|\bold{v}_1\|^2} \left(\bold{v}_2 H_{\eta} \bold{v}_2 - \bold{v}_1 H_{\eta} \bold{v}_1 - \ba_2 H_{\eta} \ba_2 \right)^2
  +
  \mathcal{O}(\varepsilon^{5}).
  \label{ede51}
\end{equation}

\begin{equation}
\frac{{\left((\bb^2_{ij}\times\bb^1_{ij+1})\cdot\bb^3_{ij}\right)}^2}{{\|\bb^2_{ij}\times\bb^1_{ij+1}\|}^2{\|\bb^3_{ij}\|}^2}
  =
\frac{ \delta_{2}^{2}\varepsilon^{4}}{4\|\bold{v}_3\|^2} \left(\bold{v}_3 H_{\eta} \bold{v}_3 + 2 \bold{v}_2 H_{\eta} \bold{v}_2 - \ba_2 H_{\eta} \ba_2 \right)^2
  +
  \mathcal{O}(\varepsilon^{5}).
  \label{ede52}
\end{equation}

\begin{equation}
  \frac{{\left((\bb^3_{ij}\times\bb^2_{ij})\cdot\bb^1_{i+1j}\right)}^2}{{\|\bb^3_{ij}\times\bb^2_{ij}\|}^2{\|\bb^1_{i+1j}\|}^2} 
  =
\frac{ \delta_{2}^{2}\varepsilon^{4}}{4\|\bold{v}_1\|^2} \left(\bold{v}_2 H_{\eta} \bold{v}_2 +2 \bold{v}_3 H_{\eta} \bold{v}_3 - \ba_1 H_{\eta} \ba_1 \right)^2
  +
  \mathcal{O}(\varepsilon^{5}).
  \label{ede53}
\end{equation}

\begin{equation}
\frac{{\left((\bb^3_{ij}\times\bb^2_{ij})\cdot\bb^2_{i+1j-1}\right)}^2}{{\|\bb^3_{ij}\times\bb^2_{ij}\|}^2{\|\bb^2_{i+1j-1}\|}^2}
  =
\frac{ \delta_{2}^{2}\varepsilon^{4}}{4\|\bold{v}_2\|^2} \left(-\bold{v}_2 H_{\eta} \bold{v}_2 + \bold{v}_3 H_{\eta} \bold{v}_3 - (\ba_1-\ba_2) H_{\eta} (\ba_1-\ba_2) \right)^2
  +
  \mathcal{O}(\varepsilon^{5}).
  \label{ede54}
\end{equation}

\begin{equation}
\frac{{\left((\bb^3_{ij}\times\bb^1_{ij})\cdot\bb^1_{i+1j}\right)}^2}{{\|\bb^3_{ij}\times\bb^1_{ij}\|}^2{\|\bb^1_{i+1j}\|}^2}
  =
\frac{ \delta_{2}^{2}\varepsilon^{4}}{4\|\bold{v}_1\|^2} \left(-\bold{v}_1 H_{\eta} \bold{v}_1 + \bold{v}_3 H_{\eta} \bold{v}_3 - \ba_1 H_{\eta} \ba_1 \right)^2
  +
  \mathcal{O}(\varepsilon^{5}).
  \label{ede55}
\end{equation}

\begin{equation}
\frac{{\left((\bb^3_{ij}\times\bb^1_{ij})\cdot\bb^2_{i+1j-1}\right)}^2}{{\|\bb^3_{ij}\times\bb^1_{ij}\|}^2{\|\bb^2_{i+1j-1}\|}^2}
  =
\frac{ \delta_{2}^{2}\varepsilon^{4}}{4\|\bold{v}_2\|^2} \left(\bold{v}_1 H_{\eta} \bold{v}_1 + 2\bold{v}_3 H_{\eta} \bold{v}_3 - (\ba_1-\ba_2) H_{\eta} (\ba_1-\ba_2)\right)^2
  +
  \mathcal{O}(\varepsilon^{5}).
  \label{ede56}
\end{equation}

By combining \eqref{ede44}--\eqref{ede56}, we get
\begin{align}
   \mathcal E_d[\boldsymbol\xi,\eta]&:=
  \sum_{i,j=1}^{N_2} \frac{3 k_d}{4\omega}
  \left[7 \left((\bold{v}_1 H_{\eta} \bold{v}_1)^2  + (\bold{v}_2 H_{\eta} \bold{v}_2)^2 + (\bold{v}_3 H_{\eta} \bold{v}_3)^2 \right) \right. \nonumber\\
&  \quad + 2 \left((\bold{v}_1 H_{\eta} \bold{v}_2)^2 + (\bold{v}_1 H_{\eta} \bold{v}_3)^2 + (\bold{v}_2 H_{\eta} \bold{v}_1)^2 +  (\bold{v}_2 H_{\eta} \bold{v}_3)^2 +  (\bold{v}_3 H_{\eta} \bold{v}_1)^2 + (\bold{v}_3 H_{\eta} \bold{v}_2)^2\right) \nonumber \\
&  \quad+ 2 \left(\bold{v}_1 H_{\eta} \bold{v}_3 \right)\left(\bold{v}_2 H_{\eta} \bold{v}_2\right)
- 2 \left(\bold{v}_2 H_{\eta} \bold{v}_1\right)\left(\bold{v}_2 H_{\eta} \bold{v}_2\right)
- 2 \left(\bold{v}_2 H_{\eta} \bold{v}_2\right)\left(\bold{v}_2 H_{\eta} \bold{v}_3\right) \nonumber \\
&  \quad+ 4  \left(\bold{v}_1 H_{\eta} \bold{v}_3\right)\left(\bold{v}_3 H_{\eta} \bold{v}_1\right)
+ 2  \left(\bold{v}_2 H_{\eta} \bold{v}_2\right)\left(\bold{v}_3 H_{\eta} \bold{v}_1\right) 
- 2  \left(\bold{v}_2 H_{\eta} \bold{v}_2\right)\left(\bold{v}_3 H_{\eta} \bold{v}_2\right)\nonumber \\
&  \quad+ 4  \left(\bold{v}_2 H_{\eta} \bold{v}_3\right)\left(\bold{v}_3 H_{\eta} \bold{v}_2\right) 
-2  \left(\bold{v}_1 H_{\eta} \bold{v}_3\right)\left(\bold{v}_3 H_{\eta} \bold{v}_3\right)
+ 2  \left(\bold{v}_2 H_{\eta} \bold{v}_1\right)\left(\bold{v}_3 H_{\eta} \bold{v}_3\right) \nonumber \\
&  \quad-2  \left(\bold{v}_2 H_{\eta} \bold{v}_2\right)\left(\bold{v}_3 H_{\eta} \bold{v}_3\right)
-2  \left(\bold{v}_2 H_{\eta} \bold{v}_3\right)\left(\bold{v}_3 H_{\eta} \bold{v}_3\right)
-2  \left(\bold{v}_3 H_{\eta} \bold{v}_1\right)\left(\bold{v}_3 H_{\eta} \bold{v}_3\right)\nonumber\\
& \quad -2  \left(\bold{v}_3 H_{\eta} \bold{v}_2\right)\left(\bold{v}_3 H_{\eta} \bold{v}_3\right)
+2 \left(\bold{v}_1 H_{\eta} \bold{v}_2\right)\left( 2\left(\bold{v}_2 H_{\eta} \bold{v}_1\right) - \bold{v}_2 H_{\eta} \bold{v}_2 + \bold{v}_3 H_{\eta} \bold{v}_3  \right) \nonumber \\
& \quad \left. - 2 \bold{v}_1 H_{\eta} \bold{v}_1 \left( \bold{v}_1 H_{\eta} \bold{v}_2 + \bold{v}_1 H_{\eta} \bold{v}_3 + \bold{v}_2 H_{\eta} \bold{v}_1 + \bold{v}_2 H_{\eta} \bold{v}_2 - \bold{v}_2 H_{\eta} \bold{v}_3 + \bold{v}_3 H_{\eta} \bold{v}_1 - \bold{v}_3 H_{\eta} \bold{v}_2 + \bold{v}_3 H_{\eta} \bold{v}_3\right) \right]\delta_{2}^{2}\varepsilon^{5} \nonumber \\
& \qquad +  \mathcal{O}(\varepsilon^{6}) \nonumber \\
& =  \sum_{i,j=1}^{N_2} \frac{3 k_d}{8\omega} \delta_{2}^{2}\varepsilon^{5}
  \left[ 7\eta_{,11}^2+4\eta_{,12}^2 + 4\eta_{,21}^2 + 8\eta_{,12}\eta_{,21} - 2\eta_{,11}\eta_{,22} + 7\eta_{,22}^2 \right] +  \mathcal{O}(\varepsilon^{6}).
 \label{ede48.1}
\end{align}

Assuming the symmetry of $H_\eta$, we get 
\begin{align}
   \mathcal E_d[\boldsymbol\xi,\eta]&:=
  \sum_{i,j=1}^{N_2} \frac{3 k_d}{4\omega}
  \left[7 \left((\bold{v}_1 H_{\eta} \bold{v}_1)^2  + (\bold{v}_2 H_{\eta} \bold{v}_2)^2 + (\bold{v}_3 H_{\eta} \bold{v}_3)^2 \right) \right. \nonumber\\
&  \quad + 8 \left((\bold{v}_1 H_{\eta} \bold{v}_2)^2 + (\bold{v}_1 H_{\eta} \bold{v}_3)^2 + (\bold{v}_2 H_{\eta} \bold{v}_3)^2 \right) \nonumber \\
& \quad+4 \left(\bold{v}_1 H_{\eta} \bold{v}_1\right) \left( -\bold{v}_1 H_{\eta} \bold{v}_2 - \bold{v}_1 H_{\eta} \bold{v}_3 + \bold{v}_2 H_{\eta} \bold{v}_3 - \bold{v}_2 H_{\eta} \bold{v}_2 - \bold{v}_3 H_{\eta} \bold{v}_3\right) \nonumber\\
&  \quad+ 4 \left(\bold{v}_2 H_{\eta} \bold{v}_2\right) \left(- \bold{v}_1 H_{\eta} \bold{v}_2+ \bold{v}_1 H_{\eta} \bold{v}_3 
- \bold{v}_2 H_{\eta} \bold{v}_3\right) \nonumber \\
&  \quad+ 4 \left(\bold{v}_3 H_{\eta} \bold{v}_3\right) \left( \bold{v}_1 H_{\eta} \bold{v}_2 - \bold{v}_1 H_{\eta} \bold{v}_3
- \bold{v}_2 H_{\eta} \bold{v}_3\right) \nonumber \\
& \left.  \quad-2  \left(\bold{v}_2 H_{\eta} \bold{v}_2\right)\left(\bold{v}_3 H_{\eta} \bold{v}_3\right) \right]\delta_{2}^{2}\varepsilon^{5} 
+  \mathcal{O}(\varepsilon^{6}) \nonumber \\
& =  \sum_{i,j=1}^{N_2} \frac{3 k_d}{8\omega}  \delta_{2}^{2}\varepsilon^{5}
  \left[ 7\eta_{,11}^2+16\eta_{,12}^2 - 2\eta_{,11}\eta_{,22} + 7\eta_{,22}^2 \right] +  \mathcal{O}(\varepsilon^{6}).
  \label{ede48.2}
\end{align}
Here, we define $\gamma_d = \frac{\sqrt{3} k_d}{2 \omega}$.
\subsection{Energies in Matrix Forms}
We define the matrix
$$V= \left(
\begin{array}{c}
 \bold{v}_1\\
 \bold{v}_2\\
 \bold{v}_3\\
\end{array}
\right)
= \left(
\begin{array}{cc}
v_{11}   &  v_{12} \\
v_{21}   &  v_{22}\\
v_{31} &  v_{32}\\
\end{array}
\right).$$
Then, 
\begin{align} V\nabla\boldsymbol\xi V^T & =\left(
\begin{array}{c}
 \bold{v}_1\\
 \bold{v}_2\\
 \bold{v}_3\\
\end{array}
\right) \nabla \boldsymbol\xi \left(
\begin{array}{ccc}
 \bold{v}_1^T &
 \bold{v}_2^T&
 \bold{v}_3^T
\end{array}
\right)\\ & = \left(
\begin{array}{c}
 \bold{v}_1\\
 \bold{v}_2\\
 \bold{v}_3\\
\end{array}
\right) \left(
\begin{array}{ccc}
\nabla\boldsymbol\xi \cdot \bold{v}_1 &
\nabla\boldsymbol\xi \cdot \bold{v}_2&
\nabla \boldsymbol\xi \cdot \bold{v}_3 
\end{array} 
\right) \\& = 
\left(
\begin{array}{ccc}
 \bold{v}_1 \cdot \nabla\boldsymbol\xi\cdot \bold{v}_1 &
 \bold{v}_1\cdot\nabla\boldsymbol\xi \cdot\bold{v}_2&
 \bold{v}_1\cdot\nabla \boldsymbol\xi\cdot \bold{v}_3 \\
  \bold{v}_2\cdot \nabla\boldsymbol\xi\cdot \bold{v}_1 &
 \bold{v}_2\cdot\nabla\boldsymbol\xi \cdot\bold{v}_2&
 \bold{v}_2\cdot\nabla \boldsymbol\xi \cdot\bold{v}_3\\
  \bold{v}_3\cdot \nabla\boldsymbol\xi\cdot \bold{v}_1 &
 \bold{v}_3\cdot\nabla\boldsymbol\xi \cdot\bold{v}_2&
 \bold{v}_3\cdot\nabla \boldsymbol\xi \cdot \bold{v}_3
\end{array} 
\right). 
\end{align}

If $A =  V\nabla\boldsymbol\xi V^T$, then 

$$ \frac{A+A^T}{2}=\left(
\begin{array}{ccc}
 \bold{v}_1 \cdot \nabla\boldsymbol\xi\cdot \bold{v}_1 &
 \frac{\bold{v}_1\cdot\nabla\boldsymbol\xi \cdot\bold{v}_2+\bold{v}_2\cdot\nabla\boldsymbol\xi \cdot\bold{v}_1}{2}&
 \frac{\bold{v}_1\cdot\nabla\boldsymbol\xi \cdot\bold{v}_3+\bold{v}_3\cdot\nabla\boldsymbol\xi \cdot\bold{v}_1}{2} \\
 \frac{\bold{v}_1\cdot\nabla\boldsymbol\xi \cdot\bold{v}_2+\bold{v}_2\cdot\nabla\boldsymbol\xi \cdot\bold{v}_1}{2} &
 \bold{v}_2\cdot\nabla\boldsymbol\xi \cdot\bold{v}_2&
  \frac{\bold{v}_2\cdot\nabla\boldsymbol\xi \cdot\bold{v}_3+\bold{v}_3\cdot\nabla\boldsymbol\xi \cdot\bold{v}_1}{2}\\
 \frac{\bold{v}_1\cdot\nabla\boldsymbol\xi \cdot\bold{v}_3+\bold{v}_3\cdot\nabla\boldsymbol\xi \cdot\bold{v}_1}{2} &
 \frac{\bold{v}_2\cdot\nabla\boldsymbol\xi \cdot\bold{v}_3+\bold{v}_3\cdot\nabla\boldsymbol\xi \cdot\bold{v}_2}{2}&
 \bold{v}_3\cdot\nabla \boldsymbol\xi \cdot \bold{v}_3
\end{array} 
\right). $$

 \begin{equation} V (\nabla \eta \otimes \nabla \eta) V^T  = 
\left(
\begin{array}{ccc}
 \bold{v}_1 \cdot  (\nabla \eta \otimes \nabla \eta)\cdot \bold{v}_1 &
 \bold{v}_1\cdot  (\nabla \eta \otimes \nabla \eta) \cdot\bold{v}_2&
 \bold{v}_1\cdot (\nabla \eta \otimes \nabla \eta)\cdot \bold{v}_3 \\
  \bold{v}_2\cdot  (\nabla \eta \otimes \nabla \eta) \cdot \bold{v}_1 &
 \bold{v}_2\cdot  (\nabla \eta \otimes \nabla \eta) \cdot\bold{v}_2&
 \bold{v}_2\cdot (\nabla \eta \otimes \nabla \eta) \cdot\bold{v}_3\\
  \bold{v}_3\cdot  (\nabla \eta \otimes \nabla \eta)\cdot \bold{v}_1 &
 \bold{v}_3\cdot  (\nabla \eta \otimes \nabla \eta) \cdot\bold{v}_2&
 \bold{v}_3\cdot  (\nabla \eta \otimes \nabla \eta) \cdot \bold{v}_3
\end{array} 
\right). 
\end{equation}

\subsection{Euler-Lagrange Equations}
\label{supp_EL}

In here we want to deduce the Euler-Lagrange equations corresponding to our continuum energy 
\begin{align}
	\mathcal F^\eps[\boldsymbol\xi,\eta] &:= 
        \frac{\eps}{2}\int_{D_1}f\left(D\left(V\nabla\boldsymbol\xi V^T \right)+\frac{\eps}{2}V\left(\nabla\eta\otimes \nabla\eta\right)V^T\right)\,d{\boldsymbol\chi}\\
       & \quad\quad +
       \gamma_d\eps^3\int_{D_1} \left[
    7\eta_{,11}^2+16\eta_{,12}^2 - 2\eta_{,11}\eta_{,22} + 7\eta_{,22}^2
    \right]\,d{\boldsymbol\chi}
        +
        \frac{1}{\varepsilon}\int_{D_1} G\left(\boldsymbol\chi,\boldsymbol\xi,\eta\right)\,d\boldsymbol\chi.
\end{align}
Recall that $D(A)=(A+A^T)/2$ for any $A\in M^{3\times 3}$ and $f\colon M^{3\times 3}\to \mathbb{R}$ is defined by
\[f\left(M\right)=\gamma_s\left(m_{11}^2+m_{22}^2 + m_{33}^2\right)+\gamma_t \left(m_{12}^2 + m_{21}^2  + m_{13}^2 + m_{31}^2 + m_{23}^2 + m_{32}^2\right),\]
for any $M=\left(
\begin{array}{ccc}
m_{11}   &  m_{12} & m_{13} \\
m_{21}   &  m_{22} & m_{23}\\
m_{31} &  m_{32} & m_{33}\\
\end{array}
\right).$

Then, we have that 
for any $\nabla f (M) = 2 \left(
\begin{array}{ccc}
\gamma_s m_{11}   &  \gamma_t m_{12} & \gamma_t m_{13} \\
\gamma_t m_{21}   &  \gamma_s m_{22} & \gamma_t m_{23}\\
\gamma_t m_{31} &  \gamma_t m_{32} & \gamma_s m_{33}\\
\end{array}
\right).$

Let $(\bar{\boldsymbol\xi},\bar\eta)$ be the minimizer of the energy $\mathcal F^\eps[\boldsymbol\xi,\eta]$. Let
$\boldsymbol{\xi}^1 = \left(\begin{array}{c} \xi_1 \\ 0 \end{array}\right)$. 
We define 
\begin{equation}
    A(\gamma) = D\left(V\nabla \left(\bar{\boldsymbol \xi} + \gamma \boldsymbol{\xi}^1 \right) V^T \right)+\frac{\eps}{2}V\left(\nabla \bar\eta\otimes \nabla \bar \eta\right)V^T, 
\end{equation}
Then, at $\gamma=0$ we have that 
$A(0) = D\left(V\nabla \bar{\boldsymbol \xi} V^T \right)+\frac{\eps}{2}V\left(\nabla \bar\eta\otimes \nabla \bar \eta\right)V^T$
and because of linearity of $D$ and $\nabla$, we have that 
$A'(0) = D\left(V\nabla \boldsymbol \xi ^1V^T \right)$. So, we can write that 
\begin{equation}
    \frac{\partial}{\partial \gamma}\left\{ \frac{\varepsilon}{2} \int_{D_1} f(A(\gamma)), dx \right\}\Big|_{\gamma=0} = \frac{\varepsilon}{2} \int_{D_1} \nabla f(A(0)):\frac{\partial A (0)}{\partial \gamma}\, dx, 
\end{equation}
where $A:B = \sum_{i}\sum_{j}A_{ij}B_{ij}$. 

Let us rewrite the term inside the integral
\begin{align}
    \nabla f(A(0)):\frac{\partial A (0)}{\partial \gamma} & =  \nabla f (A (0)) : D\left(V\nabla \boldsymbol \xi ^1V^T \right) \nonumber\\
    & = \nabla f (A (0)) : \frac{V\nabla \boldsymbol \xi ^1V^T + V\nabla \boldsymbol \xi^1V^T}{2} \nonumber\\ 
    & =  \frac{1}{2} \left( \nabla f (A (0)) : V\nabla \boldsymbol \xi ^1V^T +
     \nabla f (A (0)) : V \left(\nabla \boldsymbol \xi^1\right)^T V^T\right) \nonumber\\
    & =  \frac{1}{2} \left( \nabla f (A (0)) : V\nabla \boldsymbol \xi ^1V^T +
     \left(\nabla f (A (0))\right)^T : V\nabla \boldsymbol \xi ^1V^T\right) \nonumber\\
    & =  \frac{1}{2} \left( \nabla f (A (0)) : V\nabla \boldsymbol \xi ^1V^T +
     \nabla f (A (0)) : V\nabla \boldsymbol \xi ^1V^T\right) \nonumber\\
    & =  \nabla f (A (0)) : V\nabla \boldsymbol \xi ^1V^T \nonumber\\
    & =  V^T \nabla f (A (0)) : \nabla \boldsymbol \xi ^1V^T \nonumber\\
    & = \nabla f (A (0)) V : V \left(\nabla \boldsymbol \xi ^1\right)^T \nonumber\\
    & = V^T\nabla f (A (0)) V :\left(\nabla \boldsymbol \xi ^1\right)^T \nonumber\\
    & = V^T\nabla f (A (0)) V : \nabla \boldsymbol \xi ^1 \nonumber \\
    & = V^T\nabla f (A (0)) V :  \left(\begin{array}{c} \nabla \xi_1 \\ 0 \end{array}\right) \nonumber\\
    & = R_1( V^T\nabla f (A (0)) V ) \cdot  \nabla \xi_1
\end{align}
Returning to the integral and integrating by parts and assuming Dirichlet boundary conditions, we have that

\begin{equation}
\frac{\varepsilon}{2} \int_{D_1} R_1( V^T\nabla f (A (0)) V ) \cdot  \nabla \xi_1 dx, = - \int_{D_1} \frac{\varepsilon}{2}\mbox{div}\left( R_1( V^T\nabla f (A (0)) V )\right) \cdot  \nabla \xi_1 dx.
\end{equation}

\end{document}